\documentclass[a4paper,12pt,twoside,openany]{article}
\usepackage[tmargin=2.5cm,bmargin=3cm,rmargin=2cm,lmargin=2cm,marginratio=1:1]{geometry} 

\usepackage[pdftex]{graphicx}
\usepackage[french,english]{babel}
\usepackage[T1]{fontenc}
\usepackage{amsmath,amssymb,amsfonts,mathrsfs,upgreek,textgreek}
\usepackage{amsthm,pdfpages,tensor}
\usepackage{longtable}
\usepackage{arydshln}\usepackage{multirow}
\usepackage{braket}
\usepackage{hyperref}
\usepackage{mathtools}
\usepackage{array}
\usepackage{enumitem}
\usepackage{hhline}
\usepackage{graphicx,caption,wrapfig,float}
\usepackage[outercaption]{sidecap}  
\usepackage{footnote}
\usepackage{tablefootnote}
\usepackage[indentafter]{titlesec}
\usepackage[latin1]{inputenc} 
\usepackage[T1]{fontenc} 
\usepackage{lmodern} 

\def\bz{{\bar z}}
\def\bw{{\bar w}}

\def\p{{\partial }}
\def\12{\frac{1}{2}}
\def\32{\frac{3}{2}}

\usepackage{pict2e}
\makeatletter
\newcommand{\loplus}{\mathbin{\mathpalette\dog@lsemi{+}}}
\newcommand{\dog@lsemi}[2]{\dog@semi{#1}{#2}{270,90}}
\newcommand{\dog@semi}[3]{%
  \begingroup
  \sbox\z@{$\m@th#1#2$}%
  \setlength{\unitlength}{\dimexpr\ht\z@+\dp\z@\relax}%
  \makebox[\wd\z@]{\raisebox{-\dp\z@}{%
    \begin{picture}(1,1)
    \linethickness{\variable@rule{#1}}
    \roundcap
    \put(0.5,0.5){\makebox(0,0){\raisebox{\dp\z@}{$\m@th#1#2$}}}
    \put(0.5,0.5){\arc[#3]{0.5}}
    \end{picture}%
  }}%
  \endgroup
}
\newcommand{\variable@rule}[1]{%
  \fontdimen8  
  \ifx#1\displaystyle\textfont3\else
    \ifx#1\textstyle\textfont3\else
      \ifx#1\scriptstyle\scriptfont3\else
        \scriptscriptfont3\relax
  \fi\fi\fi
}
\makeatother
\expandafter\def\expandafter\normalsize\expandafter{%
    \normalsize
    \setlength\abovedisplayskip{10pt}
    \setlength\belowdisplayskip{10pt}
    \setlength\abovedisplayshortskip{10pt}
    \setlength\belowdisplayshortskip{10pt}
}

\usepackage{xcolor}
\definecolor{oldmauve}{rgb}{0.4, 0.19, 0.28}
\definecolor{pansypurple}{rgb}{0.47, 0.09, 0.29}
\definecolor{burgundy}{rgb}{0.5, 0.0, 0.13}
\definecolor{carminepink}{rgb}{0.92, 0.3, 0.26}
\definecolor{blue(pigment)}{rgb}{0.2, 0.2, 0.6}
\definecolor{darkseagreen}{rgb}{0.56, 0.74, 0.56}
\definecolor{darkspringgreen}{rgb}{0.09, 0.45, 0.27}
\definecolor{ceruleanblue}{rgb}{0.16, 0.32, 0.75}

\numberwithin{equation}{section} 
\usepackage{cite} 

\addto\captionsenglish{}

\usepackage{xcolor,colortbl}
\definecolor{green}{rgb}{0.1,0.1,0.1}

\newcommand{\be}{\begin{eqnarray}}
\newcommand{\en}{\end{eqnarray}}
\newcommand{\badat}{\begin{alignedat}}
\newcommand{\eadat}{\end{alignedat}}
\newcommand{\bitm}{\begin{itemize}}
\newcommand{\eitm}{\end{itemize}}

\newcommand{\ep}{\epsilon}
\newcommand{\gr}{\textbf}
\newcommand{\virg}{\hspace{1 mm}, \hspace{8 mm}}

\hypersetup{%
  colorlinks=false,
  linkbordercolor=blue(pigment) 
}

\definecolor{arsenic}{rgb}{0.23, 0.27, 0.29}
\definecolor{bluegray}{rgb}{0.4, 0.6, 0.8}
\usepackage[framemethod=tikz]{mdframed}
\definecolor{mycolor}{rgb}{0.122, 0.435, 0.698}

\newmdenv[innerlinewidth=0.5pt, roundcorner=4pt,linecolor=black,innerleftmargin=6pt,
innerrightmargin=6pt,innertopmargin=6pt,innerbottommargin=6pt]{blackbox}

\newmdenv[innerlinewidth=0.5pt, roundcorner=4pt,linecolor=mycolor,innerleftmargin=6pt,
innerrightmargin=6pt,innertopmargin=6pt,innerbottommargin=6pt]{bluebox}

\newmdenv[innerlinewidth=1pt, roundcorner=4pt,linecolor=bluegray,innerleftmargin=6pt,
innerrightmargin=6pt,innertopmargin=6pt,innerbottommargin=6pt]{arsenicbox}

\newmdenv[innerlinewidth=0.5pt, roundcorner=4pt,linecolor=darkseagreen,innerleftmargin=6pt,
innerrightmargin=6pt,innertopmargin=6pt,innerbottommargin=6pt]{greenbox}

\newmdenv[innerlinewidth=0.5pt, roundcorner=4pt,linecolor=burgundy,innerleftmargin=6pt,
innerrightmargin=6pt,innertopmargin=6pt,innerbottommargin=6pt]{burgundybox}

\newmdenv[innerlinewidth=0.5pt, roundcorner=4pt,linecolor=carminepink,innerleftmargin=6pt,
innerrightmargin=6pt,innertopmargin=6pt,innerbottommargin=6pt]{carminepinkbox}

\numberwithin{equation}{section} 

\begin{document}

\begin{titlepage}
  \thispagestyle{empty}

  \begin{center}  
  
\vspace*{5cm}
  
{\LARGE\textbf{Celestial holography:}}
\vskip0.5cm
{\LARGE\textbf{An asymptotic symmetry perspective}}

\vskip1cm
Laura Donnay\let\thefootnote\relax\footnotetext{\href{mailto:ldonnay@sissa.it}{email: ldonnay@sissa.it}}

\vskip0.5cm

\normalsize
\medskip

\textit{SISSA,
Via Bonomea 265, 34136 Trieste, Italy}

\textit{INFN, Sezione di Trieste,
Via Valerio 2, 34127, Italy \\
\vspace{2mm}
}

\vskip2cm

\begin{abstract}
We review the role that infinite-dimensional symmetries arising at the boundary of asymptotically flat spacetimes play in the context of the celestial holography program. Once recast into the language of conformal field theory, asymptotic symmetries provide key constraints on the sought-for celestial dual to quantum gravity in flat spacetimes.
\end{abstract}

\vskip2cm

\emph{Invited review for Physics Reports}
\end{center}

\end{titlepage}

\tableofcontents 
\section{Introduction}
Amongst the simplest, yet most mysterious formulae in physics is the Bekenstein-Hawking formula  
\cite{Bekenstein:1972tm,Hawking:1974sw}
\begin{equation}\label{BH}
S=\frac{A c^3}{4G \hbar},
\end{equation}
which expresses the fact that black holes have an enormous amount of entropy\footnote{The Boltzmann constant is set to one here.} ($S$) which is proportional to the area of their event horizon ($A$).
In retrospect, the formula \eqref{BH} can be seen as the first hint that gravity might be fundamentally \emph{holographic}, since all the information seems to be stored at the boundary of the black hole, namely on the surface of the event horizon. 

It seems fair to say (see \cite{Armas:2021yut} for a broad account of diverse viewpoints on that matter) that almost everything we understand about the quantum properties of black holes is, in a way or another, related to the holographic correspondence \cite{tHooft:1993dmi,Susskind:1994vu,Maldacena:1997re}. Its best developed and most celebrated implementation requires bulk spacetimes with negative cosmological constant, i.e. asymptotically Anti-de Sitter (AdS) spacetimes; see \cite{Aharony:1999ti} for a review. However, the cosmological constant $\Lambda$ of our Universe was measured to be positive~\cite{SupernovaSearchTeam:1998fmf,SupernovaCosmologyProject:1998vns}, and furthermore, realistic black holes, such as the ones we observe at the centre of galaxies \cite{Akiyama:2019cqa}, do not possess (away from the extremal case) an AdS near-horizon region.  If quantum gravity is intrinsically holographic, there should be a way to extend (at least some of) the spectacular successes of the AdS/CFT correspondence \cite{Maldacena:1997re,Aharony:1999ti} to more realistic scenarios. This report aims to provide a (very partial) account of recent efforts in this direction, with the working assumption that $\Lambda \approx 0$. That is indeed what the celestial holography program is aiming at: establishing a holographic correspondence for quantum gravity in asymptotically \emph{flat} spacetimes.\\

\begin{figure}[h!]
\begin{center}
\includegraphics[scale=0.6,trim = {0 0cm 0cm 0.5cm}]{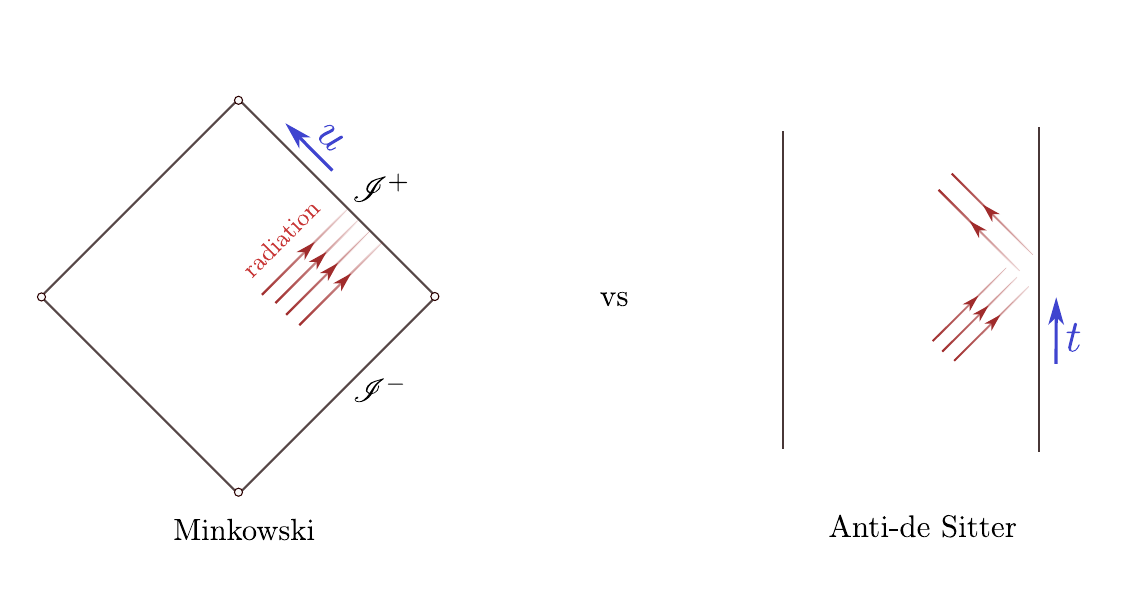}
\captionsetup{width=0.7\linewidth}
\caption{\small{Penrose (conformal) diagram of flat spacetime (on the left) versus the one of Anti-de Sitter (AdS) case. }}
\label{fig:contrast}
\end{center}
\end{figure}

Flat space holography raises novel technical and conceptual puzzles that one is not used to encounter in AdS holography, as soon noticed in the pioneer works from the late 90s-early 2000s \cite{Susskind:1998vk,Polchinski:1999ry,Giddings:1999jq,Arcioni:2003td,Arcioni:2003xx,Mann:2005yr}. To understand why flat space holography is so difficult, it is useful to look at its conformal boundary and to contrast the situation with the AdS case. The boundary of AdS space is a timelike Lorentzian boundary (depicted at the right in Figure \ref{fig:contrast}), which is naturally endowed with a clear notion of time evolution. In contrast, the conformal boundary of flat space, depicted at the left of Figure \ref{fig:contrast}, is described by two null (i.e. lightlike) hypersurfaces, future and past null infinity (denoted $\mathscr I^+$ and $\mathscr I^-$, respectively). QFTs living on such null manifolds should be of a very peculiar nature\footnote{There are known as `Carrollian QFTs' (see e.g. \cite{Donnay:2022wvx,deBoer:2023fnj} and references therein).} and it is not clear what their defining properties actually are. Moreover, the precise way the different pieces ($\mathscr I^\pm$ but also future and past timelike infinities) are connected to each other should be specified. Another difference with the AdS holographic set-up is that the field/operator map in AdS/CFT follows from imposing Dirichlet boundary conditions at the boundary. However, if one wishes to study physically relevant scenarios in flat spacetimes where gravitational radiation escapes throughout the boundary, imposing reflective boundary conditions at null infinity does not seem the best thing to do. We are thus very far from the comfortable situation typically considered in AdS where we `put gravity in a box', and this forces us to address the principle of holography beyond this set-up.
These two key features of, on the one hand, the presence of null boundaries and, on the other hand, the need to deal with leaks of gravitational radiation at infinity constitute two major difficulties that a viable proposal for a flat version of holography has to face.\\

\noindent \textcolor{blue(pigment)}{\gr{From heaven spaces\dots} \,\,} As pointed out by the authors of \cite{Adamo:2021lrv}, the very first origins of flat space holography can be traced back to the works of Newman and Penrose in the seventies. Newman was indeed already attempting to reconstruct the bulk spacetime from `cuts' of $\mathscr I$ \cite{Newman:1976gc}. This led to the construction of `Heaven-space' ($\mathcal H$-space), the four-dimensional complex manifold of asymptotically shear-free null hypersurfaces (the so-called `good cuts'). $\mathcal H$-spaces are solutions to self-dual gravity: they are naturally equipped with a (complexified, holomorphic) metric which satisfies Einstein equations and has (anti-)self-dual Weyl tensor.  In fact, all asymptotically flat (anti-)self-dual solutions can be obtained in this way (see \cite{Hspace}). More generally, by means of Penrose's non-linear graviton construction, one can construct (complex) spacetimes with self-dual curvature from deformations of the complex structure on twistor spaces \cite{Penrose:1976js,Penrose:1976jq}. When one considers specific deformations which are parametrized by the asymptotic shear of a spacetime, one then obtains the so-called `asymptotic twistor spaces'. The nonlinear graviton construction derived from these spaces reproduces Newman's $\mathcal H$-space. Cuts at $\mathscr I$   encode the `bulk' physics and the asymptotic twistor theory is thus intrinsically of a holographic nature.\\

\noindent \textcolor{blue(pigment)}{\gr{\dots to celestial holography} \,\,} The modern resurgence of the question of how to encode the physics of flat spacetimes into a theory living at the boundary came from the realization that some results in the General Relativity (GR) literature \cite{Bondi:1962px,vanderBurg,Sachs:1962zza,Sachs:1962wk,sachs_gravitational_1961} had been somehow totally overlooked \cite{Barnich:2010eb,Strominger:2013jfa}. Bondi, van der Burg, Metzner and Sachs (BMS), the authors of these works themselves, did not realize the importance of their findings. What their work showed is that the symmetry group of asymptotically flat spacetimes in GR is not the Poincar\'e
group (as they expected), but instead an infinite-dimensional extension of it. In the
BMS group, the four spacetime translations are indeed enhanced to a whole function's worth of `supertranslations' that act independently on each point of the celestial sphere (the sphere at infinity). A key realization due to Strominger was that this result, once combined with appropriate matching conditions that relate the action of BMS symmetries at $\mathscr I^+$ with the ones acting at $\mathscr I^-$, implies that those symmetries constrain the gravitational scattering problem \cite{Strominger:2013jfa}. From then on, new research avenues opened, and various works started to explore the deep connections between asymptotic symmetries and the infrared structure not only of gravity but also of gauge theories; see \cite{Strominger:2017zoo} for a review of this rich topic.

In this perspective, asymptotic symmetries have played a novel role in the quantum field theory (QFT) context as their Ward identities turn out to encode universal factorization properties of scattering amplitudes which involve zero-energy particles, the so-called `soft theorems' (see e.g. \cite{Low:1958sn,Burnett-Low,Weinberg:1965nx,Weinberg:1995mt,Gross-Jackiw,Jackiw}). These new connections have been used to enrich both the GR and QFT sides: known soft theorems predicted the existence of new asymptotic symmetries and, in turn, asymptotic symmetry enhancements led to the discovery of new soft theorems (see \cite{He:2014laa,Cachazo:2014fwa,Lysov:2014csa,Kapec:2014opa,Kapec:2015ena,Campiglia:2015yka,Campiglia:2016jdj} for early examples and \cite{Strominger:2017zoo} for further references). In particular, one of the newly discovered (subleading in the energy expansion) soft graviton theorem \cite{Cachazo:2006mq} was used to derive the existence of a soft graviton mode whose insertion into the (tree-level) $\mathcal S$-matrix was shown to obey the Virasoro-Ward identities of a CFT$_2$ stress tensor \cite{Kapec:2016jld}. This particularly interesting finding fueled the idea that there could be a way to extract scattering amplitude elements in terms of  correlators of a certain two-dimensional conformal field theory living on the celestial sphere. To highlight as much as possible the action of conformal transformations, a new basis for scattering amplitudes started to emerge \cite{Pasterski:2016qvg,Pasterski:2017kqt}. In this `conformal basis', massless plane waves are mapped into 
operators on the celestial sphere which carry a conformal dimension (the dual variable of their energy), and scattering amplitudes transform covariantly as two-dimensional conformal correlators. This approach, now known as `celestial holography' \cite{Pasterski:2021raf} (in its bottom-up version),   gradually gained more and more attention as it repeatedly showed its ability to provide a bridge between different research topics and to bring together researchers working at the forefront of conformal field theory, mathematical GR, twistor theory, asymptotic symmetries,  scattering amplitudes and gravitational wave observations (among others).\\

The goal of this report is to review aspects of the celestial holography program that are most closely connected to the question of asymptotic symmetries of flat spacetimes. By design, it should therefore not be considered as a complete overview of the program; we refer to \cite{Raclariu:2021zjz,Pasterski:2021rjz,McLoughlin:2022ljp,Pasterski:2021raf,Pasterski:2023ikd} for other reviews of celestial holography which contain complementary  material. 

The report is organized as follows: in chapter \ref{chap:bms}, we review the critical notion of asymptotically flat spacetimes, as originally proposed by Bondi and collaborators. After a discussion on the various extensions of allowed asymptotic symmetry groups, we focus on the so-called extended BMS symmetries, as the latter have played a prominent role in the context of celestial holography. The chapter ends with a short account of the appearance of BMS symmetries at spacelike and timelike infinities. In chapter \ref{chap:charges}, we present BMS surface charges and fluxes, taking into account recent prescriptions that capture important information about the vacuum structure of the gravitational phase space. It is then shown that a certain soft part of the BMS supermomentum and super angular momentum fluxes define conformal fields of definite weights on the celestial sphere. The second part of the report opens in chapter \ref{chap:celestial} with some of the fundamental aspects of the celestial holography program, including the definitions of celestial amplitudes and (massless) conformal primary wavefunctions. We also comment on the shadow transform and how the celestial holographic dictionary is expected to be implemented in the case of massive particles. We bridge together asymptotic symmetries and celestial CFTs in chapter \ref{chap:last} where it is shown that asymptotic symmetries lead to a tower of currents on the celestial sphere that encode soft graviton modes and can be neatly organized into a very rich algebra.  The report ends in chapter \ref{chap:conclusion} with a list of further reading material.

\newpage
\section{Symmetries of asymptotically flat spacetimes}
\label{chap:bms}
In the early sixties, very little was known about gravitational radiation in Einstein's theory of General Relativity (GR). There was confusion about whether it could exist a rigorous derivation of wavelike solutions to Einstein's field equations which even led to doubts about the very existence of gravitational waves (see e.g. \cite{article} or the review \cite{Frauendiener:2000mk}).  A first important development in this context was due to Trautman \cite{Trautman:1958zdi} (inspired by Pirani and Lichnerowicz's works), who searched for boundary conditions for the gravitational field that were relaxed enough to allow for gravitational radiation from an isolated source, but not to generic so that there was a sense in which one could guarantee uniqueness of the solution given some reasonable initial data.  On his side, Sachs proposed an invariant
formulation of the outgoing radiation condition for gravitational fields and obtained a characteristic fall-off behavior for the Weyl tensor, the so-called `peeling-off' properties \cite{sachs_gravitational_1961}. 

Remarkable improvements were then obtained thanks to the works of Bondi and his collaborators van der Burg and Metzner \cite{Bondi:1962px}, who proposed to tackle this question by a methodical study of null or characteristic hypersurfaces, while previous works tended to focus rather on spacelike surfaces and their associated Cauchy problem. While their first analysis restricted to axisymmetric spacetimes, Sachs soon removed this assumption\cite{Sachs:1962wk}. The novel ingredient was the use of a coordinate system involving a retarded time $u$ and a luminosity distance $r$ which is particularly well adapted to follow null geodesics. Asymptotic conditions were imposed so that the metric approaches flat spacetime at large values of $r$, assuming that the metric functions were analytic functions in $1/r$. Bondi's analysis turns out to greatly facilitate the study of field equations as the latter organize in a neat hierarchical way.  A celebrated result that was obtained is the `Bondi mass loss formula', showing that outgoing gravitational waves carry away positive energy from an isolated system and hence diminish its mass. Their analysis also confirmed that spacetimes which satisfy Einstein's vacuum field equations and the Bondi-Sachs boundary conditions also satisfy Sachs' peeling property. From then on, there was little doubt left about the fact that General Relativity predicted the existence of gravitational waves.

In parallel to these developments, Penrose \cite{Penrose1,Penrose:1965am} made the notion of `infinity' precise by means of his elegant construction of conformal compactification of spacetime, where the metric is rescaled by a conformal factor  which approaches zero asymptotically.  In this process, a boundary is attached to the physical spacetime and in the case of a vanishing cosmological constant, the boundary is a lightlike hypersurface (null infinity $\mathscr I$). If the boundary can be attached to the interior of the rescaled manifold in a regular way,
the spacetime is said to be asymptotically flat. Notice that, despite the deep physical insight underlying these developments, it was not clear for quite a long time whether or not any global (non trivial) solutions with these asymptotic actually existed. Since then,
a vast class of smooth solutions has been constructed, see the review \cite{Friedrich:2017cjg} for more details.

\begin{figure}
\begin{center}
\includegraphics[scale=0.8,trim = {0 2.5cm 0cm 2cm}]{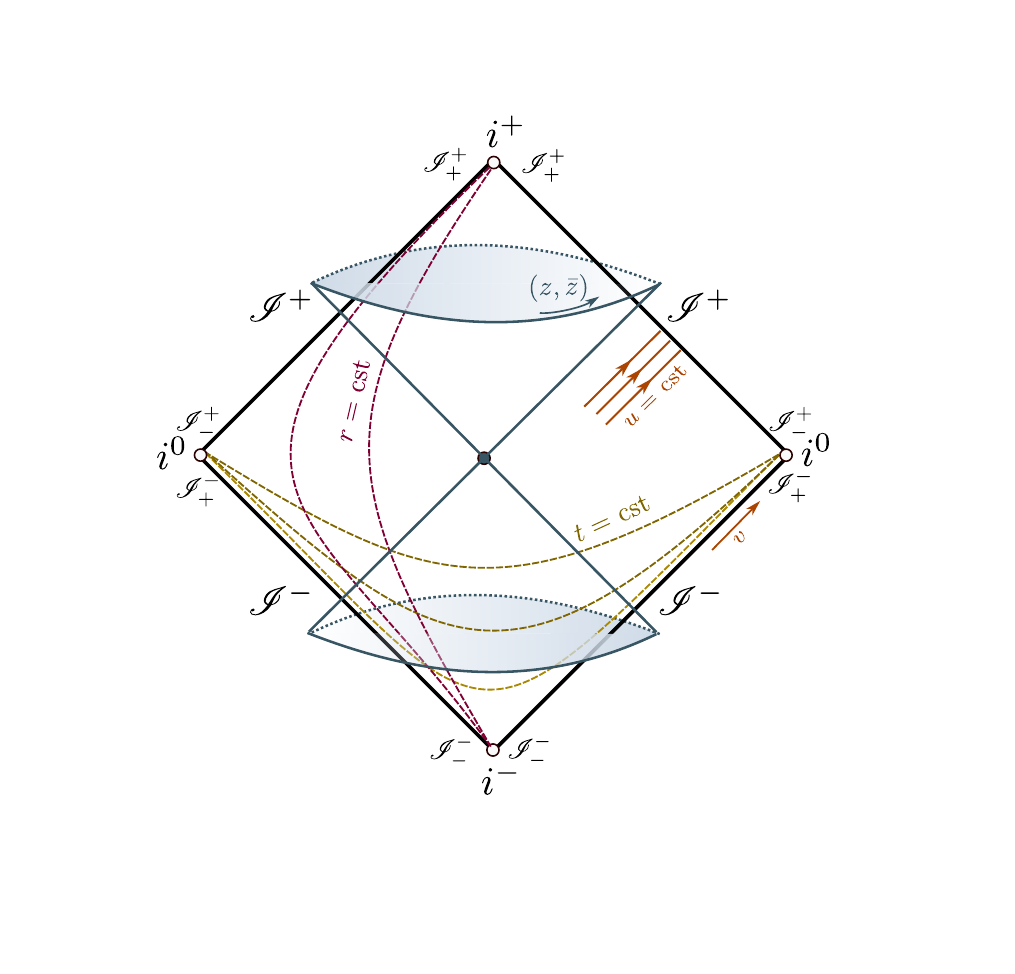}
\captionsetup{width=1\linewidth}
\caption{\small{Penrose diagram of flat spacetime. The different disconnected boundaries are null infinity (future $\mathscr I^+$ and past $\mathscr I^-$ null infinity), timelike (future $i^+$ and past $i^-$) infinity and spatial infinity $i^0$). Null geodesics follow constant $u$ trajectories and transverse coordinates are denoted by $x^A=(z,\bz)$. In this diagram, every left-right pair of points at the same $r>0$ and $t$ corresponds to a two-sphere $S^2$, each pair being exchanged under the antipodal map on $S^2$.}}
\label{fig:Penrose}
\end{center}
\end{figure}

The goal of this first section of the report is to review the notion of asymptotically flat spacetimes and their associated symmetries, with the aim of making contact with recent developments in celestial holography. For this reason, we will follow in section \ref{sec:Bondi} Bondi's approach consisting of using a well-adapted coordinate system and specifying certain fall-off conditions for the gravitational field `away from an isolated source'. Of course, this way of defining asymptotic flatness is not completely satisfactory from a geometrical point of view as, on the one hand, it raises the question to which extent it depends or not on the specific choice of coordinates taken and, on the other hand, it does not inform us on global aspects of the spacetime. This pragmatic approach is nevertheless justified since it was shown that the associated boundary conditions can in fact be rederived from Penrose's invariant definition of asymptotically simple and flat spacetimes \cite{Tamburino:1966zz,Madler:2016xju}. As far as the global aspects of the spacetimes are concerned, we will always assume in this review that they are `close' to Minkowski in the sense that they have the same topology and boundary as flat spacetime; see the Penrose diagram of Fig. \ref{fig:Penrose}. This is justified and motivated by the fact that we are interested in scattering gravitational radiation from the past to the future but are excluding a regime where black holes would form.
In section \ref{sec:BMS}, we will present the famous `Bondi-Metzner-Sachs' (BMS) symmetries, together with their modern extensions. We will then focus in section \ref{sec:extendedBMS} on a particular one, the so-called the `extended BMS' group, which comprises supertranslations together with local conformal transformations on the celestial sphere, and thus is very natural in the context of a holographic description. We will close in section \ref{sec:BMS_everywhere} by a short account of the recent understanding of the emergence of BMS symmetries at other boundaries of flat spacetime, timelike and spacelike infinities.

\label{Asymptotically flat spacetimes}
\subsection{Bondi-Sachs metric}
\label{sec:Bondi}
Bondi-Sachs coordinates $x^\mu=(u, r, x^A)$ are based on a family of outgoing null $u=$ cst hypersurfaces along which gravitational waves travel \cite{Bondi:1962px,Sachs:1962wk,Sachs:1962zza}. The transverse coordinates $x^A$ ($A,B=1,2$) are constant along null rays, only the radial coordinate $r$ (which measures the sphere radius) varies along null rays.  The requirements that the normal vector $\partial_\mu u$ is null and that $x^A$ are constant along the null rays lead to $g^{\mu \nu}(\p_\mu u) (\p_\nu u)=0=g^{\mu \nu}(\p_\mu u) (\p_\nu x^A)$; this implies the gauge fixing conditions $g^{uu}=g^{uA}=0$  or, equivalently, $g_{rr}=g_{rA}=0$. In Bondi gauge, the spacetime metric can thus be written as\footnote{We mainly follow the notations and conventions of \cite{Barnich:2010eb,Barnich:2011mi}; see also \cite{Madler:2016xju,Alessio:2017lps} for reviews on the Bondi-Sachs formalism.}
\begin{equation}
    ds^2 = e^{2\beta} \frac{V}{r} {d}u^2 - 2 e^{2\beta}{d}u {d}r + g_{AB} ({d}x^A - U^A {d}u)(dx^B - U^B {d}u),
    \label{Bondi gauge metric}
\end{equation} where $\beta$, $V$, $g_{AB}$, $U^A$ are some functions of the coordinates $(u, x^A)$. On top of that, one usually completes the gauge fixing conditions $g_{rr}=g_{rA}=0$ by adding a fourth condition, which is that the transverse metric $g_{AB}$ satisfies the determinant condition
\begin{equation}\label{det_cond}
    \partial_r \det (r^{-2} g_{AB}) = 0.
\end{equation} 
The latter fixes the radial coordinate $r$ to be a luminosity distance.
There are of course other possible choices for the gauge fixing conditions, for instance the Newman-Unti gauge imposes $g_{ur}=-1$ in place of \eqref{det_cond} (where $r$ is then the affine parameter of the null congruence)\cite{Newman:1962cia,Barnich:2011ty}.  We refer the reader to  \cite{Geiller:2022vto} for a detailed discussion of these different gauge choices and  for an analysis of asymptotics in a `relaxed' Bondi gauge. 

An \emph{asymptotically flat} spacetime in the sense of Bondi, van der Burg, Metzner and Sachs (BMS)\footnote{The acronym should have been BVMS instead (as proposed in \cite{vanderBurg}) but was unfortunately not retained.} is a spacetime of the form \eqref{Bondi gauge metric} which can be expanded in powers of $1/r$ with the following boundary conditions as one reaches future null infinity $\mathscr{I}^+$. At leading order, the transverse metric is taken to be the round sphere metric times a conformal factor
\begin{equation}
\badat{2}
& g_{AB} = r^2 \bar \gamma_{AB}    + r C_{AB} + \mathcal{O}(r^{-1})\virg \bar \gamma_{AB} dx^A dx^B = e^{2\varphi}(d\theta^2+\sin^2\theta d\phi^2)\,
    \label{sphere metric}
    \eadat
\end{equation}
and the next-to-leading piece involves $C_{AB}(u,x)$, a two-dimensional symmetric traceless tensor. 
The remaining fall-off conditions are
\begin{equation}
\badat{2}
 &   \beta = \mathcal{O}(r^{-2}), \quad \frac{V}{r} = -e^{-2\varphi} +\bar \Delta \varphi+  \mathcal{O}(r^{-1}), \quad U^A = \mathcal{O}(r^{-2}) \,,
    \label{BC Bondi gauge}
    \eadat
\end{equation}
where $\bar \Delta$ denotes the Laplacian associated to $\bar \gamma_{AB}$.
We will consider a time-independent conformal factor $\partial_u\varphi =0$ (this assumption can be relaxed as we will comment in section \ref{sec:BMS}).

As we mentioned above, these fall-off conditions are in fact locally equivalent to Penrose's conformal compactification at $\mathscr{I}^+$ (see e.g. Section 4 of \cite{Madler:2016xju}). Notice that the ansatz of making a $1/r$ expansion for the metric is tied to the `peeling property' of the Weyl tensor \cite{Newman:1961qr}, as it implies that the metric is sufficiently smooth along the asymptotic boundary. In the above, we have set the order $r^0$ in the expansion of $g_{AB}$ to zero. Turning on this term would bring some $\log r$ terms in the expansion\cite{Tamburino:1966zz, Winicour1985LogarithmicAF,Barnich:2010eb} that we will not consider here.  We refer the reader to \cite{Andersson:1993we, Chrusciel:1993hx , Andersson:1994ng, Ashtekar:1996cm , Friedrich:2017cjg , Angelopoulos:2017iop, Kroon:1998tu, Godazgar:2020peu,Geiller:2022vto} for further discussions on these polyhomogeneous spacetimes.

Solving Einstein's equations in vacuum with vanishing cosmological constant for the boundary conditions \eqref{sphere metric}, \eqref{BC Bondi gauge} yields the following expressions~\cite{Tamburino:1966zz,Barnich:2010eb}:
\begin{equation}
\begin{split}
\frac{V}{r} &= -\frac{\bar R}{2} + \frac{2M}{r} + \mathcal{O}(r^{-2}) ,\qquad 
\beta = \frac{1}{r^2} \left[ - \frac{1}{32}C^{AB} C_{AB} \right] +  \mathcal{O}(r^{-3}), \\
U^A &= -\frac{1}{2 r^2} D_B C^{AB} -\frac{2}{3} \frac{1}{r^3} \left[ N^A -\frac{1}{2} C^{AB} D^C C_{BC} \right] + \mathcal{O}(r^{-4}),
\end{split}
\label{fall-off}
\end{equation} 
where $M = M(u,x^A)$ is the Bondi mass aspect, $N_A = N_A (u,x^B)$ is the angular momentum aspect\footnote{Our convention for $N_A$ are the ones of \cite{Barnich:2010eb}, notice that it differs e.g. from \cite{Hawking:2016sgy} by $N_A^{\text{there}}= N_A^{\text{here}}-u\partial_A M+\frac{1}{4}C_{AB}D_C C^{BC}+\frac{3}{32}\partial_A (C_{BC}C^{BC}).$}. Both of these quantities appear as `integrating constants' when solving the equations of motion.

The two-sphere indices in \eqref{fall-off} are lowered and raised with $\bar \gamma_{AB}$ and its inverse, $D_A$ is the Levi-Civita connection associated to $\bar \gamma_{AB}$ and $\bar R$ denotes the associated Ricci curvature. 
The symmetric and traceless tensor $C_{AB}=C_{AB}(u,x^A)$ is called the asymptotic shear and encodes the two polarization modes
of the gravitational waves.  Its retarded time derivative $N_{AB} \equiv \partial_u C_{AB}$ is the `Bondi news tensor', whose square is proportional to the energy flux going through $\mathscr I^+$.

The Bondi mass and angular momentum aspects satisfy the time evolution equations
\begin{equation}
\begin{split}
\partial_u M &= - \frac{1}{8} N_{AB} N^{AB} +\frac{1}{8}\bar \Delta \bar R+ \frac{1}{4} D_A D_B N^{AB} , \\ 
\partial_u N_A &= D_A M + \frac{1}{16} D_A (N_{BC} C^{BC}) - \frac{1}{4} N^{BC} D_A C_{BC}  -\frac{1}{4} D_B (C^{BC} N_{AC} - N^{BC} C_{AC}) \\
&\quad - \frac{1}{4} D_B D^B D^C C_{AC}+ \frac{1}{4} D_B D_A D_C C^{BC} +\frac{1}{4}C_{AB}D^B \bar R\,.
\label{EOM1} 
\end{split}
\end{equation} 
In contrast, the time evolution of the gravitational shear is \emph{not} constraint by Einstein's equations. In a GR perspective, $C_{AB}(u,x^A)$  has to be seen as a \emph{free data} for a characteristic initial value problem. In the perspective of setting up a scattering problem, the shear is identified with an asymptotically free gravitational field (see section \ref{sec:btob}).

For many purposes, and especially in the context of celestial holography, it turns out very useful to choose the celestial sphere coordinates to be the complex stereographic coordinates $x^A = (z, \bar{z})$ given by $z=\cot \frac \theta {2}e^{i\phi}$, $\bar z=\cot \frac \theta {2}e^{-i\phi}$, in terms of which the transverse metric reads
\begin{equation}
    \bar \gamma_{AB} dx^A dx^B 
    =2 e^{2\tilde \varphi } dz d\bar{z}\,,
    \label{flat sphere}
\end{equation}
with
\begin{equation} \label{representative}
\tilde \varphi= \varphi- \ln P, \quad P = \frac{1+ z\bar{z}}{\sqrt 2} \,.
\end{equation}
Most of the time, one picks the round sphere representative, given by $\varphi=0$, or the flat representative by taking $\tilde \varphi=0$. We will adopt the latter choice here. In that case, $\bar R=0$, $D_A=\partial_A$ and the form of an asymptotically flat spacetime is thus
\begin{equation}
\badat{2}
    ds^2 = &\frac{2M}{r} {d}u^2 - 2 {d}u {d}r +2r^2{d}z{d}\bz+rC_{zz}{d}z^2+rC_{\bz\bz}{d}\bz^2\\
    &+\left[(\partial^z C_{zz}-\frac{1}{6r}\partial_z(C_{zz} C^{zz})+\frac{2}{3r}N_z)\,du dz +\text{c.c.}\right]+\dots\,,
    \eadat
    \label{Bondi gauge metric2}
\end{equation} 
where $\text{c.c.}$ stands for complex conjugate terms and the dots denote subleading terms in the large $r$ expansion that we will not consider here.

\subsection{The BMS group and its modern extensions}
\label{sec:BMS}

In recent years, there have been several motivations and claims for what are/should be the largest nontrivial set of symmetries that preserve a certain notion of asymptotic flatness which includes the one presented above. This is of course a question that is very subtle to address since boundary conditions that are considered too relaxed for certain (e.g. because they do not obey peeling properties or do not lead to finite charges) are considered too restrictive for other purposes (e.g. because they do not capture potential observables). The purpose of this section is to give an overview of the different proposals that exist in the literature regarding the different definitions or extensions of `BMS symmetries'. Notice that sometimes a given extension is referred to different names (depending on the authors), which has led to extra confusion. In all cases, the (restricted to $\mathscr I^+$) infinitesimal  generator  of all these symmetry algebras takes the following form: 
\begin{equation}
\xi_{(\mathcal{T}, Y, \mathcal W)}= \mathcal{T} \p_u+Y^A \p_A+\mathcal W(u\p_u-r\p_r)+\dots,
\label{xiTYW}
\end{equation} 
but they differ depending on which restrictions apply to the functions $\mathcal{T}$, $Y^A$ and $\mathcal W$.\\

\noindent \textcolor{pansypurple}{\gr{Standard BMS} \,\,}
This is the original group obtained in the sixties by Bondi, van der Burg, Metzner and Sachs \cite{Bondi:1962px,Sachs:1962zza,Sachs:1962wk}. It consists of an infinite-dimensional extension of the Poincar\'e group, arising from the fact that the four usual spacetime translations get enhanced to a whole smooth function's worth of what Sachs called `supertranslations',
generated by an arbitrary angle-dependent function $\mathcal{T}(z,\bar z)$ in \eqref{xiTYW}. 
Famously, the enhancement of the Poincar\'e algebra into this much bigger set was a source of great disappointment for Sachs\footnote{One can read in \cite{Sachs:1962wk}, ``The fact that $\alpha$ [here $\mathcal{T}$] is an arbitrary function of its argument is quite annoying in some respects. [...] Several other attempts to restrict or eliminate $\alpha$ by changing the coordinate and/or the boundary conditions have also failed.''}, even though he finally acknowledges that the presence of supertranslation symmetry might be ``a blessing in disguise'' \cite{Sachs:1962wk}. 
As opposed to their recent extra extensions (see below), the specificity of the standard BMS group is that vector fields  $Y^A$  are required to be \emph{globally well defined} conformal vector fields on the sphere. There are only six of them, they span the usual Lorentz group (namely the global conformal group) $\mathfrak{so}(3,1)$ (see e.g. exercise 12 in \cite{Strominger:2017zoo}). We thus have the isomorphism 
\begin{equation}
   \mathfrak{bms}^{\text{stand}}_4 \cong \mathfrak{so}(3,1) \loplus \mathfrak{s}\,,
\end{equation} 
where $\mathfrak{s}$ denotes supertranslations (the structure of the semi-direct sum will be clarified in the next section).\\

\noindent \textcolor{pansypurple}{\gr{Extended BMS \,\,}}
Instead of restricting to globally well-defined vector fields on the celestial sphere, this extension allows for possible singular violations of the conformal Killing equation for $Y^A$ \cite{Barnich:2010eb,Barnich:2009se,Barnich:2011ct}. This effectively amounts to make room for \emph{meromorphic} functions $Y^z=Y^z(z)$ and $ Y^\bz= Y^\bz(\bz)$\footnote{For example, $Y^z=\tfrac{1}{z-w}$ violates the conformal Killing equation at isolated points since $\p_\bz Y^z=2\pi\delta^{(2)}(z-w)\neq 0$.}. The latter span the so-called `superrotation' symmetries\footnote{They were dubbed super-Lorentz in \cite{Compere:2018ylh,Freidel:2021fxf}.} which form two commuting copies of the centerless Virasoro (i.e. Witt) algebra. The global conformal group is thus extended to the local conformal group, hence the extended BMS algebra is\footnote{Notice that, as a result, supertranslations are now potentially singular functions on the sphere.}
\begin{equation}
   \mathfrak{bms}^{\text{ext}}_4 \cong \text{Witt} \loplus \mathfrak{s}.
\end{equation} 
 In this case, one is then dealing with the two-punctured sphere as celestial Riemann surface $\mathcal{S} \simeq \mathscr{I}^+ / \mathbb{R}$ (see e.g. \cite{Barnich:2021dta}).   
The local violations of the standard BMS boundary conditions were physically interpreted in \cite{Strominger:2016wns} as arising from cosmic strings piercing the celestial sphere. This extension is the most natural to study in celestial holography since it readily implies the conformal symmetries on the celestial Riemann surface.  Let us nevertheless now briefly comment on further extensions of the BMS group that have subsequently appeared in the literature.\\
 
\noindent \textcolor{pansypurple}{\gr{Generalized BMS \,\,}}
This extension of the Lorentz transformations to the full diffeomorphisms of the celestial sphere was proposed by Campiglia and Laddha \cite{Campiglia:2015yka,Campiglia:2014yka}. It was originally motivated by having a bijective map between Ward identities and the subleading soft graviton theorem (see also \cite{Donnay:2020guq} for a discussion on that point). In the generalized BMS algebra, the generators $Y^A$ are no longer required to be holomorphic or meromorphic functions, they span all diff$(S^2$) transformations on the celestial sphere;
\begin{equation}
   \mathfrak{bms}^{\text{gen}}_4 \cong \mathfrak{diff}(S^2)  \loplus \mathfrak{s}.
\end{equation} 
This group was also used and promoted in \cite{Compere:2018ylh,Flanagan:2019vbl,Capone:2023roc}.\\

\noindent \textcolor{pansypurple}{\gr{Weyl BMS \,\,}}
On top of the generalized BMS symmetries, one can go even bigger and allow for local
Weyl rescaling symmetries \cite{Barnich:2010eb, Barnich:2016lyg , Barnich:2019vzx , Freidel:2021fxf,Freidel:2021qpz,Freidel:2021cjp}: 
\begin{equation}
   \mathfrak{bms}^{\text{Weyl}}_4 \cong[\mathfrak{diff}(S^2)\loplus \text{Weyl}]  \loplus \mathfrak{s}.
\end{equation} 
The Weyl rescalings are spanned by the function $\mathcal W$ in \eqref{xiTYW}.
This extension plays an important role in the so-called `corner proposal'; see \cite{Ciambelli:2022vot} for a review.\\

 We hope that the reader's confusion about the different BMS extensions and their associated nomenclatures will be dissipated once looking at Table \ref{table:BMS}.\\

\small
\begin{table}[h!]
\footnotesize
\renewcommand*{\arraystretch}{1.2}
\centering
\begin{tabular}{|c||c|c||c|c||c|}
\hline
$\mathfrak{bms}_4$  &$\delta g_{uu}$  & \multicolumn{1}{c|}{$\delta g_{AB}$}  &$\xi_Y$ &$\xi_\mathcal W$ & algebra \\
  \hline  \hline
  \small Standard &\multicolumn{1}{c|}{$\mathcal O (r ^{-1})$} & \multicolumn{1}{c|}{$\mathcal O (r )$ }&  6 Lorentz &  $\frac{1}{2}D_AY^A$ & $\mathfrak{so}(3,1) \loplus \mathfrak{s}$ \\[3pt] \cdashline{1-6}
  
  \small Extended &\multicolumn{1}{c|}{$\mathcal O (r ^{-1})$} & \multicolumn{1}{c|}{$\mathcal O (r)$ \text{locally} } &  local CKV & $\frac{1}{2}D_AY^A$ & $\text{Witt} \loplus \mathfrak{s}$\\[3pt] \cdashline{1-6}
  
  \small Generalized  &\multicolumn{1}{c|}{$\mathcal O (r^{-1})$} & \multicolumn{1}{c|}{$r^2 q_{AB}(x^C)+\mathcal O (r)$ } & $Y^A(x^B)$ & $\frac{1}{2}D_AY^A$ &  $\mathfrak{diff}(S^2)  \loplus \mathfrak{s}$ \\[3pt] \cdashline{1-6}
  
    
  \small Weyl  &\multicolumn{1}{c|}{$\mathcal O (1)$} & \multicolumn{1}{c|}{$r^2 q_{AB}(x^C)+\mathcal O (r)$}  & $Y^A(x^B)$  & \small $\mathcal W(x^A)$ & $\mathfrak{diff}(S^2)\loplus \text{Weyl}  \loplus \mathfrak{s}$\\[3pt] \cdashline{1-6}
\hline
\end{tabular}
\captionsetup{width=1\linewidth}
\caption{Summary of the different BMS group extensions, where infinitesimal BMS generators are denoted $\xi_{(\mathcal{T}, Y, \mathcal W)}= \mathcal{T} \p_u+Y^A \p_A+\mathcal W(u\p_u-r\p_r).$ All extensions contain an infinite-dimensional extension of Poincar\'e spanned by an arbitrary function on the celestial sphere $\mathcal T(z,\bar z)$ (supertranslations $\mathfrak s$). They differ depending on the possibility of deforming or not the round sphere metric $\bar \gamma_{AB}$ ($g_{AB}=r^2\bar \gamma_{AB}+\mathcal O(r)$) and including or not extra Weyl rescalings, spanned by an arbitrary angle-dependent $\mathcal W$. }
 \label{table:BMS}
\end{table}
\normalsize

\subsection{Focus on the extended BMS group}
\label{sec:extendedBMS}
In this section (mostly based on \cite{Donnay:2021wrk}) and in the rest of this report, we will focus on the extended BMS symmetries since, as already emphasized, they readily give the usual conformal transformations of a $2d$ CFT.  
The associated symmetry group is generated by residual diffeomorphisms that preserve the Bondi gauge \eqref{Bondi gauge metric} and fall-off conditions \eqref{BC Bondi gauge}, where it is understood that the celestial sphere metric is fixed locally. Recall that we adopt here the flat representative, as discussed around equation \eqref{representative}.
These asymptotic Killing vectors are generated by vectors fields $\xi = \xi^u \partial_u + \xi^r \partial_r+\xi^z \partial_z + \xi^{\bar{z}}\partial_\bz $ whose components are (see e.g. \cite{Barnich:2010eb,Kapec:2014opa} or the review \cite{Kervyn:2023adk} for detailed steps) 
\begin{equation}
\begin{split}
    &\xi^u = \mathcal{T} + u\alpha \virg \\
      &\xi^r = - (r+u)\alpha + \partial_z  \partial^z \mathcal{T}+ \mathcal{O}(r^{-1}),\\
    &\xi^z = Y^z -\frac{1}{r} \partial^z\left(\mathcal{T} +u\alpha\right) + \mathcal{O}(r^{-2}) \virg 
\xi^{\bar{z}} = Y^\bz -\frac{1}{r} \partial^\bz \left(\mathcal{T} +u\alpha\right) + \mathcal{O}(r^{-2}), \\ 
    \end{split}
    \label{AKV Bondi}
\end{equation} where $\alpha \equiv \frac{1}{2} \partial_A Y^A$, $\mathcal{T}= \mathcal{T}(z, \bar{z})$ is an arbitrary function of the angles and $Y^A=(Y^z,Y^\bz)$ is a $2d$ vector on the celestial sphere satisfying locally the conformal Killing equation
\begin{equation}\label{CKV}
    \partial_\bz Y^z = 0, \qquad \partial_z Y ^\bz = 0.
\end{equation} 
Locally, this equation admits an infinity of solutions of the form $ Y^z \sim z^n$ with possible poles at isolated points on the sphere. As we will see below, the meromorphic functions $Y^z (z)\equiv Y(z)$, $Y^{\bz}(\bz)\equiv \bar Y(\bz)$ span the Virasoro part of the $\mathfrak{bms}_4$ algebra, and thus obviously play a central role in a holographic context. We will shortly introduce the definition of conformal primary fields and conformal weights that will be used extensively in the rest of the review, but let us before present the algebra spanned by these asymptotic vector fields.\\

\noindent \textcolor{pansypurple}{\textbf{$\mathfrak{bms}^{\text{ext}}_4$ algebra\,\,}}  Using the modified Lie bracket $[\xi_1, \xi_2]_\star =[\xi_1, \xi_2] - \delta_{\xi_1}\xi_2 + \delta_{\xi_2} \xi_1$ where the last two terms take into account the field-dependence of the asymptotic Killing vectors \eqref{AKV Bondi} at subleading order in $r$ \cite{Schwimmer:2008yh , Barnich:2010eb}, the asymptotic Killing vectors \eqref{AKV Bondi} can be checked to satisfy the commutation relations
\begin{equation}
    \big[\xi (\mathcal T_1, Y^z_1, Y^{\bar z}_1), \xi (T_2, Y^z_2, Y^{\bar z}_2) \big]_\star = \xi (\mathcal{T}_{12}, Y^z_{12}, Y^{\bar z}_{12}) ,
    \label{commutation relations 1}
\end{equation} with
\begin{equation}
\begin{split}
    \mathcal{T}_{12} &= Y^z_1 \partial_z \mathcal T_2 - \frac{1}{2} \partial_z Y^z_1 \mathcal T_2 + \text{c.c.} - (1 \leftrightarrow 2)   \, , \\
    Y^z_{12} &= Y^z_1 \partial_z Y^z_2 - (1 \leftrightarrow 2)\,, \quad Y^{\bar z}_{12} = Y^{\bar z}_1 \partial_{\bar z} Y^{\bar z}_2 - (1 \leftrightarrow 2) \,.
\end{split}
    \label{commutation relations 2}
\end{equation} 
The above commutators define the  $\mathfrak{bms}^{\text{ext}}_4$ algebra.
They can equivalently be rewritten using the notations $\xi(\mathcal{T},0,0) \to \mathcal{T}$, $\xi(0, Y, 0) \to Y$, $\xi (0,0, \bar{Y}) \to \bar{Y}$, $[.,.]_\star \to [.,.]$ as well as adopting the shorthand notation $\partial_z\equiv \partial$, $\partial_\bz\equiv \bar \partial$ as
\begin{equation}
\begin{split}
&[Y_1, Y_2] = Y_1 \partial Y_2 - Y_2 \partial Y_1, \qquad [\bar{Y}_1, \bar{Y}_2] = \bar{Y}_1 \bar{\partial} \bar{Y}_2 - \bar{Y}_2 \bar{\partial} \bar{Y}_1,\qquad [T_1, T_2] = 0,\\ 
&[Y_1 , T_2] =  Y_1 \partial T_2 - \frac{1}{2} \partial Y_1 T_2 , \qquad [\bar{Y}_1, T_2] = \bar{Y}_1 \bar{\partial} T_2 - \frac{1}{2} \bar{\partial}\bar{Y}_1 T_2.
\end{split}
\end{equation} 

\noindent \textcolor{pansypurple}{\textbf{Conformal weights\,\,}}We will say that $\phi_{h, \bar{h}}(z, \bar{z})$ is a \emph{conformal primary field} of weights $(h, \bar{h})$ if it transforms under the action of superrotations as
\begin{equation}
    \delta_{Y} \phi_{h, \bar{h}} = (Y^z \partial_z + Y^\bz \partial_\bz + h \partial_z Y^z  +\bar{h}  \partial_\bz Y^\bz )\phi_{h, \bar{h}}\,.
    \label{def conformal field}
\end{equation} 
One sometimes expand a conformal field in formal series as
\begin{equation}\label{formal_series}
    \phi_{h, \bar h} (z,\bz)=  \sum_{k,\ell}a_{k,\ell}\, {}_{h,\bar{h}}Z_{k,\ell} \virg {}_{h,\bar{h}}Z_{k,\ell} =z^{-h-k}\bar z^{-\bar h-\ell}\,,
\end{equation} 
where the complex coefficients $a_{k,\ell}$ satisfy appropriate conditions (see e.g. \cite{Schottenloher}) and $k,\ell$ are taken to be integers (resp. half integers) if $h,\bar h$ are integers (resp. half integers).

From \eqref{commutation relations 2}, one sees that the supertranslation parameter ${\mathcal{T}}(z,\bar{z})$ admits to be considered as a density of weights $(-\frac{1}{2},-\frac{1}{2})$ and can be expanded in series as \cite{Barnich:2010eb}
\begin{equation}
    \mathcal{T}(z, \bar{z}) = \sum_{k,\ell} t_{k,\ell} \mathscr{T}_{k,\ell}, \qquad \mathscr{T}_{k,\ell} 
    =z^{\frac{1}{2}-k}
  \bar{z}^{\frac{1}{2}-\ell},
  \label{commu 1 prime}
\end{equation} 
with $k,\ell$ half-integers and $\bar{t}_{k,\ell} = t_{\ell,k}$ so that $\mathcal{T}$ is real. Notice that the four usual Poincar\'e translations are spanned by $\mathscr{T}_{\12,\12}$, $\mathscr{T}_{\12,-\12}$, $\mathscr{T}_{-\12,\12}$ and $\mathscr{T}_{-\12,-\12}$.

Superrotations $Y^z= Y(z)$, $Y^\bz=\bar{ Y}(\bar z)$ have weights $(-1,0)$ and $(0,-1)$, respectively and can be expanded in modes as
\begin{equation}
   Y(z) = \sum_{m} y_m Y_m, \quad Y_m = z^{1-m}, \qquad \bar{Y}(\bar{z}) = \sum_{m} \bar{y}_m \bar{Y}_m, \quad \bar{Y}_m = \bar{z}^{1-m},
    \label{commu 2 prime}
\end{equation} where $m \in \mathbb{Z}$ and $y_m, \bar{y}_m$ are complex numbers. The six global Lorentz parameters are spanned by $Y_{-1}$, $Y_{0}$, $Y_{1}$, and their complex conjugates. In terms of the modes \eqref{commu 2 prime}, the conformal transformation \eqref{def conformal field} can be rewritten as
\begin{equation}
   \delta_{Y_{m}}  \phi_{h, \bar h} =  z^{1-m} \partial \phi_{h, \bar h} + h (1-m) z^{-m} \phi_{h, \bar h}, \qquad  \delta_{\bar{Y}_{m}}  \phi_{h, \bar h} =  \bar{z}^{1-m} \bar{\partial} \phi_{h, \bar h} + \bar{h} (1-m) \bar{z}^{-m} \phi_{h, \bar h}\,.
\end{equation} 
Using the formal series expansion \eqref{formal_series}, the above transformations are also equivalent to
\begin{equation}
    \delta_{Y_m}  ({}_{h,\bar{h}}Z_{k,\ell}) = -( k + h m) \,{}_{h, \bar{h}} Z_{m+k, \ell}, \qquad \delta_{\bar{Y}_m} ({}_{h,\bar{h}}Z_{k,\ell}) = - (\ell+\bar{h}m)  \,{}_{h,  \bar{h}}Z_{k, \ell+m}.\\
     \label{commu 2 prime 2}
\end{equation} 
If the reader is more accustomed to write algebra in terms of mode expansions, they can note that, using \eqref{commu 1 prime} and \eqref{commu 2 prime}, the $\mathfrak{bms}_4$ commutation relations read \cite{Barnich:2010eb,Barnich:2017ubf}
\begin{equation}
  \begin{split}
   &[ {Y}_m, {Y}_n ] = (m-n) {Y}_{m+n} , \quad [
    {\bar{Y}}_m, {\bar{Y}}_n ] = (m-n){\bar{Y}}_{m+n}, \\ 
    &[{Y}_m ,
\mathscr{T}_{k,\ell} ] = \Big( \frac{1}{2} m - k \Big)
\mathscr{T}_{m+k,\ell}, \quad [{\bar{Y}}_m , \mathscr{T}_{k,\ell} ] =
\Big(\frac{1}{2} m - \ell\Big) \mathscr{T}_{k,m+\ell}, \\ &[ {Y}_m ,
{\bar{Y}}_n ] = 0 = [ \mathscr{T}_{k,\ell} ,
\mathscr{T}_{r,s}] \,.
\end{split} 
\end{equation}
and recognize two commuting copies of the Witt (centerless Virasoro) algebra in semi-direct sum with an abelian ideal of supertranslations.\\

\noindent \textcolor{pansypurple}{\textbf{Action of the symmetries on the phase space\,\,}} Having presented the asymptotic symmetries, let us now see how they act on the different quantities appearing in the BMS expansion \eqref{Bondi gauge metric2}. For convenience, it is useful to introduce the notation $f \equiv \mathcal{T} + u\alpha$ where we recall that $\alpha =\frac{1}{2} \partial_A Y^A$. The way that extended BMS symmetries act infinitesimally on the solution space is as follows (see e.g. \cite{Barnich:2010eb,Barnich:2016lyg} for a derivation)
\begin{equation}
\badat{4}
\delta_{\xi(\mathcal{T}, Y)}  M &= [f \partial_u + \mathcal{L}_Y + \frac{3}{2} \partial_C Y^C] M + \frac{1}{8} \partial_C \partial_B \partial_A Y^A C^{BC} + \frac{1}{4} N^{AB} \partial_A \partial_B f + \frac{1}{2} \partial_A f \partial_B N^{AB},\\
\delta_{(\mathcal{T}, Y)} N_A &= [f\partial_u + \mathcal{L}_Y + \partial_C Y^C] N_A + 3 M \partial_A f - \frac{3}{16} \partial_A f N_{BC} C^{BC}  \\
&\quad - \frac{1}{32} \partial_A \partial_B Y^B C_{CD}C^{CD} + \frac{1}{4} (2 \partial^B f + \partial^B \partial_C \partial^C f) C_{AB}  \\
&\quad - \frac{3}{4} \partial_B f (\partial^B \partial^C C_{AC} - \partial_A \partial_C C^{BC}) + \frac{3}{8} \partial_A (\partial_C \partial_B f C^{BC}) \\
&\quad + \frac{1}{2} (\partial_A \partial_B f - \frac{1}{2} \partial_C \partial^C f \bar \gamma_{AB}) \partial_C C^{BC} + \frac{1}{2} \partial_B f N^{BC} C_{AC}.  \\
\delta_{\xi(\mathcal{T}, Y)}  C_{AB} &= [f \partial_u + \mathcal{L}_Y - \frac{1}{2} \partial_C Y^C ] C_{AB} - 2 \partial_A \partial_B f + \bar \gamma_{AB} \partial_C \partial^C f,\\
\delta_{\xi(\mathcal{T}, Y)} N_{AB} &= [f\partial_u + \mathcal{L}_Y] N_{AB} - (\partial_A \partial_B \partial_C Y^C - \frac{1}{2} \bar \gamma_{AB} \partial_C \partial^C \partial_D Y^D).
\eadat
\label{transformation on the solution space}
\end{equation}
Comparing the above transformation laws with the transformation law of a conformal primary field, it might seem that extracting  any CFT-looking information from the gravitational phase space of an asymptotically flat spacetime is a  hopeless task. However, we will see in section \ref{sec:BMSfluxes} that there exist quantities built out of the Bondi mass $M$, angular momentum aspect $N_A$ and shear $C_{AB}$ which do transform as primaries under the action of superrotations.

\subsection{BMS symmetries at every corner}
\label{sec:BMS_everywhere}
The asymptotic analysis we presented at the beginning of this chapter led to one copy of the BMS group at $\mathscr I^+$. We could have equivalently performed an analogous analysis near past null infinity, and we would have found another copy of the BMS group, acting at $\mathscr I^-$. 
Now, as mentioned in the introduction, a critical point in the recent understanding of the connections between asymptotic symmetries (via the Ward identity of their charges) and scattering amplitudes is the existence of a \emph{single} BMS group which acts simultaneously on future and past null infinity~\cite{Strominger:2013jfa,He:2014laa}. Its existence stems from a set of matching conditions at spatial infinity of fields living at $\mathscr I^+$ and  $\mathscr I^-$ which are a posteriori justified by the shown equivalence to Weinberg's soft graviton theorem. These results hint at the fact that spatial infinity should admit a description for which Strominger's global BMS algebra appears as an asymptotic symmetry algebra.

With this motivation in mind, we provide in this section a short account of recent developments regarding the occurrence of BMS symmetries at \emph{other boundaries} of flat spacetimes, namely at spacelike but also at timelike infinity. Notice that in this section we will only consider the standard BMS group (six Lorentz and supertranslations), as a complete study of the extensions of BMS at timelike and spacelike infinities awaits further investigation.

\subsubsection{Spacelike infinity} 

    Spacelike infinity, $i^0$ in the Penrose diagram of Fig. \ref{fig:Penrose}, is the asymptotic region of spacelike separation.  
    While this region does not correspond to the asymptotic home of any physical particle, it has nevertheless played and keeps playing a crucial role in the context of scattering in asymptotically flat spacetime as the region that bridges together the two disconnected future and past null boundaries, $\mathscr I^+$ and $\mathscr I^-$.      It is well known that $i^0$ has to be to some extent singular in order to allow for nontrivial spacetimes \cite{Ashtekar:1978zz}, which a priori prevents a generic canonical identification between $\mathscr I^+_-$ (the far past of $\mathscr I^+$) and $\mathscr I^-_+$ (the far future of $\mathscr I^-$). Nevertheless, we will see that there exist certain conditions which allow for such an identification.

\begin{wrapfigure}{r}{7cm}
\includegraphics[scale=0.5,trim = {0.1cm 1cm 0cm 1.5cm}]
{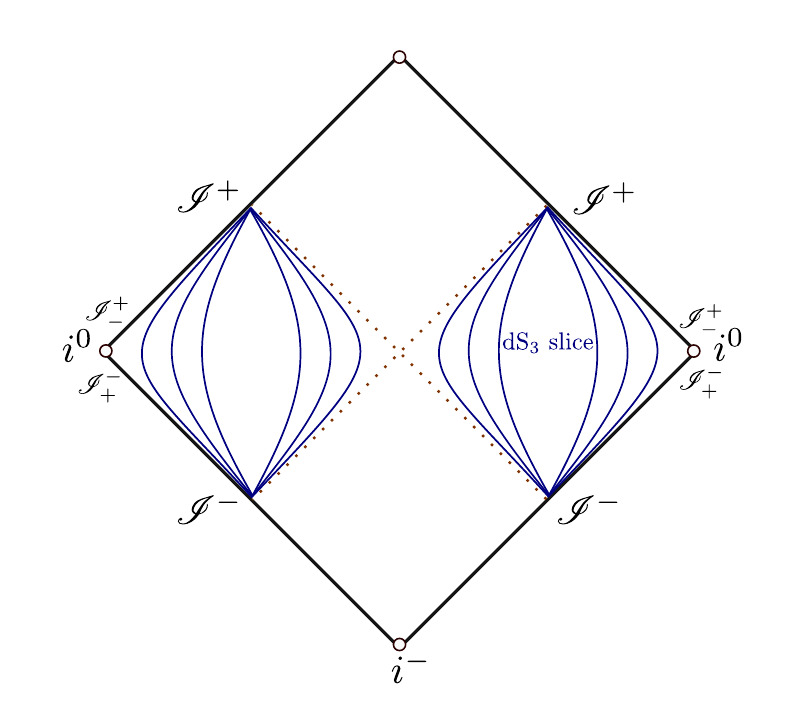}
\captionsetup{width=0.9\linewidth}
\caption{\footnotesize{Hyperbolic resolution of spatial infinity $i^0$ via dS$_3$ slices of $\mathbb R^{3,1}$.}}
\label{fig:i0}
\end{wrapfigure}

In early works on spatial infinity, Ashtekar and Hansen constructed `Spi' (standing for spatial infinity), a manifold where $i^0$ is tied to null infinity \cite{Ashtekar:1978zz}. They found that the asymptotic symmetry group near spatial infinity shared similarities with the BMS group: the so-called `Spi group' includes an abelian sub-algebra of `Spi supertranslations'. In a different construction where $i^0$ is resolved by means of a hyperbolic slicing (illustrated in Fig. \ref{fig:i0}), Beig and Schmidt provided an explicit ansatz in coordinates that defines an asymptotically flat metric near spatial infinity (see \eqref{BS} below) and studied the associated field equations.

Ashtekar and Romano then showed that spatial infinity could genuinely be considered as a boundary of spacetime by means of a geometrical construction which is very similar to Penrose's description of asymptotic flatness at null infinity \cite{Ashtekar_1992}. As opposed to \cite{Ashtekar:1978zz}, this latter approach did not rely on any assumption about $\mathscr I$ and provided an invariant definition of spatial infinity. Nicely, the latter was shown to reproduce Beig-Schmidt's ansatz upon using a hyperbolic coordinate system.

In those early works, while similarities and potential connections with asymptotic analyses at null infinity were discussed, the explicit realization of BMS symmetries at spatial infinity was absent.  It only appeared in the more recent work of Troessaert \cite{Troessaert:2017jcm}, where it was shown that the asymptotic symmetry algebra associated with the boundary conditions at spatial infinity that were proposed in \cite{Compere:2011ve} does include the (standard) BMS algebra.  In a series of works \cite{Henneaux:2018cst,Henneaux:2019yax,Henneaux:2018hdj}, Henneaux and Troessaert then showed the natural emergence of BMS symmetry at spatial infinity in the Hamiltonian formalism of GR, upon the appropriate choice of parity conditions under the antipodal map on the celestial sphere (see also \cite{Fuentealba:2022xsz}). 

The rest of this section is aimed at reviewing how BMS symmetries can be realized at spatial infinity, and to do so we will adopt the analysis that resembles the most Bondi's expansion of section \ref{sec:Bondi}, i.e. we will focus on the hyperbolic resolution of spatial infinity and the associated Beig-Schmidt's ansatz for the associated asymptotic expansion.
We extensively used Refs. \cite{Compere:2011ve,Troessaert:2017jcm,Compere:2023qoa} (where much more details can be found) to write this section.\\

\noindent \textcolor{pansypurple}{\gr{Beig-Schmidt ansatz \,\,}}  Let us consider a hyperbolic slicing of flat spacetime as depicted in Fig. \ref{fig:i0}, where 
Minkowski metric is mapped, performing the change of coordinates $t=\varrho \sinh \uptau$, $r=\varrho \cosh \uptau$, to
    \begin{equation}\label{dS}
 ds^2=d\varrho^2+\varrho^2 \hat h_{ab}d\phi^a d\phi^b, 
\end{equation}
where the unit hyperboloid $\mathcal H^0$ metric (Lorentzian de Sitter$_3$) of coordinates $\phi^a=(\uptau, x^A)$ reads as follows: 
\begin{equation}
 \hat h_{ab}d\phi^a d\phi^b=-d\uptau^2+\cosh^2\uptau \,\gamma_{AB}dx^A dx^B\,,
\end{equation}
with $\gamma_{AB}$ the unit sphere metric.
A spacetime can thus be called asymptotically flat near spatial infinity if there exists a coordinate system in which it takes the asymptotic form \eqref{dS} as $\varrho \to \infty$. A precise ansatz was given by Beig-Schmidt \cite{BeigSchmidt} (see also \cite{Compere:2011ve})
    \begin{equation}\label{BS}
    \badat{2}
 ds^2&\stackrel{\varrho \to \infty}{=}\left(1+\frac{2\sigma}{\varrho}+\frac{\sigma^2}{ 
 \varrho^2}+o(\varrho^{-2}) \right)d\varrho^2+o(\varrho^{-1}) \,d\varrho d\phi^a\\
 &\quad +\left(\varrho^2 \hat h_{ab}+\varrho (\hat k_{ab}-2\sigma \hat h_{ab})+\log \varrho\, \hat i_{ab}+\hat j_{ab}+o(\varrho^{0})\right)d\phi^a d\phi^b, 
 \eadat
\end{equation}
where the asymptotic fields $\sigma$, $\hat k_{ab}$, $\hat i_{ab}$, $\hat j_{ab}$ depend on $\phi^a$.\\

\noindent \textcolor{pansypurple}{\gr{Log and Spi supertranslations\,\,}} 
Infinitesimal transformations that preserve the ansatz \eqref{BS} take the form
\begin{equation}
\badat{2}
&\zeta^\rho=H \log \varrho + \omega +o(\varrho^0)\\
&\zeta^a=\chi^a+\frac{\log \varrho}{\varrho}\mathcal D^a H+\frac{1}{\varrho }\mathcal 
 D^a(H+\omega)+o(\varrho^{-1})\,,
 \eadat
\end{equation}
where $H$, $\omega$ are scalar functions on the hyperboloid, $\chi^a$ is a Killing vector of $\hat h$, while $\mathcal D^a$ denotes the covariant derivative associated with $\hat h$. Transformations generated by arbitrary $H(\phi^a)$ are the so-called `logarithmic supertranslations'. Allowing for generic $H$ could be used to set $\sigma=0$ but would also turn on a logarithmic term of the form $\varrho \log \varrho (\mathcal D_a \mathcal D_b H+\hat h_{ab}H)d\phi^a d\phi^b$ in \eqref{BS}. Not allowing for this term reduces $H$ down to a four-parameter family (the four solutions of $(\mathcal D_a \mathcal  D_b +\hat h_{ab})H=0$).  Logarithmic translations act on the leading metric fields as 
\begin{equation}
\delta_H\sigma=H \virg \delta_H \hat k_{ab}=0.
\end{equation}
The transformations parametrized by the scalar $\omega$ form the abelian sub-algebra of `Spi supertranslations' originally discussed in \cite{Ashtekar:1978zz}; their action reads
\begin{equation} \label{deltaomegak}
\delta_\omega\sigma=0 \virg \delta_\omega \hat k_{ab}=2(\mathcal D_a \mathcal D_b+\hat  h_{ab})\omega.
\end{equation}

\noindent \textcolor{pansypurple}{\gr{Looking for BMS\,\,}} Where are the BMS symmetries? It seems that there are actually \emph{too many} parameters: we have six Killing fields $\chi^a$, four numbers in $H$, and one free function $\omega$ on $i^0$, while the (standard) BMS group is spanned by six parameters and one free function on the sphere (generating supertranslations).  
The reduction to BMS goes essentially in two steps. The first one is to impose Comp\`{e}re-Dehouck \cite{Compere:2011ve} conditions\footnote{These conditions were motivated by having a well-defined variational principle.} on the asymptotic expansion:
\begin{equation}\label{CDe}
\mathcal D^a \hat k_{ab}=0 \virg \hat k^a_a=0\,.
\end{equation}
The divergence-free condition comes from the leading term of one of the field (Einstein) equations, while the tracefree condition on $\hat k_{ab}$ can always be reached by a change of coordinates (it can thus be seen as some partial gauge fixing for the Spi supertranslations). Using \eqref{deltaomegak}, one can see that the requirement \eqref{CDe} reduces the Spi supertranslations $\omega$ down to the ones that satisfy
\begin{equation}\label{down}
(\mathcal D^2+3)\omega=0.
\end{equation}
Now, the space of solutions of this equation can be split into two pieces, $\omega^{even}$ and $\omega^{odd}$, each of them fully characterized by a function on the sphere (see the appendix of \cite{Troessaert:2017jcm}). The superscript denotes respectively even and odd functions under a combination of time reversal $\tau \to -\tau$ and antipodal map $x^A \to -x^A$ or $(\theta,\phi) \to (\pi-\theta,\phi+\pi)$. Hence at this stage, we still have twice as much as supertranslations as we need. These should be thought of as two (so far) unrelated supertranslations of two distinct BMS groups acting at future/past null infinity.
The second step precisely amounts to killing half of them by imposing \emph{parity conditions} on $k_{ab}$:
\begin{equation}\label{parity_conditionsk}
\Upsilon^*_{\mathcal H}k_{ab}=-k_{ab}\,,
\end{equation}
where we defined as in \cite{Compere:2023qoa} $\Upsilon_{\mathcal H}(\tau,\theta,\phi)=(-\tau,\pi-\theta,\phi+\pi)$.  The odd-parity condition \eqref{parity_conditionsk} restricts to $\omega=\omega^{odd}$ ($\Upsilon^*_{\mathcal H}\omega=-\omega$) whose asymptotic behaviour as $\tau \to \pm \infty$ is found to be \cite{Compere:2023qoa}
\begin{equation}\label{omegaodd}
\badat{2}
&\omega^{odd}(\tau,x^A) \stackrel{\tau \to  +\infty}{\sim} -\frac{1}{2}e^{\tau}T(x^A),\\
&\omega^{odd}(\tau,x^A) \stackrel{\tau \to -\infty}{\sim} +\frac{1}{2}e^{-\tau}T(-x^A)\,,
\eadat
\end{equation}
where $T(x^A)$ is the function on the sphere parametrizing the odd-parity solutions of \eqref{down}.
Therefore, we see that the supertranslations at spatial infinity have reduced to a \emph{single} arbitrary function of the sphere.
The logarithmic translation degrees of freedom are removed when imposing that $\sigma$ is parity-even\footnote{In \cite{Compere:2023qoa}, they instead considered the functions $\sigma^\pm$ such that they vanish in the limit $\tau \to \pm \infty$. Such a choice is physically equivalent as the one taken here as it only differs from \eqref{parity_conditionssigma} by a  logarithmic pure gauge transformation.}, 
\begin{equation}\label{parity_conditionssigma}
\Upsilon^*_{\mathcal H}\sigma=\sigma \,,
\end{equation}
since the four $H$ are odd under time-reversal.
Finally, one can easily note that
the algebra of Killing vector fields $\chi^a$ of the hyperboloid is isomorphic to the algebra of global conformal Killing vector fields of the sphere. We have therefore now exactly the right amount of symmetries, and the commutation relations between  $\omega^{odd}$ and $\chi^a$ show that they indeed span the $\mathfrak{bms}_4$ algebra \cite{Troessaert:2017jcm}.

\subsubsection{Timelike infinity}
\begin{wrapfigure}{r}{6cm}
\includegraphics[scale=0.5,trim = {0.1cm 1cm 0cm 1.5cm}]
{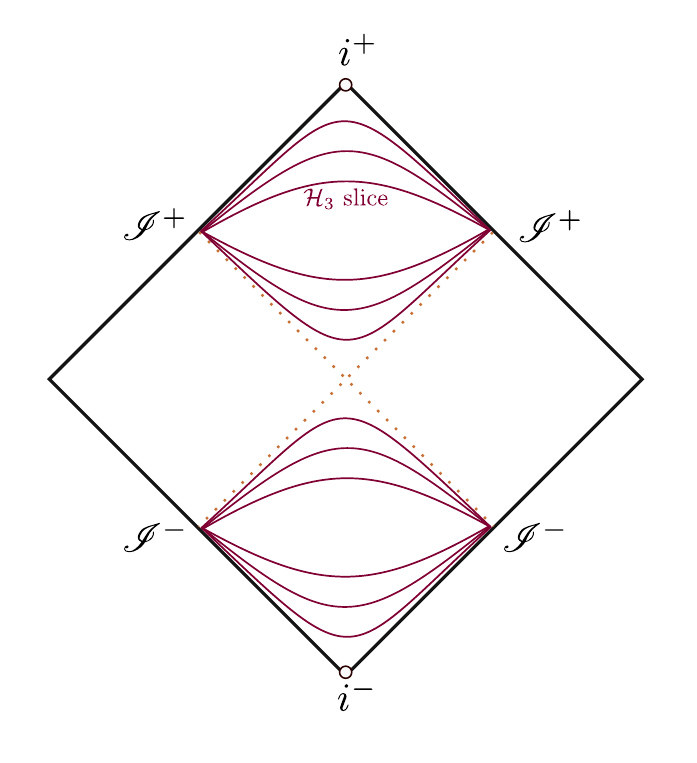}
\captionsetup{width=0.9\linewidth}
\caption{\footnotesize{Hyperbolic slicing of $\mathbb R^{3,1}$ for future (past) timelike infinity $i^+$ ($i^-$).}}
\label{fig:hyper}
\vspace{0.5cm}
\end{wrapfigure}

Timelike infinity is the location where massive particles end.  In the Penrose diagram depicted in Fig. \ref{fig:Penrose}, $i^+$  collapses to a point. One can `resolve' it in a way that is very similar to what we just did with spacelike infinity, where now the hyperbolic slicing is in terms of Euclidean AdS$_3$ slices (sometimes called Milne wedges), as illustrated in Fig. \ref{fig:hyper}.
The change of coordinates $t=\tau \cosh \rho$, $r=\tau \sinh \rho$ maps Minkowski metric to
\begin{equation}
ds^2=-d\tau^2+\tau^2h_{ab}dx^a dx^b\,,
\end{equation}
with $h_{ab}$ the metric of the unit hyperboloid $\mathcal H^+$ (Euclidean AdS$_3$) of coordinates $x^a=(\rho,x^A)$
\begin{equation}\label{i+}
h_{ab}dx^a dx^b=d\rho^2+\sinh^2\rho \,\gamma_{AB}dx^A dx^B.
\end{equation}
We will denote by $D^a$ the derivative compatible with $h_{ab}$.
As $\tau \to \infty$, massive free particles of a given momentum asymptote to constant $x^a$ and are assigned to a point on the hyperboloid.
Notice that the expression \eqref{i+} is related to the one for spatial infinity \eqref{dS} by analytic continuation:
\begin{equation}
\varrho=i\tau  \virg  \hat \tau=\rho-\frac{i\pi}{2} \virg \hat h=-h\,.
\end{equation}

From this analytic continuation, one can prescribe an asymptotic expansion near $i^+$ which is the timelike infinity analog of the Beig-Schmidt ansatz \eqref{BS}
\cite{Compere:2023qoa,Chakraborty:2021sbc}:
    \begin{equation}\label{BStimelike}
    \badat{2}
 ds^2&\stackrel{\tau \to \infty}{=}\left(-1-\frac{2\upsigma}{\tau}-\frac{\upsigma^2}{\tau^2}+o(\tau^{-2}) \right)d\tau^2+o( 
 \tau^{-1})\, d\tau dx^a\\
 &+\left(\tau^2 h_{ab}+\tau (k_{ab}-2\upsigma  h_{ab})+\log \tau\, i_{ab}+j_{ab}+o(\tau^{0})\right)dx^a dx^b\,,
 \eadat
\end{equation}
with $\upsigma$, $k_{ab}$, $i_{ab}$, $j_{ab}$ fields on $\mathcal H^+$.

Similarly with the previous analysis, infinitesimal transformations preserving \eqref{BStimelike} are found to be
\begin{equation}
\badat{2}
&\xi^\tau=\hat H \log \tau + \hat \omega +o(\tau^0)\\
&\xi^a=\psi^a-\frac{\log \tau}{\tau}D^a H-\frac{1}{\tau } 
 D^a(\hat H+\hat \omega)+o(\tau^{-1})\,,
 \eadat
\end{equation}
where $\hat \omega(x^a)$ is a free function on $i^+$, $\hat  H(x^a)$ satisfies $(D_a D_b-h_{ab})\hat H=0$ and generates a four-parameter family of transformations, while $\psi^a$ are the 6 Killing vectors on the hyperboloid. We thus find ourselves in a situation which is very similar to what was discussed above for $i^0$.

The recovery of the BMS group follows in this case from the following extra assumptions  \cite{Compere:2023qoa}: i) imposing $ \underset{\rho \to \infty}{\lim}\upsigma=0$ removes the logarithmic translations $H$\footnote{It is also useful in order to match with $\mathscr I^+$; see \cite{Compere:2023qoa}.};
ii) setting the trace of $k$ to vanish which in turn implies,
\begin{equation}\label{omegai+}
    (D^2-3)\hat \omega=0\,.
\end{equation}
The latter equation gives a solution that is determined by a single function on the sphere.
It is interesting to contrast the situation here with the $i^0$ case discussed previously. While at spatial infinity the equation \eqref{down} allowed for \emph{two} branches of solutions and we had to restrict to the odd-parity branch, at timelike infinity all regular solutions to \eqref{omegai+} are in one-to-one correspondence with (smooth) functions on the sphere. Notice that this difference has nothing to do with the sign difference ($+$3 vs $-3$) in \eqref{down} and \eqref{omegai+} but is due to the difference of signature of the two metrics. The latter changes the nature of the differential equation, which is hyperbolic in \eqref{down} but elliptic in \eqref{omegai+}. 

 Having found a single function on the sphere, the $\mathfrak{bms}_4$ algebra can then be checked to be spanned by the latter, together with the Killing vectors $\psi^a$ of the hyperboloid \cite{Chakraborty:2021sbc,Compere:2023qoa}.
Notice that it would have reduced to the Poincar\'e algebra if one would have adopted the too-restrictive condition $k_{ab}=0$.

\subsubsection{On matching conditions}

As we have just seen, some reasonable assumptions on the Cauchy data near spacelike and timelike infinities can be identified so that the asymptotic symmetric group reduces to the BMS group. It is therefore reasonable to expect that one could recover in this way Strominger's diagonal BMS group \cite{Strominger:2013jfa}. For spatial infinity, expressions \eqref{omegaodd} were already strongly suggesting an antipodal map between the supertranslation parameter at $\mathscr I^+$ and $\mathscr I^-$. And indeed, one can explicitly check that antipodal relationships hold around spatial infinity, for the symmetry generators but also between fields (mass aspect and shear)~\cite{Troessaert:2017jcm,Compere:2023qoa}. 

In the set-up described above, the matching between all pieces, $i^+$ with $\mathscr I^+$, matching across $i^0$ and  matching between $i^-$ and $\mathscr I^-$, requires enforcing certain assumptions about the behavior of Bondi and Beig-Schmit quantities at large $u$ and large $\rho$; see \cite{Compere:2023qoa} for detail. In general, the question of matching conditions in asymptotically flat spacetimes is extremely delicate and several important questions remain open. In particular, proving that such matching conditions arise from an evolution of characteristic data at scri would be of great relevance. We refer the interested reader to references \cite{Herberthson:1992gcz,Friedrich:2002ru,Troessaert:2017jcm,Henneaux:2018cst,Mohamed:2021rfg,Prabhu:2019fsp,Prabhu:2021cgk,Capone:2022gme,Compere:2023qoa,Henneaux:2023neb} for further material of this important topic.

\newpage

\section{BMS charges and fluxes}
\label{chap:charges}
\begin{wrapfigure}{r}{6cm}
\includegraphics[scale=0.5,trim = {0.1cm 1cm 0cm 1.5cm}]
{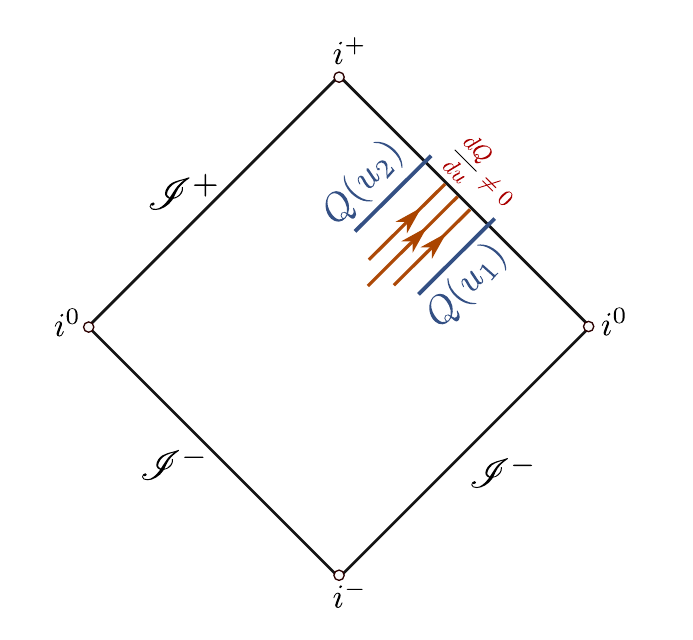}
\captionsetup{width=0.9\linewidth}
\caption{\footnotesize{Non-conservation of surface charges $Q$ due to outgoing flux of radiation at $\mathscr I^+$ \cite{Donnay:2022wvx}.}}
\label{fig:fluxes}
\end{wrapfigure}

Noether's theorems establish a profound and elegant correspondence between symmetries and conservation laws. The construction of conserved quantities in General Relativity with boundary conditions at null infinity is a notorious subtle task, as in general symplectic current will be radiated away and quantities defined at $\mathscr I$ will not be conserved; see Fig. \ref{fig:fluxes}.
This contrasts with the case of spatial infinity, where conserved quantities can be constructed in a clear and straightforward manner
from the Hamiltonian formulation of GR. Since the seminal work of Wald and Zoupas \cite{Wald:1999wa}, much more has been understood about the status of `conserved quantities' at generic spacetime boundaries, including null surfaces at infinity but also at finite distance; we refer the reader to \cite{Ciambelli:2022vot} for a fairly complete list of references on the matter.

Surface charges associated with (extended) BMS symmetries were first constructed in the pioneer work \cite{Barnich:2011mi}. There, they were shown to realize the extended $\mathfrak{bms}_4$ algebra up to a field-dependent central extension\footnote{The same conclusion was obtained in \cite{Distler:2018rwu} from an analysis of consecutive double soft limits of graviton tree-level amplitudes.}.   Since this generalized cocycle vanishes when restricting to globally well-defined (i.e. standard) BMS algebra, its presence is tied to the inclusion of superrotation symmetries. This in turn implied that, in order to find a realization of the extended BMS algebra as a symmetry of quantum gravity, some subtle aspects regarding the definition of superrotation charges had to be addressed.

The results we will present in this chapter came out as the output of the efforts of many different works from different perspectives. Rather than attempting to describe the somehow complex road that has led to them, we will present what we believe is the main output needed in order to bridge between asymptotic symmetries and key objects in celestial holography.
We will start in section \ref{sec:charges} with a presentation of BMS surface charges, which are defined at a cut of null infinity.  The existence of Goldstone modes associated with supertranslation and superrotation symmetries will then lead us in section \ref{sec:vacuum} to a detailed analysis of the vacuum structure. The latter will be used in section \ref{sec:sympl} to define a symplectic structure associated with an extended phase space including superrotation symmetries. We will finish this chapter with a presentation in section \ref{sec:BMSfluxes}  of BMS fluxes as well as their splitting into their so-called hard and soft pieces and show that the latter separately form a representation of the extended $\mathfrak{bms}_4$ algebra without any cocycle. We will moreover see that they give rise to conformal primaries of definite (half-)integer weights of the celestial sphere.
The main references used in this chapter are \cite{Campiglia:2021bap,Donnay:2021wrk,Donnay:2022hkf}.

\subsection{Surface charges and fluxes}
\label{sec:charges}

BMS surface charges are defined at a constant-$u$ cut of $\mathscr I^+$\footnote{As usual, an analogous analysis holds for past null infinity $\mathscr I^-$.} and pair asymptotic symmetry parameters $\xi=\xi(\mathcal{T},Y)$, the supertranslations and superrotations of section \eqref{sec:extendedBMS} together with functions of the gravitational phase space. Following up on earlier works \cite{Barnich:2011mi,Barnich:2011ty,Kapec:2014opa,Flanagan:2015pxa}, a `good prescription' for these charges has emerged in the literature \cite{Compere:2018ylh,Campiglia:2020qvc,Compere:2020lrt,Compere:2021inq,Fiorucci:2021pha,Donnay:2021wrk,Freidel:2021qpz,Freidel:2021dfs,Freidel:2021ytz,Donnay:2022hkf} 
\begin{equation}
    Q_\xi= \frac{1}{16\pi G} \int d^2 z\, \big(4 \mathcal{T} \mathscr M + 2 Y^A\mathscr N_A\big) , \label{grav charge}
\end{equation}
where the mass $\mathscr M$ and angular momentum aspect $\mathscr N_A$ are given by
\begin{equation}
\begin{split}
   \mathscr  M = M &+ \frac{1}{8}N_{AB}C^{AB} , \\
    \mathscr N_A = N_A &- u\, \partial_A    \mathscr  M  + \frac{1}{4}C_{AB} \partial_C C^{BC} +\frac{3}{32}\partial_A (C_{BC}C^{BC}) \\
    &+ \frac{u}{4}  \partial^B \big[(\partial_B \partial_C - \frac{1}{2}N_{BC}) {C_A}^C \big] - \frac{u}{4} \partial^B \big[(\partial_A \partial_C - \frac{1}{2}N_{AC}) {C_B}^C\big] . \label{BMS momenta}
    \end{split}
\end{equation}
The above quantities entering the surface charges might look rather mysterious, as expressed in  \eqref{BMS momenta} in terms of the Bondi asymptotic expansion functions.  They in fact admit a much more elegant expression in terms of Newman-Penrose coefficients
\cite{Newman:1961qr,Newman:1962cia} as the compact expressions (see appendix A of \cite{Donnay:2022hkf} for details)
\begin{equation}
\badat{2}
&  \mathscr  M = -\frac{1}{2}(\Psi_2^0+\bar\Psi_2^0)\\
&\mathscr  N_z = -\Psi_1^0+u\,\partial_z \Psi_2^0
\eadat
\end{equation}and $\mathscr  N_{\bar z} = \mathscr N_z^*$. These expressions also coincide with the `covariant observables' that were identified in \cite{Freidel:2021qpz,Freidel:2021dfs,Freidel:2021ytz}, where they nicely provide a definition of conserved charges parametrizing the non-radiative corner phase space.

As mentioned in the beginning of this section, because of gravitational leaks through the boundary, surface charges are \emph{not} conserved in time (see Fig. \ref{fig:fluxes}), but their evolution is encoded into `flux-balance laws' 
\begin{equation}
    \frac{d Q_\xi}{d u} = \int d^2 z\, F_{\xi(\mathcal T,Y)} \neq 0 , \label{flux balance BMS 1}
\end{equation}
where the (local) fluxes $F_\xi$ can be obtained from the constraint equations \eqref{EOM1}.

Writing separately the fluxes associated with supertranslations and superrotations, one finds, using \eqref{EOM1} for the flat sphere representative as wel as \eqref{grav charge}, \eqref{BMS momenta}, we can write\footnote{See \cite{Donnay:2022wvx,Agrawal:2023zea} for expressions with the inclusion of matter fields.}
\begin{equation}
    \badat{2}
    &F_{\xi(\mathcal{T},0)} = \frac{1}{16\pi G}\,\mathcal{T}\left[\partial_z^2 N_{\bar z\bar z} + \frac{1}{2} C_{\bar z\bar z}\partial_u N_{zz}+\text{c.c.} \right] 
    \\
    &F_{\xi(0,Y)} = \frac{1}{16\pi G}\,Y^z \left[-u\partial^3_z N_{\bar z\bar z} + C_{zz}\partial_z N_{\bar z\bar z} - \frac{u}{2}\partial_z C_{zz}\partial_u N_{\bar z\bar z} - \frac{u}{2}C_{zz}\partial_z\partial_u N_{\bar z\bar z}\right]
  +\text{c.c.}  \,
    \label{total_fluxes}
    \eadat
\end{equation}
As a subset of the relationships \eqref{flux balance BMS 1}, one can easily recover for $\mathcal{T}=1$ the famous mass loss formula \cite{Bondi:1962px}
\begin{equation}
    \frac{d m_B}{d u} = -\frac{1}{8}\int d^2 z\,N_{AB}N^{AB}, \label{mass_loss}
\end{equation}
for the Bondi mass $m_B(u)=\int d^2 z\, M(u,z,\bar z)$. What we have now on top of this formula is an infinite tower of flux-balance laws, one for each supertranslation (and also superrotation) parameter. Importantly, BMS flux-balance laws provide consistency constraints on the waveforms that are generated by black hole binary mergers (see \cite{Compere:2019gft} and references therein).

\subsection{Goldstone and radiative modes}
\label{sec:vacuum}
Of chief importance in understanding the infrared (IR) structure of gravity is the realization that, because there is no preferred Poincar\'e subgroup inside the BMS group, \emph{the vacuum in gravity is infinitely degenerate} \cite{Ashtekar:1987tt,Ashtekar:2014zsa,Strominger:2014pwa,Strominger:2017zoo}.
An easy way to see the origin of the vacuum degeneracy is the following: suppose you start with Minkowski spacetime, which is trivially described by the Bondi expansion \eqref{Bondi gauge metric2} with $M=N_z=C_{zz}=0$ and that you act on it with a supertranslation. Naively, one might just say that since we are acting with a mere diffeomorphism, we will land again on flat spacetime. This is correct, the mass or energy will not be changed, \emph{however},  the action of BMS supertranslation will move us to a different vacuum! Indeed, the vanishing of the curvature requires that the supertranslated Minkowski space now has\cite{Strominger:2017zoo}
\begin{equation}
    C_{zz}=-2\partial_z^2 C(z,\bar z) \neq 0,
\end{equation}
with $C(z,\bar z)$ some function on the celestial sphere. $C$ thus admits to be  interpreted as a `Goldstone boson' as it encodes the spontaneous breaking of supertranslation symmetry\footnote{Since the ordinary four spacetime translations are precisely those which satisfy $\partial_z^2 \mathcal T=0$, we see that they are not broken.} and transforms under an infinitesimal supertranslation as a pure inhomogeneous shift (see the third expression in \eqref{transformation on the solution space}):
\begin{equation}
    \delta_{\mathcal T}C(z,\bz)=\mathcal T(z,\bz)\,.
\end{equation}
 Notice that nothing in the above argument depends on the choice of coordinates. There exists a geometrically precise and invariant manner to understand the `space of gravity vacua' as the space of flat normal tractor connections; see \cite{Herfray:2020rvq,Herfray:2021xyp,Herfray:2021qmp}.
Crucially, this phenomenon of vacuum transition by the action of BMS supertranslation gives rise to a low energy \emph{observable} effect. It is  indeed by now well understood that the difference between the value of the asymptotic shear at late and early time is in one-to-one correspondence with the gravitational (displacement) memory effect \cite{Ashtekar:2014zsa,Strominger:2014pwa,Strominger:2017zoo}. \\

\noindent \textcolor{pansypurple}{\gr{Goldstone modes\,\,\,}}In order to encompass those important features, one is forced to take into account the behavior of the radiative field as one approaches $\mathscr I^+_-$ (the $u\to -\infty$ limit of $\mathscr I^+$) and $\mathscr I^+_+$ (the $u\to +\infty$ limit of $\mathscr I^+$), depicted at the corners of Fig. \ref{fig:Penrose}. In this report, we will adopt the following early/late time conditions\footnote{This can be relaxed to allow for gravitational tails; see \cite{Compere:2023qoa,Agrawal:2023zea}.}
\cite{Campiglia:2021bap,Campiglia:2020qvc,Compere:2020lrt,Donnay:2021wrk}:
\begin{equation}
    C_{zz}\stackrel{u\to \pm \infty}{=}  - 2 \partial_z^2 C_\pm + (u+C_\pm)N^{vac}_{zz} +o(u^{-1}),\quad N_{zz} \stackrel{u\to \pm \infty}{=} N_{zz}^{vac} + o(u^{-2})\,, \label{falloff in u for SR}
\end{equation}
and similarly for the $\bar z \bar z$ components. $C_\pm(z,\bar z)$ denotes the value of the supertranslation field at $\mathscr I^+_\pm$, respectively; it transforms under the extended BMS symmetries as \cite{Compere:2016jwb,Himwich:2020rro}
\begin{equation}
\badat{2}
   & \delta_{\xi(\mathcal T, Y)} C_{\pm} =  \left(Y^z \partial_z + Y^\bz \partial_\bz - \frac{1}{2} \partial_z Y^z - \frac{1}{2} \partial_\bz Y^\bz \right) C_\pm + \mathcal T \,.
    \label{transfo Cpm}
    \eadat
\end{equation} 
Defining the sum and difference of these boundary values at future null infinity by
\begin{equation}\label{C and N}
   \mathscr C\equiv \frac{1}{2}(C_++C_-) \virg  \mathscr N\equiv \frac{1}{2}(C_+-C_-)\,,
\end{equation}
we see from \eqref{transfo Cpm} that they both transform as $(-\12,-\12)$ primaries; see Table \ref{table:weights}.  The Goldstone mode $\mathscr C$ transforms with an extra supertranslation shift, while $\mathscr N$ is a difference between the early and late time value of the field and is sometimes referred to as the `memory mode'\footnote{See \cite{Nande:2017dba} for the QED analogs of these modes.}.

Let us now discuss the $N_{zz}^{vac}$ appearing in \eqref{falloff in u for SR}. In the same way that supertranslations move us from one vacuum to another, different superrotation vacua are labeled by the so-called `vacuum news tensor' $N_{zz}^{vac}$~\cite{Compere:2016jwb,Compere:2018ylh,Campiglia:2020qvc}\footnote{$N_{zz}^{vac}$ is nothing but the tracefree part of the Geroch tensor  \cite{Campiglia:2020qvc,1977asst.conf....1G} (see also \cite{Nguyen:2020hot}).}. It is built out of a holomorphic potential $\varphi(z)$, sometimes called the `superboost' or `Liouville' field
\begin{equation}
    N_{zz}^{vac} = \frac{1}{2}(\partial_z\varphi)^2 - \partial_z^2 \varphi \,,
    \label{Nvac}
\end{equation}
whose transformation rule is (see \cite{Compere:2018ylh} for details)
\begin{equation}
    \delta_{\xi(\mathcal T, Y)} \varphi = Y^z \partial_z \varphi + \partial_z Y^z\,,
\end{equation}
    so that
\begin{equation}
    \delta_{\xi(\mathcal T, Y )} N_{zz}^{vac} = (Y^z \partial_z + 2 \partial_z Y^z) N_{zz}^{vac} - \partial_z^3 Y^z\,.
    \label{transfo Nvac}
\end{equation} 
This is the transformation rule of a $2d$ stress tensor of unit central charge\footnote{The relationship between soft superrotation dynamics and the Lyapunov behaviour of CCFT was discussed in \cite{Pasterski:2022lsl}.}.
Notice that the vacuum news can be written in a manner that is reminiscent of the AdS$_3$ literature using the Schwarzian derivative  $\{f,z\}=\frac{f'''}{f'}-\frac{3}{2}\left(\frac{f''}{f'}\right)^2$ as
\begin{equation}
 N_{zz}^{vac}=-\{G(z),z\}
\end{equation} 
with $G'(z)=e^{\varphi}$.
As always, there is also an anti-holomorphic piece $ N_{\bz \bz}^{vac}$ built out of $\bar{\varphi}(\bz)$ that we will often omit writing. 

The Liouville fields $\varphi$, $\bar{\varphi}$  can be used to build the superrotation-covariant derivative operators~\cite{Barnich:2021dta,Campiglia:2020qvc,Donnay:2021wrk}
\begin{equation}
    \begin{split}
       \mathscr{D}_z\phi_{h, \bar{h}} = [\partial_z - h \partial_z \varphi] \phi_{h, \bar{h}}\,,  \qquad 
       \mathscr{D}_\bz \phi_{h, \bar{h}} =  [\partial_\bz - \bar{h} \partial_\bz \bar{\varphi}] \phi_{h, \bar{h}}\,,
    \end{split}  \label{derivative operators conformal}
\end{equation} which have the nice feature of proliferating primary descendants out of a conformal field $\phi_{h, \bar{h}}$. Indeed, from the defining property \eqref{def conformal field},  $\mathscr{D}_z^n\phi_{h, \bar{h}}$, $ \mathscr{D}_\bz ^n\phi_{h, \bar{h}}$ can be checked to transform as primary fields of weights $(h+n, \bar{h})$ and $(h, \bar{h}+n)$, respectively. Notice that they also satisfy $[\mathscr{D}_z, \mathscr{D}_\bz] \phi_{h, \bar{h}} = 0$.

Using those derivatives, one can construct the second descendants of the modes \eqref{C and N} which are called, respectively, the Goldstone current $\mathscr C_{zz}$ and the leading soft graviton operator $\mathscr N^{(0)}_{zz}$ \cite{He:2014laa} :
\begin{equation}\label{Nzzzero}
  \mathscr C_{zz}=-2\mathscr D_z^2   \mathscr C \virg  \mathscr N^{(0)}_{zz}=-4\mathscr D_z^2  \mathscr N\,.
\end{equation}
Their transformation rules under superrotations are the ones of $(\32,-\12)$ primaries as expected:
\begin{equation}
    \begin{split}
       &\delta_{\xi(\mathcal T, Y)} {\mathscr C}_{zz} =  \left(Y^z \partial_z + Y^\bz \partial_\bz + \frac{3}{2} \partial_z Y^z - \frac{1}{2} \partial_\bz Y^\bz \right) \mathscr C_{zz} -2 \mathscr{D}_z^2 \mathcal T \,, \\
       &\delta_{\xi(\mathcal T, Y)} \mathscr{N}^{(0)}_{zz} = \left(Y^z \partial_z + Y^\bz \partial_\bz + \frac{3}{2} \partial_z Y^z - \frac{1}{2} \partial_\bz Y^\bz \right) \mathscr{N}^{(0)}_{zz} \,.\\
    \end{split} 
    \label{variations C}
\end{equation}
We will see later in section \ref{sec:sympl} that these two modes are symplectically paired with each other, see Table \ref{table:Goldstone}. \\

\noindent \textcolor{pansypurple}{\gr{Radiative modes \,\,}} Having singled out above the zero modes, we can now define the  `shifted' shear and news tensors,
\begin{equation}
  \hat C_{zz}\equiv C_{zz}-  \mathscr C_{zz}-uN_{zz}^{vac} \virg  \hat N_{zz}\equiv N_{zz}-N_{zz}^{vac}\,,
\end{equation}which are such that they transform homogeneously under the action of BMS. Indeed, from \eqref{transformation on the solution space}, \eqref{variations C} and \eqref{transfo Nvac}, one finds
\begin{equation}
    \begin{split}
    &\delta_{\xi(\mathcal T, Y)} \hat{C}_{zz} =  \left(Y^z \partial_z + Y^\bz \partial_\bz + \frac{3}{2} \partial_z Y^z - \frac{1}{2} \partial_\bz Y^\bz \right) \hat{C}_{zz} + \left( \mathcal T + u\alpha \right) \hat{N}_{zz} \,, \\
    &\delta_{\xi(\mathcal T, Y)} \hat{N}_{zz} =  \left(Y^z \partial_z + Y^\bz \partial_\bz + 2 \partial_z Y^z  \right) \hat{N}_{zz} + \left(\mathcal T + u \alpha \right) \partial_u \hat{N}_{zz} \,,
    \end{split} 
    \label{variations physical}
\end{equation}
with $\alpha=\frac{1}{2}\partial_A Y^A$. These variables will sometimes be called `hard' (as opposed to `soft') variables.
The $u$-fall off conditions \eqref{falloff in u for SR} imply that the asymptotic shear is purely `electric'\footnote{This condition is equivalent to setting the magnetic mass aspect to zero: Im$\Psi^{(0)}_2|_{\mathscr{I}^+_\pm}=0$.} at early and late times
\begin{equation}
\begin{split}\label{elec}
\left(\mathscr D_z^2 \hat{C}_{\bz \bz}-\mathscr D_\bz^2 \hat{C}_{zz}\right)|_{\mathscr{I}^+_\pm}=0\,,
\end{split}
\end{equation}
and that the spacetime is non-radiative at the corners $\hat{N}_{zz}|_{\mathscr{I}^+_\pm} = 0$. The electricity condition \eqref{elec}
is solved by
\begin{equation}
\begin{split}
     &\hat{C}_{zz}|_{\mathscr{I}^+_\pm} = \mp 2  \mathscr{D}_z^2 \mathscr N\,, 
\end{split}
\end{equation}
with $\mathscr N$ given in \eqref{C and N}.
Finally, in the same way that, at quantum level, \eqref{Nzzzero} plays the role of the leading soft graviton operator, the subleading soft graviton is denoted by $\mathscr N^{(1)}_{zz}$; they can both be expressed as a $u$-integral of the shifted news as \cite{He:2014laa,Kapec:2014opa}
\begin{equation}\label{soft_operators}
   \mathscr N_{zz}^{(0)}=\int_{-\infty}^{+\infty}du \, \hat N_{zz}(u,z,\bz) \virg  \mathscr N_{zz}^{(1)}=\int_{-\infty}^{+\infty}du \,u \hat N_{zz}(u,z,\bz) \,.
\end{equation}

\begin{table}[h]
\begin{center}
\renewcommand*{\arraystretch}{1.2}
  \begin{tabular}{  c | c | c }
& Supertranslation & Superrotation \\
  \hline 
Goldstone current & $\mathscr C_{zz}$ & $N_{zz}^{vac}$\\
\hline
Symplectic partner & $\mathscr N^{(0)}_{zz}$ & $\Pi_{zz}$\\
\hline
    \end{tabular}
    \caption{Supertranslation and superrotation Goldstone currents \eqref{Nzzzero} and \eqref{soft_operators} with below their corresponding symplectic partners defined in \eqref{Nvac} and \eqref{pi}.}
  \label{table:Goldstone}
\end{center}
\end{table}

\subsection{Symplectic structure}
\label{sec:sympl}
As we have seen in the previous section, the gravitational phase space is composed of a `hard' sector $\Gamma^H$ which collects the (shifted) shear and news tensors: 
\begin{equation}
    \Gamma^{H} = \{\hat C_{zz},\hat C_{\bar z\bar z},\hat N_{zz},\hat N_{\bar z\bar z}\}. \label{Gamma hard}
\end{equation}
On the other hand, the `soft' sector $\Gamma^S$ is composed of variables associated to the vacuum structure:
\begin{equation}
    \Gamma^{S} = \{\mathscr C_{zz},\mathscr C_{\bar z\bar z}, \mathscr N^{(0)}_{zz},\mathscr N^{(0)}_{\bar z\bar z},\Pi_{zz},\Pi_{\bar z\bar z},N^{vac}_{zz},N^{vac}_{\bar z\bar z}\}\label{Gamma soft}\,,
\end{equation}
where $\Pi_{zz}$ is the composite field identified in \cite{Campiglia:2021bap}:
\begin{equation}\label{pi}
    \Pi_{zz}\equiv 2\mathscr N_{zz}^{(1)}+\mathscr 
 C \mathscr N_{zz}^{(0)}\,,
\end{equation}
whose transformation law can be easily checked to be
\begin{equation}
    \begin{split}
       &\delta_{(\mathcal T, \mathcal Y)} \Pi_{zz} = \left(\mathcal Y^z \partial_z + \mathcal Y^\bz \partial_\bz + \partial_z \mathcal Y^z - \partial_\bz \mathcal Y^\bz \right)  \Pi_{zz} - \mathcal T \mathscr N_{zz}^{(0)}\,.\\
    \end{split} 
    \label{deltaPi}
\end{equation}
A symplectic form $\Omega$ on this extended BMS phase space was proposed by Campiglia and Laddha \cite{Campiglia:2021bap}. 
It decomposes accordingly into a hard and a soft sector and is given in terms of the above-defined quantities by
\begin{equation}
\begin{split}
    &\Omega =  \Omega^{hard} + \Omega^{soft} \,,\\
    &\Omega^{hard} = \frac{1}{32\pi G} \int_{\mathscr{I}^+} du d^2 z \left[ \delta \hat{N}_{zz} \wedge \delta \hat{C}_{\bar{z}\bar{z}}  + c.c.\right] \,, \\
    &\Omega^{soft} = \frac{1}{32\pi G} \int  d^2 z \left[\delta \mathscr{N}^{(0)}_{zz} \wedge \delta \mathscr C_{\bar{z}\bar{z}}  + \delta \Pi_{zz}  \wedge \delta N_{\bar{z}\bar{z}}^{vac} +c.c.   \right] \,.
\end{split}
\label{symplectic structure full}
\end{equation} 
One can notice that this symplectic structure generalizes the one introduced in the early works of Ashtekar-Streubel (AS)
 \cite{Ashtekar:1978zz,Ashtekar:1981bq,Ashtekar:1981sf}
\begin{equation}
    \Omega_{AS} = \frac{1}{32\pi G} \int_{\mathscr{I}^+} du d^2z \left[\delta N_{zz} \wedge \delta C_{\bar{z}\bar{z}} + c.c. \right]\,,
\end{equation} 
to include superrotation vacua\footnote{$\Omega$ and $\Omega_{AS}$ can be seen to differ by a corner term ambiguity, see \cite{Donnay:2022hkf}.}.

The Poisson bracket associated with the above symplectic structure is defined such that for two functions $f$, $g$ on the phase space, we have $\{ f, g \} = i_{X_f} i_{X_g} \Omega$, where the Hamiltonian vector field $X_f$ satisfies $i_{X_f} \Omega = \delta f$. One can thus extract from the symplectic structure \eqref{symplectic structure full} that the non-vanishing Poisson brackets are given by~\cite{Campiglia:2021bap,Donnay:2022hkf}
\begin{equation}
    \begin{split}  
    &\{ \hat{N}_{zz}(u), \hat{C}_{\bar{w}\bar{w}}(u') \} = -16\pi G\,  \delta^{(2)} (z-w) \delta (u-u')\,,\\
    & \{ \mathscr{N}^{(0)}_{zz} , \mathscr C_{\bar{w}\bar{w}}\} = -16\pi G\,  \delta^{(2)}(z-w) \,,\\
     & \{ \Pi_{zz}, N_{\bar{w}\bar{w}}^{vac}  \} =- 16 \pi G\, \delta^{(2)}(z-w)\,.
    \end{split}
    \label{non vanishing brackets}
\end{equation}
This shows that $\mathscr C_{\bar{z}\bar{z}}$ is the symplectic partner of the leading soft graviton operator, while the vacuum news $N_{\bar{z}\bar{z}}^{vac}$ is paired with $\Pi_{zz}$; see Table \ref{table:Goldstone}. Notice that setting to zero the Liouville fields $\varphi=0=\bar \varphi$ one recovers the Dirac brackets earlier obtained in \cite{He:2014laa}.

\subsection{BMS hard and soft fluxes}
\label{sec:BMSfluxes}
The BMS fluxes 
\begin{equation}
  \mathcal F_\xi =\int du d^2z F_\xi \equiv \int d^2 z \,\left( \mathcal T \mathcal P + \mathcal Y^A \mathcal J_A \right)
\end{equation}
introduced in section \ref{sec:charges} can be derived from the symplectic structure \eqref{symplectic structure full} presented above. To do so, one has to contract the latter together with the Hamiltonian vector fields generating BMS symmetries:
\begin{equation}
    i_{\delta_{\xi(\mathcal{T}, Y)}} \Omega^{soft} = \delta \mathcal F^{soft}_{\xi(\mathcal{T}, Y)}\,, \qquad i_{\delta_{\xi(\mathcal{T}, Y)}} \Omega^{hard} = \delta \mathcal F^{hard}_{\xi(\mathcal{T}, Y)}\,,
    \label{canonical generators}
\end{equation}
with $\mathcal F^{soft}_\xi +\mathcal F^{hard}_\xi=\mathcal F_\xi$.
After some lengthy computations which make use of the variations \eqref{variations C}, \eqref{variations physical}, \eqref{transfo Nvac}, \eqref{deltaPi}, one can derive that the hard and soft parts of the supertranslation flux $\mathcal F_{\mathcal{T}}\equiv \mathcal F_{\xi(\mathcal{T},0)}$ are  \cite{Donnay:2022hkf}
\begin{equation}
\begin{split}
     &\mathcal F_{\mathcal{T}}^{hard} = - \frac{1}{16\pi G} \int_{\mathscr{I}^+} du d^2z\, \mathcal T \left[ \hat{N}_{zz}  \hat{N}_{\bar{z}\bar{z}} \right] \,,\\
     &\mathcal F_{\mathcal{T}}^{soft} = \frac{1}{8\pi G} \int d^2 z\, \mathcal T \left[ \mathscr{D}_z^2\mathscr{N}^{(0)}_{\bar{z}\bar{z}} \right]\,.
\end{split} \label{supertranslation fluxes}
\end{equation}
Superrotation fluxes $\mathcal F_{Y}\equiv \mathcal F_{\xi(0,Y)}$ are in turn given by
\begin{equation}
\begin{split}
     &\mathcal F_{Y^z}^{hard} = \frac{1}{16\pi G} \int_{\mathscr{I}^+} du d^2 z\,  Y^z \left[\frac{3}{2}\hat C_{zz} \partial_z \hat N_{\bar z \bar z}+ \frac{1}{2}\hat N_{\bar z \bar z} \partial_z \hat C_{zz}+\frac{u}{2} \partial_z (\hat N_{zz} \hat N_{\bar z \bar z})\right] \,,\\
     &\mathcal F_{Y^z}^{soft} = \frac{1}{16\pi G}\int d^2 z\,  Y^z  \left[-{\mathscr D}_z^3 \mathscr N_{\bar z \bar z}^{(1)} + \frac{3}{2} \mathscr C_{zz} \mathscr{D}_z \mathscr {N}^{(0)}_{\bar z \bar z}+ \frac{1}{2}\mathscr {N}^{(0)}_{\bar z \bar z} \mathscr{D}_z \mathscr C_{zz}\right] \,,
\end{split}
\label{superrotation fluxes}
\end{equation}
together with a similar expression for the fluxes associated with anti-holomorphic superrotations $Y^{\bar z}$.\\

\noindent \textcolor{pansypurple}{\gr{Supermomenta as primaries\,\,}} Anticipating section \ref{chap:last}, we will see that the conformally soft sector of celestial holography is described by conformal primaries of integer dimension. Since these CFT-like currents are intrinsically living on the celestial sphere, it is natural that, from the bulk point of view, they are related to BMS fluxes. In particular, writing 
\begin{equation}
  \mathcal F_\xi^{soft} = \int d^2 z \,\left( \mathcal T \mathcal P^{soft}  + \mathcal Y^A \mathcal J_A^{soft}  \right)\virg  \mathcal F_\xi^{hard} = \int d^2 z \,\left( \mathcal T \mathcal P^{hard}  + \mathcal Y^A \mathcal J_A^{hard}  \right)\,,
\end{equation}
 we can single out the celestial primary fields $\mathcal P$, $\mathcal J_A$ whose both hard and soft parts  transform separately in the coadjoint representation of BMS \cite{Barnich:2021dta,Donnay:2021wrk}:
\begin{equation}
\badat{2}
\label{delta P and J}
        &\delta_{(\mathcal{T}, Y)} \mathcal{P}^{soft}  = \left(Y^z \partial_z + Y^\bz \partial_\bz + \frac{3}{2} \partial_z Y^z + \frac{3}{2} \partial_\bz Y^\bz\right)  \mathcal{P}^{soft} \,, \\
        &\delta_{(\mathcal{T}, Y)} \mathcal{J}^{soft}_z = \left( Y^z \partial_z + Y^\bz \partial_\bz + 2 \partial_z Y^z +  \partial_\bz Y^\bz \right) \mathcal{J}^{soft}_z  + \frac{1}{2} \mathcal{T} \bar{\partial} \mathcal{P}^{soft}  + \frac{3}{2} \bar{\partial} \mathcal{T} \mathcal{P}^{soft}  \,,
\eadat
\end{equation}
and similarly for the hard part. Focusing on the soft pieces, we thus conclude that BMS symmetries single out a $(\frac{3}{2}, \frac{3}{2})$ primary, a `supermomentum flux' which can be read from the second expression in \eqref{supertranslation fluxes} 
\begin{equation}\label{PDN}
    \mathcal P^{soft} = \frac{1}{8\pi G}  \mathscr{D}_z^2\mathscr{N}^{(0)}_{\bar{z}\bar{z}},
\end{equation}
and is thus the second descendant of the leading soft graviton operator. Similarly, the soft piece of the `super-angular momentum flux' is a $(2,1)$ primary and reads, from \eqref{superrotation fluxes},
\begin{equation}\label{angular_flux}
    \mathcal J_z^{soft} = \frac{1}{16\pi G} \left(-{\mathscr D}_z^3 \mathscr N_{\bar z \bar z}^{(1)} + \frac{3}{2} \mathscr C_{zz} \mathscr{D}_z \mathscr {N}^{(0)}_{\bar z \bar z}+ \frac{1}{2}\mathscr {N}^{(0)}_{\bar z \bar z} \mathscr{D}_z \mathscr C_{zz}\right);
\end{equation}
while $\mathcal J_\bz^{soft}$ carries weights $(1,2)$; see Table \ref{table:weights}. We will see at the end of section \ref{chap:last} that this current gives rise to a stress tensor on the celestial sphere.\\

\begin{table}[h!]
\begin{center}
\renewcommand*{\arraystretch}{1.2}
  \begin{tabular}{ | c || c | c || c || c | c | c | c || c | c | c | }
  \hline  & $
    {\mathcal{T}}$ & $ {Y^z} $ &  $\mathscr{D}_z$& $\mathscr{C} \,,\, \mathscr N$  & $\mathscr C_{zz}\,,\, \mathscr{N}_{zz}^{(0)}$ & $P_z$ & $\mathcal P$ &$\mathscr{N}_{zz}^{(1)}, \Pi_{zz}$ & $N^{vac}_{zz}\,,\,T_{zz}$ & 
     $\mathcal J_z$ \\ 
    \hline \hline  
    $h$ & $-\frac{1}{2}$ & $-1 $ &
    $1$ & $-\frac{1}{2}$  & $\frac{3}{2}$ & $\frac{3}{2}$ & $\frac{3}{2}$ & $1$ & $2$ & $2$  \\ \hline
    $\bar h$ & $-\frac{1}{2}$ & $0$
    & $0$ & $-\12$ & $-\frac{1}{2}$ & $\12$ & $\frac{3}{2}$ & $-1$ & $0$ & $1$   \\ \hline \hline
    $\Delta$ & $0$ & $-1$ & $1$ & $-1$ & $1$ & $2$ & $3$ & $0$ & $2$ & $3$ 
 \\ \hline
 $J$ & $0$ & $-1$ & $1$ & $0$ & $2$ & $1$ & $0$ & $2$ & $2$ & $1$  \\ \hline \end{tabular} \caption{Summary of the conformal weights $(h, \bar{h})$, or,  equivalently, of the conformal dimension $\Delta=h+\bar h$ and spin $J=h-\bar h$ carried by the different fields.} \label{table:weights}
\end{center}
\end{table}

\noindent \textcolor{pansypurple}{\gr{BMS flux algebra \,\,}}
We end this section by presenting the essential property of the symplectic structure \eqref{symplectic structure full}, which is that the soft (hard) fluxes act independently on the soft \eqref{Gamma soft} (hard \eqref{Gamma hard}) phase space, namely
\begin{equation}
    \begin{split}
        &\big\{ \mathcal F^{hard}_{\xi},\Gamma^{H} \big\} = \delta_{\xi}\Gamma^{H},\quad \big\{ \mathcal F^{hard}_{\xi},\Gamma^{S} \big\} = 0,\\
        &\big\{ \mathcal F^{soft}_{\xi},\Gamma^{S} \big\} = \delta_{\xi}\Gamma^{S} , \quad \big\{ \mathcal F^{soft}_{\xi},\Gamma^{H} \big\} = 0 \,.
    \end{split} \label{variations generated}
\end{equation}
One can moreover show that they form separately two representations of the extended BMS algebra \eqref{commutation relations 1}--\eqref{commutation relations 2}
\cite{Donnay:2021wrk,Donnay:2022hkf}
\begin{equation}
  \badat{2}\label{flux algebra}
        &\big\{ \mathcal F^{hard}_{(\mathcal T_1,\mathcal Y_1)},\mathcal F^{hard}_{(\mathcal T_2,\mathcal Y_2)}\}=- \mathcal F^{hard}_{[(\mathcal T_1,\mathcal Y_1),(\mathcal T_2,\mathcal Y_2)]}\\
              &\big\{ \mathcal F^{soft}_{(\mathcal T_1,\mathcal Y_1)},\mathcal F^{soft}_{(\mathcal T_2,\mathcal Y_2)}\}=- \mathcal F^{soft}_{[(\mathcal T_1,\mathcal Y_1),(\mathcal T_2,\mathcal Y_2)]},\\
                    &\big\{ \mathcal F^{hard}_{(\mathcal T_1,\mathcal Y_1)},\mathcal F^{soft}_{(\mathcal T_2,\mathcal Y_2)}\}=0\,.
\eadat
\end{equation}
In other words, we thus conclude that the algebra spanned by BMS fluxes closes the BMS algebra, without central extension.

\newpage
\section{Celestial holography}
\label{chap:celestial}
``If you look up at the
sky on a clear cloudless night, you appear to see a hemispherical dome above
you, punctuated by myriads of stars'' (Penrose, \emph{The road to reality}, 2004, p. 428). What you see is the celestial sphere ($\mathcal {CS}$) or, rather, half of it (but you would see it all if you were floating in space). More precisely, looking at the night sky, one observes (half of) all light rays that make up the light cone centred at a given event. Celestial holography proposes that the celestial sphere is a natural host for a conformal field theory which encodes in a holographic manner quantum gravity in the bulk of spacetime.

\begin{figure}[h!]
\centering
\includegraphics[scale=0.5]{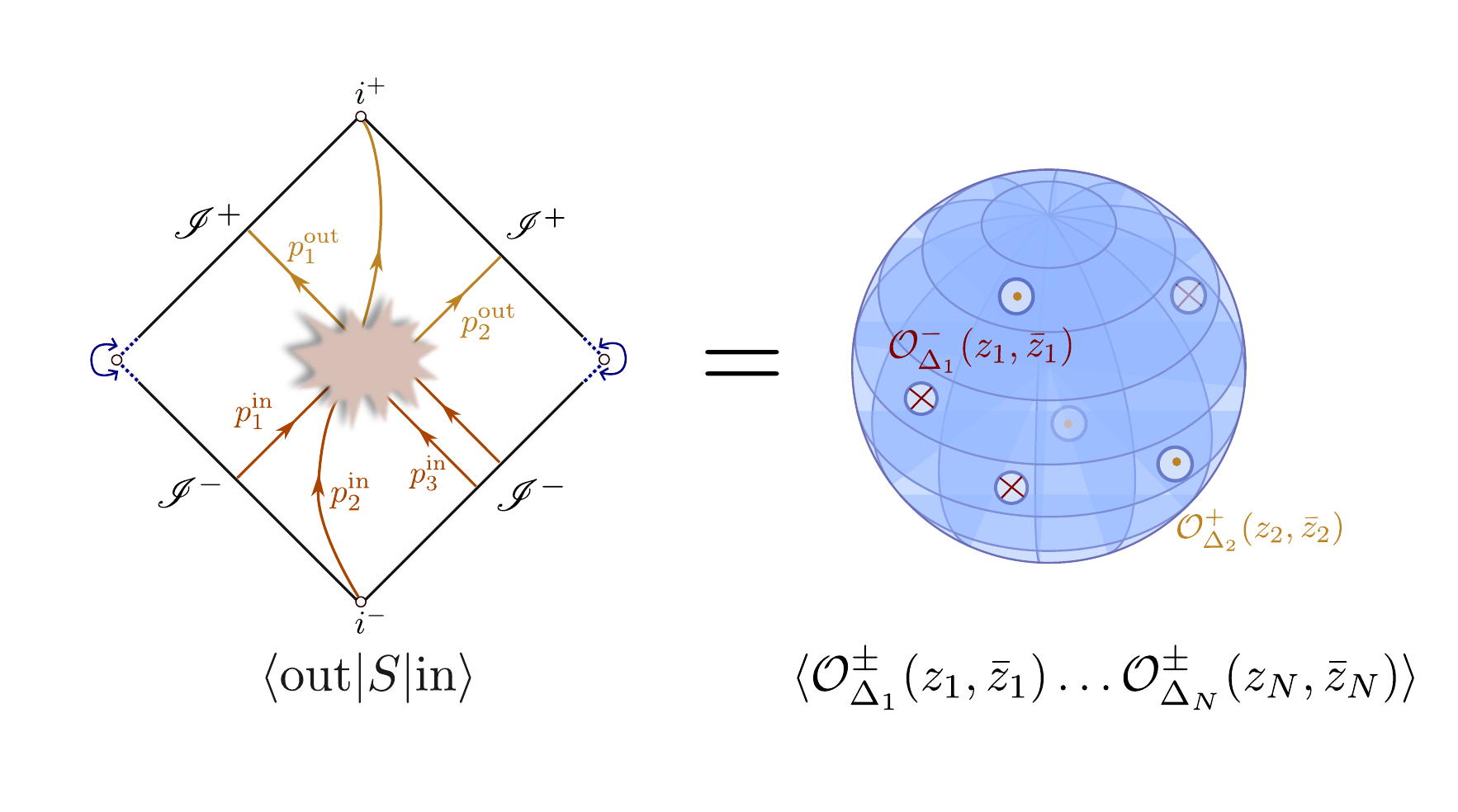}
\captionsetup{width=1\linewidth}
\caption{\small{In celestial holography, $N$-particle scattering amplitudes in $4d$ flat spacetime are encoded into correlation functions of a $2d$ 
 celestial conformal field theory~\cite{Strominger:2017zoo}. Each (incoming or outgoing) particle of momentum $p_i$ is associated with a celestial operator $\mathcal O_i$ which depends on a conformal dimension $\Delta_i$ and the point $(z_i,\bar z_i)$ where the particle enters or exits the spacetime. Notice that the angles on $\mathscr I^-$ and $\mathscr I^+$ are antipodally identified such that a free massless particle enters and exists at the same point on the celestial sphere.}}
\label{fig:cele}
\end{figure}

The core idea of celestial holography is that the most basic observables, scattering amplitudes, can be mapped to conformal correlators on the celestial sphere \cite{deBoer:2003vf,He:2015zea,Strominger:2017zoo,Pasterski:2016qvg,Pasterski:2017ylz}; see Fig. \ref{fig:cele}. In this picture, the sought-for holographic dual of quantum gravity in $d+2$ spacetime dimensions in 
flat space is therefore realized by a (perhaps exotic) $d$-dimensional conformal field theory living on the celestial sphere: a `celestial conformal field theory' (CCFT). Concretely, each incoming ($-$) or outgoing  ($+$) particle\footnote{It is understood that we are talking about stable particles, unstable particles are not part of the celestial sphere data as they do not reach infinity.} is described by a CCFT operator $\mathcal O^\pm_\Delta(z,\bar z)$ which can be obtained from plane wave packets. The specific form of the map that relates celestial operators with single particle states depends on whether the particle is massless or massive. In both cases, the conventional three-momentum $p$ of the particle is 
replaced by a conformal dimension $\Delta$ and the point $(z, \bar z)$ at which the asymptotic particle pierces the celestial sphere (coming (exiting) from past (future) null or timelike infinities).

We will start this chapter by collecting in section \ref{sec:sphere} a set of useful facts and formulae about the celestial sphere and the parametrization for the momentum of a massless particle heading towards null infinity. The fact that the Lorentz group acts as the global conformal group on the celestial sphere will motivate in section \ref{sec:amplitudes} to recast scattering amplitudes in a boost basis (rather than the usual momentum basis). Section \ref{sec:operators} contains an introduction to conformal primary wavefunctions and celestial operators while their interesting shadow transforms will be presented in section \ref{sec:shadow}. Finally, we will end this chapter in section \ref{sec:massive} by discussing the case of massive particle, for which the celestial machinery also applies.

\subsection{Celestial sphere as a holographic screen}
\label{sec:sphere}
Among the recurrent questions regarding celestial holography is: why does one want to formulate the dual theory on the celestial sphere? Indeed, from what we know from the AdS/CFT correspondence \cite{Maldacena:1997re}, the quantum field theory lives on a co-dimension one boundary, not co-dimension two. However, as we have tried to point out in the introduction, one does not expect that the way flat space holographic could work should necessarily proceed in the exact same way as it is implemented for its negative curvature counterpart. That being said, it does not mean that it would be impossible or not interesting to investigate a holographic correspondence where the dual theory would live on the co-dimension one boundary, e.g. at null infinity. In fact this latter perspective has also attracted attention and goes under the name of `Carrollian\footnote{The name Carrollian refers to the geometry associated to null hypersurfaces.} holography' and provides a complementary road to celestial holography (see comments on that in section \ref{chap:conclusion}).

What really makes the celestial sphere a particularly interesting candidate for a holographic screen is the very simple observation that the $4d$ Lorentz group $SL(2, \mathbb C)$ acts as the $2d$
global conformal group on the celestial sphere (see also \eqref{SL2} below):
\begin{equation}\label{}
	z \mapsto z'= \frac{az+b}{cz+d}\,,\qquad \bz \mapsto \bar z'= \frac{\bar a \bar z+\bar b}{\bar c \bar z+\bar d}\,,
\end{equation}
with $a,b,c,d \in \mathbb C$ and $ad-bc=1$. Making the $SL(2, \mathbb C)$ transformation manifest at the level of scattering amplitudes seems a very natural thing to do for holographic purposes. Indeed the $4d$ scattering amplitudes will then have by construction the same conformal covariance as correlators of a CFT$_2$. What is more (and as we will in section \ref{sec:CCFT_currents}), the surprising existence of a $2d$ current which obeys the Virasoro-Ward identity of a stress-tensor \cite{Kapec:2016jld} has further fueled this idea as it showed that quantum gravity amplitudes in the $4d$ bulk are in certain representations of the \emph{full} (infinite-dimensional) local conformal group.

This conformal presentation of QFT amplitudes employed in celestial holography can be traced back to a pretty old idea of Dirac \cite{Dirac:1936fq}, who observed that a very natural host for the conformal group $SO(d+1,1)$ is the (fictitious) embedding space $\mathbb R^{d+1,1}$ where the conformal group is realized in a linear way.
This is also the same spirit which is followed in the so-called embedding space formalism, which has been used extensively in the CFT (bootstrap) literature (see e.g. \cite{Costa:2011dw} and references therein). Celestial holography employs the embedding formalism in a very literal manner, since for us the embedding space is not an abstract spacetime but instead the very physical location where the scattering of particles occurs \cite{Pasterski:2017kqt}. 

The rest of this section is devoted to a brief presentation of how the embedding of the celestial sphere into the lightcone of Minkowski space is performed in practice in most of the celestial literature, following mostly Refs \cite{Pasterski:2017kqt} and section 2 of \cite{Donnay:2022ijr}. We will work with a $d$-dimension celestial sphere, even though we will then specify to the $d=2$ case in subsequent sections.\\

\noindent \textcolor{pansypurple}{\gr{Projective light-cone\,\,}} 
In order to parametrize the light-cone, defined by $
p^2=0$,
it is useful to fix a reference chart of the form
\begin{equation}\label{plambdaq}
	p^\mu = \lambda\, q^\mu(w),
\end{equation}
where $\lambda$ and $w^a$ ($a=1,\dots,d$), are real coordinates. As we will see later, one often selects a specific form for the null vector $q^\mu$. Once such a choice  has been made, it induces tangent vectors $e^\mu_a$ and a metric $h_{ab}$
\begin{equation}\label{eamu}
	e_a^\mu(q)=\partial_aq^\mu \virg h_{ab}(q)=e_a^\mu(q) \,\eta_{\mu \nu}\,e_a^\nu(q)
\end{equation} 
on the $\lambda=1$ cross section it describes.  
One can define the projective light-cone by identifying vectors that differ only by an overall rescaling:
\begin{equation}\label{plc}
	p^2=0\,,\qquad p^\mu \sim \lambda\, p^\mu\,. 
\end{equation}
One can observe that the expression 
\begin{equation}\label{epsilon1}
	\epsilon_a^\mu(q) = \partial_a \left(
	\frac{q^\mu}{u\cdot q}
	\right)
	=    \frac{1}{u\cdot q}\left(
	\partial_a q^\mu - \frac{u\cdot \partial_a q}{u\cdot q}\,q^\mu
	\right)\,,
\end{equation}
is invariant under rescalings
\begin{equation}\label{}
	q^\mu \to \frac{q^\mu}{u\cdot q}\,,
\end{equation}
with a suitable reference vector $u^\mu$. Namely, by construction,
\begin{equation}
	\epsilon_a^\mu(\lambda\, q) = \epsilon_a^\mu(q)\,.
\end{equation}
and its associated metric
\begin{equation}\label{gammaab}
	\gamma_{ab}(q)=\epsilon_a^\mu(q) \, \eta_{\mu\nu}\,\epsilon_b^\nu(q) = \frac{1}{(u\cdot q)^2}\, e_a^\mu(q) \, \eta_{\mu\nu}\,e_b^\nu(q) =\frac{h_{ab}(q)}{(u\cdot q)^2}\,,
\end{equation}
is also manifestly invariant under local rescalings.
 \eqref{epsilon1} will constitute our preferred tangent vectors.
Under a generic change of coordinates $w=w(w')$, the transformation laws are\footnote{We use interchangeably $q^\mu$ or $w$ to denote points on the projective light-cone, writing for instance $
	\epsilon_a^\mu(w)
	=
	\epsilon_a^\mu (q(w))
$.}
\begin{equation}
	\epsilon'^\mu_{a}(w')  =
	\frac{\partial w^b}{\partial w'^a}\, \epsilon^\mu_{b}(w)\,,
	\qquad
	\gamma'_{ab}(w')=\frac{\partial w^c}{\partial w'^a}\,  \gamma_{cd}(w)\, \frac{\partial w^d}{\partial w'^b}\,.
\end{equation}

Let us now consider a Lorentz transformation $\Lambda\indices{^\mu_\nu}\in SO(d+1,1)$, which acts via 
\begin{equation}
    q^\mu \mapsto \Lambda\indices{^\mu_\nu} \, q^\nu .
\end{equation}
Since $\Lambda\indices{^\mu_\nu}q^\nu(w)$ is null, the effect of the Lorentz transformation can be also be expressed by a mapping $w\mapsto w'(w)$ which is such that
\begin{equation}\label{Lorentzw}
	\Lambda\indices{^\mu_\nu}\,q^\nu(w) = \alpha_\Lambda (w) q^\mu(w')\,,
\end{equation}
for a suitable factor $\alpha_\Lambda$. This factor acts as a   conformal factor for $h_{ab}$. Indeed, taking a derivative of \eqref{Lorentzw} with respect to $w'^a$ gives
\begin{equation}\label{}
	\Lambda\indices{^\mu_\nu} \frac{\partial w^c}{\partial w'^a} e_c^\nu(w) = \alpha_\Lambda (w) e_a^\mu(w')
	+
	\frac{\partial \alpha_\Lambda}{\partial w'^{a}}\, q^\mu(w'),
\end{equation}
and using $q^2=0=e^\mu_a q_\mu$, the ``square'' of this identity reads
\begin{equation}\label{alphaconf}
	\frac{\partial w^c}{\partial w'^a} 
	\, h_{cd}(w)\, \frac{\partial w^d}{\partial w'^b}= \alpha^2_\Lambda(w) h_{ab}(w')\,.
\end{equation}
The determinant of this relation gives
\begin{equation}\label{myalpha}
	\alpha^{d}_\Lambda(w)=
	\sqrt{\frac{h(w)}{h(w')}}\
	\mathrm{det}\left(\frac{\partial w^a}{\partial w'^b}
	\right),
\end{equation}
where $h(w)=\mathrm{det}(h_{ab}(w))$.
One can verify that $\epsilon^\mu_a$ behaves as follows under Lorentz transformations
\begin{equation}\label{}
	\epsilon_a^\mu(\Lambda q) \equiv \frac{\partial}{\partial w'^a} \left(
	\frac{q^\mu(w')}{u\cdot q(w')}
	\right)
	=
	\epsilon_a^\mu(w')\,.
\end{equation}
Consequently, it obeys the nonlinear transformation law 
\begin{equation}\label{projepsilon}
	\left(
	\Lambda\indices{_\nu^\mu}-\frac{q^\mu}{u\cdot q}\, \Lambda\indices{_\nu^\rho}\, u_\rho
	\right) \epsilon^\nu_a(\Lambda q) = \frac{u\cdot q}{u\cdot \Lambda q}\,\frac{\partial w^b}{\partial w'^a}\, \epsilon^\mu_b(q)\,.
\end{equation}
Now, making the dependence on the reference vector $u^\mu$ explicit (writing for instance $\epsilon_a^\mu(u;q)$), one finds the simpler transformation rule
\begin{equation}\label{simpleeps}
	\epsilon^\mu_a(\Lambda u;w')=
	\frac{\partial w^b}{\partial w'^a}\,
	\Lambda\indices{^\mu_\nu}
	\epsilon_b^\nu(u;w)\,.
\end{equation}
The polarization vectors constructed above can be related to the ones usually built in terms of little-group elements (e.g.~\cite{Weinberg:1964kqu,Weinberg:1964ev}). In particular, the explicit relation between their transformation laws on the celestial sphere and standard Wigner rotations were detailed in \cite{Donnay:2022ijr} (see also \cite{Banerjee:2018gce,Pasterski:2021rjz}).\\

\noindent \textcolor{pansypurple}{\gr{Standard parametrizations\,\,}}
A very convenient choice which is taken in most of celestial holography literature (e.g. \cite{Pasterski:2017ylz,Pasterski:2017kqt}) for $q^\mu(w)$ is 
\begin{equation}\label{parconv}
	q^\mu(w)= \left(
	1+|w|^2, 2w^a, 1-|w|^2
	\right)
\end{equation}
which amounts to taking
\begin{equation}\label{}
	\lambda = p^0+p^{d+1}\,,\qquad w^a = \frac{p^a}{p^0+p^{d+1}}\,,
\end{equation}
in \eqref{plambdaq}.
This choice corresponds to the flat representative $h_{ab} = \delta_{ab}$ and the coordinates $w^a$ cover Euclidean space $\mathbb{R}^{d}$. 
Moreover, identifying $u^\mu=-n^\mu=(-1,0,\ldots,0,1)$, one has $u\cdot q(w)=1$ for any $w^a$, and hence $e_a^\mu(q) = \epsilon_a^\mu(q)$.

Under the action of Lorentz transformation via \eqref{Lorentzw}, one gets
\begin{equation}\label{Lorentzwpc}
	-n_\mu\Lambda\indices{^\mu_\nu}\,q^\nu(w) = \alpha_\Lambda (w)\,,
\end{equation}
while \eqref{alphaconf} and \eqref{myalpha} become
\begin{equation}\label{alphaconfpc}
	\frac{\partial w^c}{\partial w'^a} 
	\, \frac{\partial w^c}{\partial w'^b}= \alpha^2_\Lambda(w) \delta_{ab} \virg \alpha^{d}_\Lambda(w)=
	\mathrm{det}\left(\frac{\partial w^a}{\partial w'^b}
	\right)\,.
\end{equation}
Finally, the transformation rule \eqref{projepsilon} simplifies to
\begin{equation}\label{projepsilonpc}
	\left(
	\Lambda\indices{_\nu^\mu}-\frac{q^\mu}{n\cdot q}\, \Lambda\indices{_\nu^\rho}\, n_\rho
	\right) e^\nu_a(\Lambda q) = \frac{1}{\alpha_\Lambda(w)}\,\frac{\partial w^b}{\partial w'^a}\, e^\mu_b(q)\,.
\end{equation}

When working in $d=2$, one often switches to complex coordinates $(w,\bar w)$
and the parametrization \eqref{parconv} can be written as
\begin{equation}\label{qmu}
	q^\mu(w, \bar w) = \big(1+w\bar w, w+\bar w,-i(w-\bar w),1-w\bar w
	\big)\,,
\end{equation}
so that $h_{w\bar w}=1$, $h_{ww}=0=h_{\bar w \bar w}$ and
\begin{equation}\label{pola}
 \epsilon_w^\mu = \partial_w q^\mu=\left(
	\bar w, 1, -i, -\bar w
	\right)\,,\qquad
	 \epsilon_{\bar w}^\mu= \partial_\bw q^\mu = \left(
	w, 1, i, -w
	\right)\,.
\end{equation}
Finite Lorentz transformations act on the celestial complex variables as
\begin{equation}\label{SL2}
	w \mapsto w'= \frac{aw+b}{cw+d}\virg \bw \mapsto \bw'= \frac{\bar a\bw+\bar b}{\bar c\bw+\bar d}\,,
\end{equation}
with $a,b,c,d \in \mathbb C$ and $ad-bc=1$, and
the Jacobian in \eqref{myalpha} reads
\begin{equation}\label{stdalpha}
	\frac{\partial w'}{\partial w} =(cw+d)^{-2}\,,\qquad
	\alpha_\Lambda (w,\bar w) = (cw+d)(\bar c \bar w + \bar d)\,. 
\end{equation}
The basic transformation rule \eqref{simpleeps} for the polarization vector is then
\begin{equation}\label{}
	\epsilon^\mu_w(\Lambda u;w',\bar w')=
	(cw+d)\,
	\Lambda\indices{^\mu_\nu}
	\epsilon_w^\nu(u;w,\bar w)\,,\qquad
	\epsilon^\mu_{\bar w}(\Lambda u;w',\bar w')=
	(\bar c \bar w+\bar d)\,
	\Lambda\indices{^\mu_\nu}
	\epsilon_{\bar w}^\nu(u;w,\bar w)\,.
\end{equation}

\subsection{Celestial amplitudes}
\label{sec:amplitudes}
Scattering amplitudes are the main observables\footnote{See \cite{Caron-Huot:2023vxl} for further discussion on asymptotic observables.} in flat spacetimes. We are used to write them down in the momentum space, where incoming and outgoing states are parametrized by their momentum $p^\mu$ and their helicity $\ell =\pm s$ ($s$ denoting the spin of the particle). For the scattering of massless particles (the case of massive particles will be treated in section \ref{sec:massive}), one can make use of the embedding presented above to parametrize on-shell momenta as $p^\mu=\pm\omega q^\mu(z,\bar z)$, with $q$ the null vector typically taken as in \eqref{qmu} and the sign distinguishes whether the particle is incoming $(-)$ or outgoing $(+)$. A one-particle state $\ket{\omega,z,\bar z,\ell}$ thus represents a particle of light-cone energy $\omega$ and helicity $\ell$ entering asymptotically flat spacetime at the point $(z,\bar z)$ on the celestial sphere. Scattering amplitudes in momentum space involving $N$ massless particles -- say with $n$ of them which are incoming -- are thus given by the $\mathcal{S}$-matrix elements 
\begin{equation}
   \mathcal A_N(p_1,\ell_1;\dots; p_N,\ell_N): = \braket{\text{out}|\mathcal S|\text{in}}
\end{equation}with
\begin{equation}
    \ket{\text{in}} = a^\dagger_{p_1}\dots a^\dagger_{p_n} |0\rangle \virg \ket{\text{out}} = a^\dagger_{p_{n+1}}\dots a^\dagger_{p_N} |0\rangle \,.
\end{equation}

Now, nothing can prevent one to re-express $\mathcal{S}$-matrix elements in another basis which, instead of making the translation invariance manifest, makes the action of $SL(2,\mathbb C)$ manifest; this is the so-called `conformal basis'~\cite{deBoer:2003vf,Pasterski:2016qvg,Cheung:2016iub,Pasterski:2017kqt}. Each one-particle state in momentum basis is mapped to a boost eigenstate via the Mellin transform
\begin{equation}
\ket{\Delta, z,\bar z} = \int_0^{+\infty} d\omega \,\omega^{\Delta-1} \ket{\omega,z,\bar z}\label{Mellin}\,,
\end{equation}
whose purpose is to trade $\omega$ for a Rindler energy $\Delta$. This new basis allows to interpret $\Delta$ as the conformal dimension of a celestial CFT operator (see below). The $4d$ helicity $\ell$ (which we dropped in \eqref{Mellin}) is simply identified with a $2d$ spin commonly denoted by $J$.
 The so-called `celestial amplitudes'
\begin{equation}
\mathcal M_N := \braket{\text{out}|\mathcal S|\text{in}}_{\text{boost}}\,
\end{equation}
are thus obtained by taking as many Mellin integral transforms as there are massless particles, namely 
\begin{equation}
    \mathcal M_N  =\prod_{i=1}^N \int_0^{+\infty} d\omega_i \, \omega_i^{\Delta_i-1}  \mathcal A_N (p_1,\ell_1;\dots; p_N,\ell_N).
    \label{Celestial S-matrix}
\end{equation}
By design, celestial amplitudes are $SL(2,\mathbb C)$ covariant; it is therefore natural \cite{deBoer:2003vf,He:2015zea,Strominger:2017zoo,Pasterski:2016qvg,Pasterski:2017ylz} to identify them with a correlator of a putative (Euclidean) $2d$ conformal field theory living on the celestial sphere, a celestial conformal field theory (CCFT):
\begin{equation}\label{cele}
  \braket{\text{out}|\mathcal S|\text{in}}_{\text{boost}} = \big\langle \mathcal{O}_{\Delta_1,J_1}(z_1, \bar{z}_1) \dots \mathcal{O}_{\Delta_{N},J_N}(z_{N}, \bar{z}_{N}) \big\rangle_{\text{CCFT}}\,.
\end{equation}
We will detail in the next section how celestial operators $\mathcal O_{(\Delta,J)}$ can be explicitly constructed out of massless bulk wavefunctions and see that they exactly transform as primaries of conformal weights $(h,\bar h)=(\frac{\Delta+J}{2},\frac{\Delta-J}{2})$ in a CFT$_2$. The massive analogue of the celestial map \eqref{Celestial S-matrix} will be presented in section \ref{sec:massive}.

At this stage, a sceptical reader might be thinking that all this is just a change of basis. While there is nothing incorrect in this statement, what would be wrong would be to underestimate that aspect. A change of basis could have been in principle a rather useless thing to do, were the symmetries of the $\mathcal S$-matrix not what they actually are. As we will see, one of the powers of celestial amplitudes comes precisely from the fact that they allow us to identify an organizing principle for the incredibly rich and repeatedly overlooked symmetries of asymptotically flat spacetimes, which we have only partially covered in chapters \ref{chap:bms} and \ref{chap:charges}.
The relationship   \eqref{cele} is thus at the core of the celestial holography program in its bottom-up approach: it defines for us, without any underlying assumption about the precise nature of the UV completion of gravity, what the holographic dual field theory should be, given the tower of infinite-dimensional asymptotic symmetry constraints. We will explore the constraints implied by the latter on CCFT correlators in chapter \ref{chap:last}.

\subsection{Celestial operators}
\label{sec:operators}
Let us now give more detail on how one constructs in practice celestial operators $\mathcal O_{(\Delta,J)}$ from an asymptotically free massless field in the $4d$ 
 bulk. It is in fact rather simple: they can be simply obtained from the Mellin transform of the usual creation and annihilation operators that appear in a plane wave mode expansion (up to normalization). To illustrate this, we will first focus on a scalar field operator, for which $J=0$, and then present analogous expressions for a spin-two field. The main references used for this section are \cite{Pasterski:2017kqt,Donnay:2018neh,Donnay:2020guq, Donnay:2022ijr}.

\subsubsection{Scalar conformal primary wavefunctions}
The starting point is the familiar QFT textbook expression for a plane wave mode expansion of a scalar field operator:
\begin{equation}\label{waveexp}
	\begin{split}
	    \Phi(X) &= \int  \delta(p^2) \theta(p^0)\, \frac{d^4p}{(2\pi)^3}\left[
	e^{ip\cdot X} a(p)
	+
	e^{-ip\cdot X} a^\dagger(p)
	\right]\,, 
	\end{split}
\end{equation}
where $X$ denotes a point in the $4d$ bulk spacetime and the creation and annihilation operators obey the commutation rules
\begin{equation}\label{laddercomm}
	[a(p),a^\dagger(p')] = 2|\vec p|(2\pi)^{3}\delta^{(3)}(\vec p-\vec p')\,.
\end{equation}
Using the parametrization $p^\mu=\omega q^\mu(\vec w)$, this is equivalent to
\begin{equation}\label{ModeExpansion}
	\Phi(X) = \frac{1}{2(2\pi)^{3}}\int d^2\vec w
	\int_0^\infty \omega\,
	d\omega\left[
	e^{i\omega q(\vec w\,)\cdot X} a(\omega\,q(\vec w\,))
	+
    e^{-i\omega q(\vec w\,)\cdot X} a^\dagger(\omega\,q(\vec w\,))\right].
\end{equation}
Now, using the inverse Mellin transform formula (see e.g. \cite{titchmarsh1948introduction} for details) 
\begin{equation}\label{InverseScalar}
    \int_{c-i\infty}^{c+i\infty}  \frac{\omega^{-\Delta}\,\Gamma(\Delta)}{(\epsilon \mp iq\cdot X)^\Delta} \frac{d\Delta}{i2\pi} = e^{\pm i\omega q\cdot X}\,,
\end{equation}
with $\epsilon >0$ a regulator,
one can alternatively decide to expand the scalar field in the boost basis as \begin{equation}\label{DeltaExpansion}
	\Phi(X) = \frac{1}{2(2\pi)^{3}} \int d^2\vec w
    \int_{c-i\infty}^{c+i\infty} \frac{d\Delta}{i2\pi} \left[
    \phi^+_\Delta(X;\vec w\,)a_{2-\Delta}(\vec w\,)
    +
     \phi^-_\Delta(X;\vec w\,)a^\dagger_{2-\Delta}(\vec w\,)
    \right],
\end{equation}
where we defined as in \cite{Pasterski:2021dqe}\footnote{They obey
$
    [a_\Delta(\vec w\,),a^\dagger_{\Delta'}(\vec w\,')]
    =
    (2\pi)^{4} 2q^0 \,\delta(q,q') 
	\delta(i(\Delta+\Delta'^\ast-2))$ with the generalized delta function defined in \cite{Donnay:2020guq}.}
\begin{equation}\label{scalaranalogy}
    a_\Delta(\vec w\,) := \int_0^\infty  \omega^{\Delta-1}a(\omega\,q(\vec w\,))\,d\omega \virg a^\dagger_\Delta(\vec w\,) \equiv \int_0^\infty  \omega^{\Delta-1}a^\dagger(\omega\,q(\vec w\,))\,d\omega\,.
\end{equation}
The wavefunctions $\phi^\pm_\Delta$ appearing in \eqref{DeltaExpansion} are by construction the Mellin transforms of (incorming or outgoing) plane wave packets\cite{deBoer:2003vf,Cheung:2016iub,Pasterski:2016qvg,Pasterski:2017kqt}
\begin{equation}\label{conf0}
	\phi^\pm_\Delta(X;\vec w):=
	\int_0^\infty d\omega \, \omega^{\Delta-1}\, e^{i(\pm q^\mu X_\mu+i\epsilon)\omega}\,.
\end{equation}
The integral in the rhs of \eqref{conf0} gives $\phi^\pm_\Delta=(\mp i)^\Delta\Gamma(\Delta)\varphi^\pm_\Delta$, where
\begin{equation}\label{varphi}
	\varphi^\pm_\Delta(X;\vec w)
	=\frac{1}{(-q \cdot X_\pm )^\Delta}\,.
\end{equation}
We adopted the notation $X^\mu_\pm:=(X^0\mp i\epsilon, X^1,X^2,X^3)$, where the $i\epsilon$ prescription circumvents the singularity at the light sheet where $q\cdot X=0$.
Crucially and as promised, the wavefunction\footnote{We drop  the $\pm$ subscript when not needed.}
$\varphi_\Delta(X;\vec w)$ defines a scalar \emph{conformal primary wavefunction} (CPW) of conformal dimension $\Delta$ \cite{Pasterski:2016qvg,Pasterski:2017kqt}. Indeed, under the mapping \eqref{Lorentzw}, it transforms as
\begin{equation}\label{scalarprimarytransf}
	\varphi_\Delta(\Lambda X;\vec w')
	=
	\alpha_\Lambda^\Delta (\vec w) \varphi_\Delta(X;\vec w)\,,
\end{equation}
with $\alpha_\Lambda(\vec w)$ given in \eqref{myalpha}.
In terms of the complex coordinates \eqref{stdalpha}, this reads
\begin{equation}\label{stdscalartransf}
	\varphi_\Delta(\Lambda X;w',\bar w')
	=
	\left(\frac{\partial w'}{\partial w} \right)^{-h} \left(\frac{\partial \bar w'}{\partial \bar w} \right)^{-\bar h} \varphi_\Delta(X;w,\bar w)\,,
\end{equation}
which coincides with the transformation law of a scalar conformal primary of dimension $\Delta=h+\bar h$ ($h=\bar h$ for a scalar). Alternatively, one can express this in terms of the usual $\mathfrak{sl}(2,\mathbb C)$ generators $L_n$, $\bar L_n$ ($n=0,1,-1$) as \cite{Cotler:2023qwh}
\begin{equation}\label{alter}
L_n \varphi_\Delta=\left(\frac{\Delta}{2}(n+1) w^n+w^{n+1}\p_w \right)\varphi_\Delta \virg \bar L_n \varphi_\Delta=\left(\frac{\Delta}{2}(n+1) \bar w^n+\bar w^{n+1}\p_\bw \right)\varphi_\Delta\,.
\end{equation}
Moreover, $\varphi_{\Delta}(X;w,\bar w)$ satisfies the Klein--Gordon equation with respect to the bulk,
\begin{equation}\label{boxphi}
	\Box \varphi_{\Delta}(X;w,\bar w)=0\,.
\end{equation}
Equations \eqref{stdscalartransf},  \eqref{boxphi} are the two defining properties of a scalar CPW.

Using the standard Klein-Gordon inner product\footnote{$
	\left(f, f'\right) =  -i \int_\Sigma d\Sigma^\mu
	\left(
	f \partial_\mu f'^\ast-f'^\ast \partial_\mu f
	\right)\,.
$}, one can extract the Mellin operators 
\begin{equation}\label{extract}
    \left(\Phi(X),
    \phi_\Delta^+(X;w,\bar w)
    \right)
    =a_\Delta(w,\bar w) \virg  \left(\Phi(X),
    \phi_\Delta^-(X;w,\bar w)
    \right)
    =a^\dagger_\Delta(w,\bar w) 
\end{equation}
so that 
\begin{equation}
a_\Delta(w,\bar w\,) \Phi(X) |0\rangle
=
\phi_\Delta^-(X;w,\bar w)|0\rangle\,\virg
a_\Delta^\dagger(w,\bar w\,) \Phi(X) |0\rangle
=
\phi_\Delta^+(X;w,\bar w)|0\rangle\,.
\end{equation}
One can also check that the above Mellin operators inherit the conformal primary transformation laws:
\begin{equation}\label{loren}
a_{\Delta}(w',\bar w') = \left(\frac{\partial w'}{\partial w}\right)^{ -h}\left(\frac{\partial \bar w'}{\partial \bar w}\right)^{-\bar h} a_{\Delta}(w,\bar w)\,.
\end{equation}
They thus define the (spinless) celestial operators $\mathcal O_{\Delta,J=0}(w,\bw)$.
Under the action of Poincar\'e translations $t_\mu$ they transform as \cite{Stieberger:2018onx,Donnay:2022wvx}
\begin{equation}\label{poin}
a_{\Delta}(w',\bar w') = e^{-i q^\mu(w,\bar w)t_\mu \hat\partial_{\Delta}}a_{\Delta}(w,\bar w) = \sum_{n=0}^{+\infty} \frac{\left(-i q^\mu(w,\bar w)t_\mu\right)^n}{n!} a_{\Delta+n}(w,\bar w) ,
\end{equation}
with the `shifting dimension' operator defined as $\hat\partial_\Delta F(\Delta) \equiv F(\Delta+1)$. The infinitesimal version of \eqref{loren} and \eqref{poin} combined reads
\begin{equation}
\begin{split}
    &\delta_{(\mathcal{T},Y)} a_{\Delta}(w,\bar w) = \left(-i\mathcal T\hat\partial_{\Delta} + Y^w \partial_{w} + Y^{\bar w} \partial_{\bar w} + h \partial_{w} Y^w  + \bar h \partial_{\bar w} Y^{\bar w}\right) a_{\Delta}(w,\bar w)\,.
\end{split}
\label{a delta transfo poincare inf}
\end{equation}
\subsubsection{Spinning conformal primary wavefunctions}
The case of spinning celestial operators $\mathcal O_{\Delta,J}$ and their associated primary wavefunctions can be analyzed in a similar fashion, upon the addition of two small technical ingredients. The first is that there is a now a polarization tensor ($\epsilon_{\mu \nu}$ for gravitons) that accompanies the mode expansion \eqref{waveexp} whose transformation under the action of $SL(2,\mathbb C)$ must be accounted for, see section \ref{sec:sphere} above. The second is that, in place of the scalar Klein-Gordon inner product, one must consider the generalization of the inner product to non-zero spins. These details have been treated with care in \cite{Pasterski:2017kqt,Donnay:2018neh,Donnay:2020guq} for spin one and two. Here we will simply collect some of the relevant formulae which pertain to the gravity case and refer the reader to \cite{Pasterski:2017kqt,Donnay:2018neh,Donnay:2020guq} for more detail. Other generalizations of conformal primary wavefunctions (including half-integer spins, $p$-forms) have been studied in \cite{Fotopoulos:2020bqj,Iacobacci:2020por,Narayanan:2020amh,Pasterski:2020pdk,Pano:2021ewd,Pasterski:2021fjn,Pasterski:2021dqe,Donnay:2022ijr}.

For positive helicity graviton $J=+2$, the polarization tensor can be taken to be  $\epsilon^{\mu\nu}_{(+)}=\epsilon^\mu_w \epsilon^\nu_w$, with $\epsilon^\mu_w$ given by \eqref{pola}. The outgoing and incoming massless spin-two CPW are given by\cite{Pasterski:2017kqt}
\begin{equation}\label{CPWgravi}
h_{\mu\nu}^{\Delta,\pm}(X;w)=\frac{\Delta-1}{\Delta+1}\epsilon_{\mu \nu(+)}\varphi_{\Delta}^\pm+\nabla_{(\mu}\,\xi_{\nu)\,(+)}^{\Delta,\pm}\,,
\end{equation}
with $\varphi^\pm_\Delta$ the scalar CPW \eqref{varphi} and where 
\begin{equation}
\label{resi}
\xi^{\Delta,\pm}_{\mu(+)}=\frac{1}{2(\Delta+1)}\left(\frac{\p_w q_\mu (\p_w q\cdot X_\pm)}{(-q\cdot X_\pm)^\Delta}+\frac{1}{2}\frac{q_\mu (\p_w q\cdot X_\pm)^2}{(-q\cdot X_\pm)^{\Delta+1}}\right)\,.
\end{equation} 
The negative helicity graviton $J=-2$, has a similar expression where now the polarization tensor is  $\epsilon^{\mu\nu}_{(-)}=\epsilon^\mu_{\bar w}\epsilon^\mu_{\bar w}$ and  $\xi_{\mu(-)}$ is given by \eqref{resi} with $w$ replaced by $\bar w$. 

Those spin-two conformal primary wavefunctions satisfy harmonic and radial gauge conditions\footnote{We omit spectator indices for notation simplicity.}
\begin{equation}\label{spin2gaugecond}
\eta^{\mu \nu}h_{\mu \nu}^{\Delta}=0 \,, \quad \nabla^\mu h_{\mu \nu}^{\Delta}=0 \,, \quad X^\mu h_{\mu \nu}^{\Delta}=0\,,
\end{equation}
and solve vacuum linearized Einstein equations, which reduce to
$
\Box  h_{\mu \nu}^{\Delta}=0\,.
$
They also transform as both a four-dimensional traceless symmetric rank-two tensor and as two-dimensional spin-two conformal primaries with conformal dimension $(h,\bar h)=\frac{1}{2}(\Delta+J,\Delta-J)$ with $J=\pm 2$ under an $SL(2,\mathbb{C})$ Lorentz transformation \cite{Pasterski:2017kqt}:
\begin{equation}
h_{\mu \nu}^{\Delta}\left(\Lambda X;w',\bar w'\right)=\left(\frac{\partial w'}{\partial w} \right)^{-h} \left(\frac{\partial \bar w'}{\partial \bar w} \right)^{-\bar h} 
 \Lambda_{\mu}^{\,\, \rho}\Lambda_{\nu}^{\,\, \sigma}h_{\rho 
\sigma}^{\Delta}(X;w,\bw)\,.
\label{SL2Cspin2}
\end{equation}

Similarly to the scalar case (see \eqref{extract}), spin-two celestial operators $\mathcal O_{\Delta,\pm 2}(z,\bar z)$ can be extracted by taking the inner product between a spin-two bulk operator $h_{\mu \nu}(X)$ and  spin-two conformal primary wavefunctions \eqref{CPWgravi}, i.e. schematically $\mathcal O_{\Delta,\pm 2}(z,\bz) \sim (h(X),h^{\Delta}(X,z,\bar z))_\Sigma $~\cite{Donnay:2020guq}.

We can observe from \eqref{CPWgravi} that something special happens when $\Delta=1$, as the wavefunction reduces to a pure residual diffeomorphism. While less obvious, the spin-two CPW in fact also reduces to a pure diffeomorphism when $\Delta=0$ \cite{Pasterski:2017kqt}. These `pure-gauge' (or Goldstone) conformal primary wavefunctions have been studied in detail in \cite{Donnay:2018neh,Donnay:2020guq,Pasterski:2021dqe,Donnay:2022sdg}(together with their memory partners). We will come back to that in section \eqref{sec:CCFT_currents} when will be see that indeed some specific integer values of the conformal dimension $\Delta$ carry important information about the soft sector of CCFTs.

\subsubsection{On the range of $\Delta$}
As we have seen, the conformal dimension $\Delta$ in celestial holography arises from the Mellin transform \eqref{Mellin} as 
a trade-off for the lightcone energy of a massless particle. The situation is very different from the AdS/CFT dictionary, where the conformal dimension of the dual operator is fixed by the mass of the particle in the bulk. Here the conformal dimension can take any (a priori complex) value, but it would then seem that the set of all conformal primary wavefunctions span a basis which is overcomplete. 
It is well understood that the set of conformal wavefunctions with $\Delta=1+i\mathbb R$ form a complete basis on the space of square normalizable wavefunctions \cite{deBoer:2003vf,Pasterski:2017kqt}. This can be shown by observing that the usual statement that plane wave packets form a delta-function normalizable basis gets mapped, upon Mellin transforms, to the restriction that $\Delta$ lies on the principal series. Celestial operators with $\Delta=1+i\mathbb R$ thus capture bulk scattering states of finite energy. However, as we have seen in \eqref{poin}, the action of translations shift the conformal dimension of celestial operators by one and it is far from obvious that the principal series is the optimal basis for holographic purposes\footnote{In \cite{Donnay:2020guq}, conformal primaries with analytically continued dimension on the complex plane were understood as certain contour integrals on the principal series.}. Recently, another basis, spanned by $\Delta \in \mathbb Z$ has been put forward. In \cite{Freidel:2022skz}, a complete integer basis was built in terms of a tower of memory and Goldstone modes. A different approach was taken in \cite{Cotler:2023qwh}, where a complete integer basis was constructed by means of irreducible integer representations of the Poincar\'e group. 

\subsection{Shadow operators}
\label{sec:shadow}
 In the CFT literature (old and recent), the so-called `shadow transforms' provide a powerful tool to calculate conformal blocks \cite{Ferrara:1972uq,Ferrara:1972xe,Ferrara:1973vz,Ferrara:1972ay,SimmonsDuffin:2012uy,Osborn:2012vt}. In a $2d$ CFT, a shadow transform maps a quasi-primary field of dimension $\Delta$ to another quasi-primary field of dimension $2-\Delta$\footnote{More generally, in dimensions $d$, the conformal dimension of the shadow operator is $d-\Delta$.}.
As it turns out, shadow operators are ubiquitous in celestial holography.
Indeed, while it was not immediately recognized as such, the shadow transform already appeared in the very early works analysing the Ward identities associated to superrotation symmetry. In \cite{Kapec:2016jld}, the $2d$ stress tensor appeared to be expressed as a non-local integral on $\mathcal{CS}^2$ of the news tensor zero mode. It was later understood that  this non-local transformation defining the celestial stress tensor was in fact nothing but the shadow transform of the subleading (conformally) soft graviton operator \cite{Donnay:2018neh,Fotopoulos:2019vac}. Notice that this is equally true for the leading soft graviton (and soft photon) operator, but in a somehow hidden way since those operators carry conformal weight $\Delta=1$, which coincides with their shadow dimension in $d=2$. In higher dimensions $d>2$, the fact that conserved currents are built out of the shadow transform of soft operators is more transparent and was pointed out in \cite{Kapec:2017gsg}.

In celestial holography, low point correlators often feature the presence of contact terms rather than the more familiar power laws of a CFT$_2$. Shadow transforms can be used as a way of taming the delta-functions which arise in celestial two-point functions. In this section, we briefly review the definition of those transformations and how they act on conformal primary wavefunctions. See Refs. \cite{Pasterski:2017kqt,Fan:2021isc,Fan:2021pbp,Crawley:2021ivb,Guevara:2021tvr,Kapec:2021eug,Kapec:2022axw,Banerjee:2022wht,Chang:2022jut,Jorstad:2023ajr} for further reading material on the role of shadow transforms in CCFT.

\subsubsection{Shadow primaries}
The shadow transform of a scalar primary operator $\mathcal O_\Delta(\vec w)$ is defined as (see \cite{Ferrara:1972uq,Ferrara:1972xe} for early works and \cite{SimmonsDuffin:2012uy,Osborn:2012vt,Pasterski:2017kqt,Kapec:2017gsg} for a more recent literature)
\begin{equation}\label{shadow0def}
 \widetilde{\mathcal O_\Delta}(\vec w\,) = k_\Delta \int \frac{d^2\vec w\,'}{|\vec w-\vec w\,'|^{2(2-\Delta)}}\,\mathcal O_{\Delta}(\vec w\,')\,,
\end{equation}
with $k_\Delta$ a normalization factor.

This non-local integral transform can be applied on the scalar conformal primary wavefunction $\phi_\Delta$ \eqref{conf0} to obtain a new CPW $\widetilde{\phi_\Delta}$ of conformal dimension $2-\Delta$.
The expression for the shadow conformal primary can be computed by using the identity (see e.g. \cite{SimmonsDuffin:2012uy})
\begin{equation}
\int d^d\vec w\,' \frac{1}{|\vec w-\vec w\,'|^{2(2-\Delta)}}\frac{1}{(-q(\vec w')\cdot X)^{\Delta}}=\frac{\pi}{(\Delta-1)}\frac{(-X^2)^{1-\Delta}}{(-q(\vec w)\cdot X)^{2-\Delta}}\,.
\end{equation}
For the normalization choice
\begin{equation}
   k_\Delta= \frac{\Gamma(2-\Delta)}{\pi\Gamma(\Delta-1)}\,,
\end{equation}
one thus finds \cite{Pasterski:2017kqt}
\begin{equation}\label{shadow_CPW}
 \widetilde{{\phi}^\pm_{\Delta}}(X;\vec w\,) =c_\pm(-X^2)^{1-\Delta}\phi^\pm_{2- \Delta}(X;\vec w\,)\,,
\end{equation}
with $c_\pm=(\mp i)^{2\Delta-2}$.
The shadow primary $ \widetilde{{\phi}_\Delta}$ satisfies the same defining properties as the original primary, namely \eqref{stdscalartransf} (or equivalently \eqref{alter}) and \eqref{boxphi}, but now with conformal dimension $2-\Delta$, i.e. if $\phi$ carries weights $(h,\bar h)$, then its shadow $\widetilde \phi$ has weights $(1-h,1-\bar h)$.

\subsubsection{Shadow plane waves}
\begin{figure}[h!]
\centering
\includegraphics[scale=0.8]{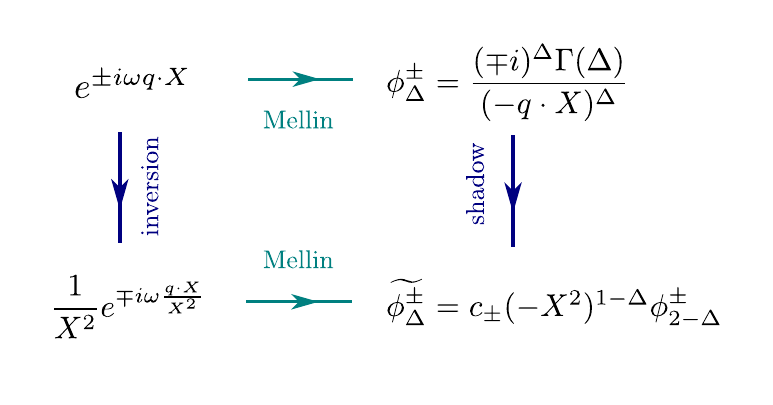}
\captionsetup{width=1\linewidth}
\caption{\small{Mellin and shadow transforms of plane waves. The Mellin transform of plane wave packets define (here scalar) conformal primary wavefunctions (CPW) \eqref{conf0}, which by design transform as primaries of dimension $\Delta$. Equally good CPW are provided by their shadow transform $\widetilde{\phi_\Delta}$, of dimension $2-\Delta$. The latter can be seen to be the Mellin transform of `shadow plane waves', which are related to the plane wave by a bulk spacetime inversion.}}
\label{fig:shadow}
\end{figure}

Since the shadow primary wavefunctions are as good as unshadowed primaries, one can as well decide to expand a bulk field into a shadow boost basis. This raises the question of what is the analog of the shadow transform in momentum basis: namely, what are `shadow plane wave packets'? A simple way to see that is by performing an inverse Mellin transform on the shadow conformal primary wavefunctions $\widetilde\phi$ given in \eqref{shadow_CPW}; using \eqref{InverseScalar}, this gives \cite{DonnayValsesia,Jorstad:2023ajr}
\begin{equation}
    \int_{c-i\infty}^{c+i\infty} \frac{d\Delta}{i2\pi} \frac{\,\Gamma(2-\Delta)(\mp i)^\Delta(-X^2)^{1-\Delta}}{(- q\cdot X)^{2-\Delta}}\omega^{\Delta-2}  =\frac{1}{X^2} e^{\mp i\omega \frac{q\cdot X}{X^2}}\,.
\end{equation}
We thus see that, from the bulk perspective, the shadow transform corresponds to implementing a coordinate inversion~\cite{Ferrara:1972uq,DonnayValsesia,Jorstad:2023ajr,Chen:2023tvj} 
\begin{equation}
X^\mu \to  - \frac{X^\mu}{X^2}\,,
\end{equation}
which exchanges null infinity with the light cone of the origin.
The different sequences of shadow and Mellin transforms on plane waves are summarized in Fig. \ref{fig:shadow}.

\subsection{Massive particles}
\label{sec:massive}
The reader will have noticed a bias in favor of massless particles in this report. However, most of the celestial machinery can be applied to massive case as well. This section collects some of the useful facts and formulae about massive conformal primary wavefunctions and how the celestial dictionary is implemented for the scattering of massive particles. The main references used are \cite{Pasterski:2016qvg,Pasterski:2017kqt}.

Massive particles follow timelike geodesics and exit (enter) flat space at future (past) timelike infinity $i^+$ ($i^-$) (see \cite{Winicour:1988aq} for a discussion). 
Let us denote $(y, z,\bar z)$ the Poincar\'e coordinates  of a three-dimensional hyperbolic space $H_{3}$ of metric\footnote{The coordinates considered here and differ from the ones used in section \ref{sec:BMS_everywhere}.}
\begin{equation}\label{Hd}
    ds^2_{H_3}=\frac{dy^2+dz d\bz }{y^2}\,,
\end{equation}
with $y>0$. The $SL(2,\mathbb C)$ isometry action on these coordinates is \cite{Pasterski:2016qvg}
\begin{equation}
\badat{3}
	&z \to z'=\frac{(az+b)(\bar c \bz + \bar d)+a \bar cy^2}{|cz+d|^2+|c|^2y^2}\,,\\
 &\bar z \to \bar z'=\frac{(\bar a\bz+\bar b)( c z+  d)+\bar a cy^2}{|cz+d|^2+|c|^2y^2}\,,\\
  &y \to  y'=\frac{y}{|cz+d|^2+|c|^2y^2}\,.
 \eadat
\end{equation}

At the boundary, located at $y=0$, the $H_3$ slice asymptotes the celestial sphere.
For an outgoing particle, one can choose an embedding map $\hat p^\mu:H_{3} \to \mathbb R^{3,1}$ for the upper hyperboloid via
\begin{equation}\label{phat}
	\hat p^\mu(y, z,\bz) = \big(\frac{1+y^2+| z|^2}{2y},\frac{\text{Re}(z)}{y},\frac{\text{Im}(z)}{y},\frac{1-y^2-|z|^2}{2y}
	\big)\,,
\end{equation}
where $\hat p^\mu$ is a unit timelike vector $\hat p^2=-1$. The isometry of \eqref{Hd} acts on $\hat p^\mu$ in a linear way, $\hat p^\mu(y', z',\bz')=\Lambda^\mu_{\,\,\nu}\hat p^\nu$.

The conformal primary wavefunction for a massive scalar\footnote{Spinning massive conformal primary wavefunctions have been studied in \cite{Law:2020tsg,Narayanan:2020amh,Iacobacci:2020por}.} is \cite{Pasterski:2016qvg} 
\begin{equation} \label{cpwmassive}
	\phi^\pm_{\Delta,m}(X;w,\bw)=\int_0^\infty \frac{dy}{y^{3}}\int d^2z \, G_{\Delta}(\hat p;w,\bw) \exp[\pm im \hat p\cdot X]\,,
\end{equation}
where $(w,\bw)$ denotes a point on the boundary of $H_{3}$ and the plus (minus) sign corresponds to the outgoing (incoming) case. $G_{\Delta}(\hat p; w,\bw)$ is the familiar bulk-to-boundary propagator that plays a key role in the AdS/CFT dictionary \cite{Witten:1998qj}
\begin{equation}
 G_{\Delta}(\hat p;w,\bw)=\left(\frac{y}{y^2+| z-w|^2}\right)^\Delta\,.
\end{equation}
 One can check that the massive primary \eqref{cpwmassive} satisfies the Klein-Gordon equation $(\Box-m^2)\phi_{\Delta,m}=0$ and transforms covariantly as a conformal primary. While these expressions suggestively remind AdS/CFT, it is important to notice the key difference with the field/operator map in AdS. Here, the mass $m$ is \emph{not} fixed by the conformal dimension $\Delta$.

Since \eqref{cpwmassive} provides us with the map from a single massive plane wave to a conformal wavefunction, we can now use it to recast any $N$-point scattering amplitude in momentum space $\mathcal A_N(p_j)$ of scalars of mass $ m_j$ into the celestial basis as follows \cite{Pasterski:2016qvg} :
\begin{equation}
    \mathcal M_N (\Delta_i,w_i,\bw_i) =\prod_{i=1}^N \int_0^\infty \frac{dy_i}{y_i^{3}}\int d^2z_i \, G_{\Delta_i}(\hat p_i; w_i,\bw_i)   \, \mathcal A_N ( m_j\hat p_j)\,,
    \label{massive_map}
\end{equation}
where $\hat p^\mu_j(y_j,z_j,\bz_j)$ is parametrized as in  \eqref{phat}.
This is the massive analog of the celestial map \eqref{Celestial S-matrix}. Each integral transform collects contributions from the mass shell data $\hat p_i$ and returns an insertion localized on the celestial sphere (at $(w_i,\bw_i)$). By construction, the celestial massive amplitude \eqref{massive_map} transforms as a  CFT$_2$ correlator of $N$ scalar primaries of conformal dimensions $\Delta_i$, i.e. under \eqref{SL2},
\begin{equation}
    \mathcal M_N (\Delta_i,w'_i,\bw'_i) =\prod_{i=1}^N  |cw_i+d|^{2\Delta_i}  \mathcal M_N (\Delta_i, w_i,\bw_i) \,.
\end{equation}

While the integral representation \eqref{cpwmassive} for a massive CPW is in practice the most convenient one to use, it is formally divergent for real mass $m$. To obtain a closed-form expression, one should evaluate the integral for a purely imaginary mass and then analytically continue it to real mass; this leads to \cite{Pasterski:2016qvg}
\begin{equation} \label{cpwmassive2}
	\phi^\pm_{\Delta,m}(X; w,\bw)=\frac{4\pi}{im} \frac{(\sqrt{-X^2})^{\Delta-1}}{(-q( w,\bw)\cdot X\mp i\ep)^{\Delta}}\, K_{\Delta-1}(im\sqrt{-X^2})\,,
\end{equation}
where $K_\alpha(x)$ denotes the modified Bessel function of the second kind. The above expression corresponds to a linear combination of two modified Bessel functions of the first kind which is such that it dies off exponentially as $e^{-m\sqrt{X^2}}$ for large $X^2$. It hence corresponds to a normalizable solution of the massive Klein-Gordon equation (i.e. it gives a finite Klein-Gordon norm).
Finally, notice that one can recover the massless conformal primary expression \eqref{conf0} by taking the $m\to 0$ limit of \eqref{cpwmassive2} (assuming Re$(\Delta)>1$)~\cite{Pasterski:2016qvg}.

\subsection{More on celestial amplitudes}
In the past years, a very large amount of celestial amplitudes have been studied, in several contexts: for massive and massless scalars~\cite{Pasterski:2016qvg,Cardona:2017keg,Nandan:2019jas,Law:2019glh,Law:2020xcf,Atanasov:2021cje,Chang:2021wvv,Guevara:2021tvr,Mizera:2022sln,Ren:2022sws,Ball:2023ukj}, gluons and gravitons~\cite{Schreiber:2017jsr,Nandan:2019jas,Banerjee:2019prz,Law:2019glh,Pate:2019lpp,Fotopoulos:2019vac,Banerjee:2020kaa,Fan:2020xjj,Casali:2020vuy,Law:2020xcf,Banerjee:2020zlg,Ebert:2020nqf,Banerjee:2020vnt,Arkani-Hamed:2020gyp,Kalyanapuram:2020aya,Fan:2021isc,Magnea:2021fvy,Gonzalez:2021dxw,Chang:2021wvv,Sharma:2021gcz,Himwich:2021dau,Fan:2021pbp,Guevara:2021tvr,Kapec:2021eug,Jiang:2021csc,Nastase:2021izh,Strominger:2021mtt,Mago:2021wje,Fan:2022vbz,Mizera:2022sln,Ghosh:2022net,Ren:2022sws,Fan:2022kpp,Monteiro:2022lwm,Hu:2022bpa,Bu:2022iak,Stieberger:2022zyk,Stieberger:2023fju,Ball:2022bgg,Adamo:2022wjo,Jorge-Diaz:2022dmy,Chang:2022seh,Banerjee:2023rni,Ren:2023trv}, with supersymmetry \cite{Fotopoulos:2020bqj,Jiang:2021xzy,Brandhuber:2021nez,Hu:2021lrx,Ferro:2021dub,Himwich:2021dau,Jiang:2021csc,Ahn:2021erj,Hu:2022bpa}, and in string theory \cite{Stieberger:2018edy,Chang:2021wvv,Donnay:2023kvm}. Celestial amplitudes in other signatures and backgrounds have been studied in \cite{Atanasov:2021oyu,Gonzo:2022tjm,deGioia:2022fcn,Crawley:2023brz} and loop corrections in \cite{Albayrak:2020saa,Gonzalez:2020tpi,Garcia-Sepulveda:2022lga,Costello:2022upu,Gu:2022maz,Bhardwaj:2022anh,Bittleston:2022jeq,Fernandez:2023abp,He:2023lvk,Donnay:2023kvm}. As we will comment at the end of this report, several interesting connections between celestial amplitudes and (ambi)twistor constructions have been investigated~\cite{Adamo:2014yya, Geyer:2014lca,Adamo:2019ipt,Bu:2021avc,Adamo:2021lrv,Adamo:2021zpw,Casali:2020uvr, Monteiro:2022xwq,Brown:2022miw,Mason:2022hly,Bittleston:2023bzp,Bu:2023cef}.
See also \cite{deBoer:2003vf,Cheung:2016iub,Pasterski:2016qvg,Pasterski:2017kqt,Cardona:2017keg,Ball:2019atb,Iacobacci:2022yjo,deGioia:2022fcn,Casali:2022fro,Gonzo:2022tjm,deGioia:2022fcn,Hu:2022txx,Costello:2023hmi,Sleight:2023ojm,Hu:2023geb,Hao:2023wln,Bu:2023cef,Chen:2023tvj,Jorstad:2023ajr} for connections between celestial and AdS correlation functions and \cite{Ogawa:2022fhy} for a relation to wedge holography.

\newpage
\section{Celestial CFT from asymptotic symmetries}
\label{chap:last}
Chapters \ref{chap:bms} and \ref{chap:charges} of this report were geared towards very classical subjects of General Relativity as they dealt with asymptotic structures of spacetimes, asymptotic symmetries, their associated Noether charges and flux-balance laws. In contrast, the main ingredients of a celestial holographic dictionary where presented in chapter \ref{chap:celestial} by means of the QFT language of scattering amplitudes. It is a priori not obvious that these two languages can be unified and how.  As a matter of fact, they lived a somewhat separate academic life for more than fifty years. The groundbreaking realization \cite{Strominger:2013jfa} that provided a bridge between these two topics can be summarized in the following few words: BMS symmetries constrain the gravitational $\mathcal S$-matrix. The canonical realization of this statement is the proof that the Ward identity associated with supertranslation symmetries is equivalent to Weinberg's (leading) soft graviton theorem \cite{He:2014laa}. In QFT, soft theorems provide universal factorization properties of scattering amplitudes which involve a soft (i.e. zero-energy) particle \cite{Low:1958sn,Burnett-Low,Weinberg:1995mt,Weinberg:1965nx,Gross-Jackiw,Jackiw}. This deep relationship between soft theorems and asymptotic symmetries carries over different types of theories and holds in different spacetime dimensions; see the reviews \cite{Strominger:2017zoo,McLoughlin:2022ljp} and references therein.

The goal of this chapter is to show how soft symmetries are realized at the level of the celestial CFT. More precisely, we will see how the conformal primaries we presented in sections \ref{sec:vacuum} and \ref{sec:BMSfluxes} and which carry information about extended BMS symmetries precisely lead to a set of strong constraints on celestial CFTs. To do so, we will start in section \ref{sec:btob} by specifying the relationship between asymptotically free quantum field operators and the radiative data of asymptotically flat spacetimes. We will then see in section \ref{sec:CCFT_currents} that the supermomentum and super angular momentum flux can be interpreted as $2d$ currents for the CCFT, among which one can find a celestial stress tensor. We will express the (OPE) constraints they implement on the `conformally soft' sector of the celestial hologram, which is defined by operators of integer conformal dimension. Finally, we will see in section \ref{sec:winfinity} that the story gets even richer, as an infinite tower of celestial currents can be concisely described by a certain $w-$symmetry group.

\subsection{From bulk to boundary fields (and back)}
\label{sec:btob}
In this section, we start by relating the gravitational fields of the Bondi expansion of section \ref{sec:Bondi} with momentum eigenmodes of the field operators. As anticipated, upon an expansion near $\mathscr I$, the radiative part of an asymptotically free graviton can be identified with the gravitational shear. The Kirchhoff-d'Adh\'emar formula allows the reverse procedure: it reconstructs a bulk field from its boundary value at null infinity.

\subsubsection{From bulk to boundary: large $r$ expansion}
Let us analyze the late-time behavior of a free massless scalar field close to future null infinity $\mathscr I^+$. The recipe to obtain its boundary value is simple (it has been used abundantly in the literature; see the review \cite{Strominger:2017zoo}):
start from the Fourier expansion of the bulk field $\Phi(X)$ \eqref{ModeExpansion} and rewrite it using retarded Bondi coordinates $(u,r,z,\bar z)$. Then evaluate the expression in the limit $r\to \infty$ with the stationary phase approximation. The dominant contribution forces the direction in momentum to be parallel to the position space direction towards which the particle is scattering\footnote{See e.g. exercise 3 in \cite{Strominger:2017zoo} for details.}. The resulting expression for a massless\footnote{See e.g. \cite{Campiglia:2015kxa} for the analogous formula near timelike infinity for a massive scalar.} scalar is $\Phi(X)\stackrel{r \to \infty}{\sim}r^{-1}\phi(u,z,\bar z)$, with the boundary value of the field given by
\begin{equation}\label{boundary field}
\phi(u,z,\bar z)= -\frac{i}{2(2\pi)^2 } \int_{0}^{+\infty}d \omega  \left[ a(\omega,z,\bar z) e^{ -i\omega u} - a(\omega,z,\bar z)^\dagger e^{+i\omega u}\right]\,.
\end{equation}
More generally, for a massless particle of spin $s$, the large $r$ behaviour goes like $\sim r^{s-1}$ (see e.g. \cite{Donnay:2022wvx} or Penrose's rest mass field equations \cite{Penrose:1965am}). In particular, following the same procedure for a spin-two field yields $h_{zz}\sim  r C_{zz}$ with \cite{He:2014laa}
\begin{equation}\label{saddleC}
C_{zz}(u,z,\bar z)= -\frac{i \kappa}{2(2\pi)^2 } \int_{0}^{+\infty}d \omega  \left[ a_+(\omega,z,\bar z) e^{ -i\omega u} - a_-(\omega,z,\bar z)^\dagger e^{+i\omega u}\right]\,,
\end{equation}
where $\kappa=\sqrt{32\pi G}$ and $C_{\bar z \bar z}=(C_{zz})^\dagger$ is obtained by replacing $a_\pm \to a_\mp$ in \eqref{saddleC}.
The same analysis can be performed for operators at the vicinity of past null infinity $\mathscr I^-$, mutatis mutandis. Notice that one can also perform a large-$r$ expansion on the conformal primary wavefunctions \eqref{varphi} (see \cite{Donnay:2018neh, Pano:2021ewd, Donnay:2022sdg} and \cite{Donnay:2022ijr} for detailed expressions obtained using the methods of regions).

Equation \eqref{saddleC} provides the bridge between the notions of a quantum field living at $\mathscr I^+$ and the asymptotic free data in the Bondi expansion of section \eqref{sec:BMS}. 
At this stage, a careful reader might notice that we have labelled the quantum field by the same name as the asymptotic shear $C_{zz}$, even though the latter was obtained from an asymptotic expansion in \emph{Bondi} gauge \eqref{Bondi gauge metric}, which is not compatible with the de Donder gauge $\Box h_{\mu \nu}=0$. While the map between Bondi and de Donder gauge is required in order to relate all metric components at different orders in the expansion (it has been explicitly done in \cite{Blanchet:2020ngx}), the important point for us is that their leading free data indeed coincides. In a $\mathscr I$-based approach to flat space holography, the boundary fields \eqref{boundary field}, \eqref{saddleC} are identified with `conformal Carrollian primaries'; see \cite{Donnay:2022wvx}. We will refer to them below as Carrollian operators.

\subsubsection{From boundary to bulk:  the Kirchhoff-d'Adh\'emar formula}
The Kirchhoff-d'Adh\'emar formula appeared in early works by Penrose \cite{Penrose1980GoldenON, Penrose:1985bww} and describes how to reconstruct the bulk free field from its boundary value. 
Following Ref. \cite{Donnay:2022wvx}, we invert the Fourier transform \eqref{boundary field} 
\begin{equation}
    \begin{split}
        a(\omega,z,\bar z) & = 4\pi i \int_{-\infty}^{+\infty} d u\, e^{i \omega u}\, \phi(u, z, \bar z) 
    \end{split} \label{fourier transform formula at scri}
\end{equation}
and inserting it into \eqref{ModeExpansion}; this gives
\begin{align}
    \Phi(X) &= \frac{i}{4\pi^2}\int \omega\, d\omega\, d^2 w\, \int_{-\infty}^{+\infty} d\tilde u\left[e^{i\omega(q\cdot X+\tilde u)} - e^{-i\omega(q\cdot X+\tilde u)}\right]\phi(\tilde u,w,\bar w) + \text{c.c.} \nonumber \\
    &= \frac{i}{4\pi^2}\int d^2 w\, \int_{-\infty}^{+\infty} d\tilde u \int_{-\infty}^{+\infty}d\omega\,\omega \, e^{i\omega(q\cdot X+\tilde u)}\,\phi(\tilde u,w,\bar w) + \text{c.c.} 
\end{align}
Using the identity 
$
    \int_{-\infty}^{+\infty} dx\, x\, e^{ ipx} = - 2\pi i\,  \partial_{p}\delta(p),
$
one obtains
\begin{align}
    \Phi(X) &=  \frac{1}{2\pi} \int\,d^2 w\, d \tilde u   \;   \Big[   \partial_{\tilde u}\delta\left(\tilde u+q\cdot X\right) \phi( \tilde u, w,\bar w) \Big] + \text{c.c.}
\end{align}
In modern terms, one would then say that the role of `the boundary-to-bulk propagator' is played by $\mathcal P(X;\tilde u,w,\bar w) =\partial_{\tilde u}\delta(q\cdot X + \tilde u)$. Integrating out the delta-function distribution, we obtain the (generalized) Kirchhoff-d'Adh\'emar formula \cite{Penrose:1985bww,Penrose1980GoldenON}
\begin{equation}\label{KA}
    \Phi(X) =  -\frac{1}{2\pi} \int\,d^2 w  \;   \;  \partial_{\tilde u}\phi(\tilde u = -q\cdot X, w,\bar  w)  + \text{c.c.}
\end{equation} 
which allows to reconstruct a bulk field $\Phi(X)$ (satisfying $\Box \Phi=0$) from its boundary value at $\mathscr{I}^+$. A similar relation holds for the boundary value at $\mathscr{I}^-$ and \eqref{KA} can be easily generalized to any spin \cite{Donnay:2022wvx}. 

\subsubsection{Null infinity and the celestial sphere}
We have expressed in equation \eqref{boundary field} above the relation between momentum basis operators and position basis (namely Carrollian) operators. Given the celestial map which exchanges the energy $\omega$ for the conformal dimension $\Delta$ via the Mellin transform, one can also directly relate the Carrollian operators to the celestial ones by composing a Fourier and a Mellin transform. This leads to\footnote{We drop here numerical factors and the $i\epsilon$ regulator for simplicity.}\cite{Pasterski:2021dqe,Donnay:2022aba,Donnay:2022wvx}
\begin{equation}
\badat{3}
\mathcal O_\Delta(z,\bar z) &=  \int_{-\infty}^{+\infty} du\, u^{-\Delta}\, \phi(u,z,\bar z) ,
\eadat \label{Btransform in two steps}
\end{equation}
where $\Delta = c+i\nu$, $c>0$. We have thus three equivalent ways of expressing boundary operators, depending on which of the three bases (momentum, position or boost) is selected \cite{Donnay:2022sdg}.

\subsection{CCFT currents}
\label{sec:CCFT_currents}
We are now ready to discuss an important set of operators in celestial holography,  the so-called `conformally soft' operators \cite{Donnay:2018neh}. As we will see shortly, the latter play a key role as they give rise to $2d$ `currents' for the celestial CFT. Conformally soft operators are celestial operators $\mathcal O_{\Delta}(z,\bz)$ (we drop the spin index here for simplicity) characterized by the fact that their conformal dimension takes specific, integer values
\begin{equation}
    \Delta \in 1-\mathbb Z_+\,.
\end{equation}
What kind of physics do those operators capture? Unless the reader has already transmuted all their physical intuitions into the celestial basis, it might be convenient to go back to the usual momentum basis. Consider an energy expansion $\mathcal O(\omega)\sim \mathcal O_n\, \omega^n$ for some $n$ as $\omega\to 0^+$. Then, using the celestial map and introducing  an upper cutoff $\Lambda$ in the intermediate steps, one can write \cite{Arkani-Hamed:2020gyp}
\begin{equation}
    \mathcal O_\Delta 
    =\int_0^\infty \omega^{\Delta-1}\mathcal O(\omega)\,d\omega
    =\mathcal O_n\int_0^\Lambda \omega^{\Delta+n-1}d\omega+\int_\Lambda^\infty \omega^{\Delta-1}\mathcal O(\omega)\,d\omega =\mathcal O_n\, \frac{\Lambda^{\Delta+n}}{\Delta+n}+\text{regular}
\end{equation}
and thus
\begin{equation}\label{conf_soft}
    \lim_{\Delta \to -n}(\Delta+n)\mathcal O_\Delta  = \mathcal O_n\,.
\end{equation} 
Therefore, the $\Delta \to -n$ limit of a celestial operator selects the $\mathcal O(\omega^n)$ term in an expansion around zero energy.

In particular, considering the emission of a massless scalar particle in a given scattering event, if in the energetically soft limit $\omega \to 0$ the (leading) soft theorem gives a behaviour like $1/\omega$ for the emission amplitude, then the conformally soft limit is $\Delta \to 1$.
More generally, sub$^{n}$-leading (with $n\neq 0$) soft theorems single out the conformal dimension $\Delta=0,-1,-2,\dots$ (see \cite{Guevara:2019ypd,Guevara:2021abz,Nandan:2019jas,Pate:2019mfs,Adamo:2019ipt,Puhm:2019zbl}). There are several ways of seeing the existence of these operators; approaches based on null states and global conformal multiplets have been presented in \cite{Pasterski:2021fjn,Pasterski:2021dqe,Banerjee:2020vnt,Banerjee:2020kaa,Banerjee:2020zlg,Banerjee:2019aoy,Banerjee:2019tam}.
The goal of this section is to show that the first celestial currents in this tower are naturally singled out by the soft BMS fluxes which were presented in section \ref{sec:BMSfluxes}.  

\subsubsection{Supertranslation current}
As already mentioned, one of the foundational results that set the stage for celestial holography was the realization that Weinberg's leading soft graviton theorem could be reformulated as the Ward identity\footnote{The equivalence between Ward identities associated to asymptotic symmetries and soft graviton theorems is reviewed in \cite{Strominger:2017zoo,Pasterski:2021rjz,Raclariu:2021zjz,McLoughlin:2022ljp}.} arising from the insertion of a (somehow peculiar) Kac-Moody current $P_z(z,\bar z)$, called the `supertranslation current' \cite{Strominger:2013jfa,He:2014laa}. In the conventional momentum basis, this Ward identity is
\begin{equation}
\big\langle \mathrm{out} | P_z^+ \mathcal S -\mathcal S P_z^-|\mathrm{in} \big\rangle = \sum_{k=1}^{n} \frac{\,\omega_k}{z-z_k} \big\langle \mathrm{out} |\mathcal S |\mathrm{in} \big\rangle \,,
\label{WI_Pz}
\end{equation}
where $P_z^\pm$ denotes the $\mathscr I^\pm$ part of the current. Focusing on the $\mathscr I^+$ part (similar expressions hold for $P_z^-$), 
$P_z\equiv P_z^+$ was identified in \cite{Strominger:2013jfa,He:2014laa} to be the first descendant of the leading soft graviton operator $\mathscr N^{(0)}_{zz}$ in \eqref{soft_operators}, which we recall is a primary of conformal weights $(h,\bar h)=(\frac{3}{2},-\frac{1}{2})$. Making use of the superrotation covariant derivative \eqref{derivative operators conformal}, we can write
\begin{equation}
P_z=\frac{1}{16\pi G}\mathscr D_\bz \mathscr N^{(0)}_{zz}\,
\end{equation}
and check that it transforms as a $(\frac{3}{2},\frac{1}{2})$ conformal primary under the action of extended symmetries \cite{Himwich:2020rro,Donnay:2021wrk}
\begin{equation}
      \delta_{(\mathcal{T}, Y)} P_z= \left(Y^z \partial_z + Y^\bz \partial_\bz + \frac{3}{2} \partial_z Y^z + \frac{1}{2} \partial_\bz Y^\bz\right)  P_z\,;
\end{equation}
see Table \ref{table:weights} for a summary of conformal weights. Its complex conjugate $P_{\bar z}$ has weights $(\frac{1}{2},\frac{3}{2})$ and corresponds to the insertion of a soft graviton with opposite helicity.  Comparing with \eqref{PDN}, one deduces that the relationship with the soft supermomentum flux is
\begin{equation}
\mathcal P^{soft}=2\mathscr D_\bz P_z=2\mathscr D_z P_\bz\,.
\end{equation}

From the argument given above (around \eqref{conf_soft}), we have seen that the leading soft pole as $\omega \to 0$ corresponds on the celestial operator side to the conformally soft limit $\Delta \to 1$. One can therefore expect that the operator 
\begin{equation}
P_z(z,\bz)=\lim_{\Delta \to 1}(\Delta-1)\partial_\bz \mathcal O_{\Delta,+2} (z,\bz)
\end{equation}
should encode Weinberg's leading soft graviton theorem on the CCFT~\cite{Donnay:2018neh}. Indeed, it has been checked explicitly that celestial amplitudes obey the following `conformally soft theorem'\footnote{See \cite{Pate:2019mfs} and references thereof for conformally soft theorems in gauge theory.} associated with supertranslations~\cite{Fotopoulos:2019vac,Puhm:2019zbl,Adamo:2019ipt} ()
\begin{equation}
\big\langle P_z(z,\bar{z}) \,\mathcal O_1\dots \mathcal O_n\big\rangle = \sum_{k=1}^{n} \frac{1}{z-z_k} \big\langle \mathcal O_1\dots\mathcal O_{\Delta_k+1}(z_k,\bar z_k)\dots  \mathcal O_n \big\rangle \,,
\label{leading soft}
\end{equation}
where we used the shorthand notation $\mathcal O_i \equiv \mathcal O_{\Delta_i,J_i}(z_i,\bar z_i)$. The shift of one unit for $\Delta_k$ in the right-hand side (rhs) of \eqref{leading soft} simply results from the presence of $\omega$ in the rhs of \eqref{WI_Pz} (after Mellin transform) and is expected from the way translations act on celestial operators (see \eqref{poin}). While unfamiliar from a standard CFT perspective, we see that the presence of supertranslation symmetry imposes that celestial correlators should obey the constraints \eqref{leading soft}, where the insertion of the supertranslation current leads to flow $(h,\bar h) \to (h+\frac{1}{2},\bar h+\frac{1}{2})$.

\subsubsection{Celestial stress tensor}
The subleading soft graviton theorem that was inferred from the Ward identity associated with BMS superrotation symmetries \cite{Kapec:2014opa,Cachazo:2014fwa} famously gave rise to the identification of $(2,0)$ current on the CCFT, a `celestial stress tensor'. As already mentioned at the opening of section \ref{sec:shadow}, it was first identified to be a certain non-local expression of the subleading soft graviton operator $\mathscr N^{(1)}$ \cite{Kapec:2016jld}. This non-local integral was later identified with the shadow transform of a conformally soft operator of dimension $\Delta=0$~\cite{Donnay:2018neh,Fotopoulos:2019vac}, namely
\begin{equation}
T_{zz}(z)=\int d^2w \frac{1}{(z-w)^4}\lim_{\Delta \to 0}\Delta \mathcal O_{\Delta,-2} (w,\bw)\,.
\end{equation}
Since it is the shadow transform of a $(h,\bar h)=(-1,1)$ primary, it carries indeed
conformal weights $(1-h,1-\bar h)=(2,0)$.
 CCFT Virasoro-Ward identities implied by superrotations were shown to give
\cite{Kapec:2016jld,Cheung:2016iub,Fotopoulos:2019tpe,Fotopoulos:2019vac}:
\begin{equation}
\big\langle T(z)\,\mathcal O_1\dots \mathcal O_n\big\rangle  =\sum_{k=1}^n \Big[ \frac{\partial_{z_k}}{z-z_k}  + \frac{h_k}{(z-z_k)^2} \Big] \, \big\langle \mathcal O_1\dots \mathcal O_n \big\rangle\,, \label{TO}
\end{equation} 
where $T(z)\equiv T_{zz}$. A similar Ward identity holds for the anti-holomorphic stress tensor $\bar T(\bar z)\equiv T_{\bz\bz}$ of weights $(0,2)$, which arises from the shadow transform of a $\Delta=0$, $J=+2$ celestial operator.
The fundamental consequence of the Ward identity \eqref{TO} is that celestial operators $\mathcal O_i$ are now promoted to Virasoro primary fields!

The object of the gravitational phase space which gives rise to the celestial stress tensor is the super-angular momentum flux $\mathcal J_z$ given in \eqref{angular_flux}. Looking at Table \ref{table:weights}, one recalls that the latter carries conformal weights $(2,1)$.  The celestial stress tensor arises from the super-angular momentum flux via~\cite{Donnay:2021wrk,Donnay:2022hkf}
\begin{equation}
\label{celestial T}
T(z)= \frac{i}{8\pi G}\int d^2 w\,  \frac{1}{z-w}  \left(-{\mathscr D}_w^3 \mathscr N_{\bar w \bar w}^{(1)} + \frac{3}{2} \mathscr C_{ww} \mathscr{D}_w \mathscr{N}^{(0)}_{\bar w \bar w}+ \frac{1}{2}\mathscr{N}^{(0)}_{\bar w \bar w} \mathscr{D}_w \mathscr C_{ww} \right)\,.
\end{equation}
Setting the Goldstone modes $N^{vac}_{AB}$ and $\mathscr C$ to zero, one recovers the original expression given in \cite{Kapec:2016jld} (namely the first term in the rhs of \eqref{celestial T} with  $\mathscr D \to \partial$). The new terms appearing in \eqref{celestial T}  take into account the vacuum structure in the presence of superrotations. Those subtleties are more than a mere technicality: indeed, it was shown \cite{Donnay:2022hkf,Pasterski:2022djr} that these corrections to the stress tensor exactly account for one-loop corrections of the subleading soft graviton theorem computed by Bern et al.~\cite{Bern:2014oka}!

\subsubsection{CCFT OPEs from symmetries}
We now turn to the constraints implied by asymptotic symmetries on operator product expansions (OPEs) in celestial CFT.  
The special kinematic configuration of identical momentum directions ($p_1 \parallel p_2$) corresponds to the insertion of operators at coincident points on the celestial sphere (since $p_1\cdot p_2 \propto z_{12}\bz_{12}$ with $z_{ij}=z_i-z_j$). Therefore,  collinear limits of celestial amplitudes carry information about CCFT OPEs. Such collinear (and also multi-collinear) limits of celestial amplitudes and OPE coefficients have been studied in~\cite{Fotopoulos:2019tpe,Pate:2019lpp,Fan:2019emx,Fotopoulos:2019vac,Fotopoulos:2020bqj,Banerjee:2020kaa,Fan:2020xjj,Ebert:2020nqf,Himwich:2021dau,Atanasov:2021cje,Guevara:2021tvr,Banerjee:2020zlg,Banerjee:2020vnt,Banerjee:2021dlm,Ren:2022sws,Ball:2023sdz}. Here we will rather focus on what we can already learn about OPE coefficients from the perspective of (asymptotic) symmetries.\\
\newpage

\noindent \textcolor{pansypurple}{\gr{OPEs of supertranslations and superrotations\,\,}} The celestial Ward identities \eqref{leading soft} and \eqref{TO} give rise to the following OPEs for the celestial theory\footnote{We denote here $P(z,\bar z)=P_z(z,\bar z).$}:
\begin{equation}\label{OPE1}
 \badat{2}
  & P(z, \bar{z}) \mathcal O_{h, \bar{h}}(w, \bar{w}) \sim \frac{1}{(z-w)} \mathcal O_{h+ \frac{1}{2}, \bar{h}+ \frac{1}{2}}(w, \bar{w})\,,\\
        &T(z) \mathcal O_{h, \bar{h}}(w,\bar{w}) \sim \frac{h}{(z-w)^2} \mathcal O_{h, \bar{h}}(w,\bar{w})+\frac{1}{(z-w)} \partial_w \mathcal O_{h, \bar h}(w,\bar{w}), 
\eadat
\end{equation} 
together with the antiholomorphic counterpart. As we have seen, they encode the leading and subleading conformally soft graviton theorems and arise from the soft part of BMS super (angular) momentum fluxes. 

The OPEs between these CCFT currents can be derived from the BMS flux algebra that we have derived in section \ref{sec:BMSfluxes}. Using standard arguments (see e.g. \cite{Schottenloher}), one can indeed deduce the singular parts of the OPEs between the operators that are associated with BMS soft fluxes from their commutation relations. Using the flux algebra \eqref{flux algebra}, one can derive (see \cite{Donnay:2021wrk} for details)
\begin{equation}
    \badat{2}
      &  P(z, \bar{z}) P(w, \bar{w})  \sim 0, \\
      &  T(z) P(w, \bar{w}) \sim \frac{3/2}{(z-w)^2} P(w, \bar{w}) +\frac{1}{(z-w)} \partial_w P(w, \bar{w}),  \\
     &   \bar{T}(\bar{z})  P(w, \bar{w}) \sim  \frac{1/2}{(\bar z-\bar w)^2} P(w, \bar{w})+\frac{1}{(\bar{z}-\bar w)} \partial_{\bar{w}} P(w, \bar{w}), \\
   &     T(z) T( w)\sim  \frac{2}{(z-w)^2}T(w)+\frac{1}{(z-w)} \partial_w T(w), \\
      &  \bar{T}(\bar{z}) T( w) \sim 0.
        \label{OPE2}
\eadat
\end{equation}
The above OPEs \eqref{OPE1}, \eqref{OPE2} can alternatively be obtained from collinear and conformally soft limits of celestial amplitudes \cite{Fotopoulos:2019tpe,Fotopoulos:2019vac}, showing the consistency of the framework. \\

\noindent \textcolor{pansypurple}{\gr{Graviton OPEs\,\,}} Remarkably, asymptotic symmetries fix the leading celestial OPE coefficients of gravitons~\cite{Pate:2019lpp}. One considers positive-helicity gravitons and the `holomorphic collinear limit' $z_{12}\to 0$ with $\bar z_1, \bar z_2$ fixed\footnote{This amounts to treating $z$ and $\bz$ as independent variables.}. The starting point is the following OPE expansion
\begin{equation}\label{OPEgravi}
 \badat{2}
        & \mathcal O_{\Delta_1, +2}(z_1,\bz_1) \mathcal O_{\Delta_2, +2}(z_2,\bz_2) \sim D(\Delta_1,\Delta_2) \frac{\bz_{12}}{z_{12}} \mathcal O_{\Delta_1+\Delta_2, +2}(z_2,\bz_2) + \dots\,,
\eadat
\end{equation} 
where the dots include contributions from $SL(2,\mathbb C$) descendants and primaries appearing at sub-leading order in the collinear expansion. The form of \eqref{OPEgravi} is fixed by $SL(2,\mathbb C$) and the leading soft graviton theorem (second line of \eqref{OPE1}). As shown in \cite{Pate:2019lpp} (see also \cite{Himwich:2021dau}), the sub-subleading soft graviton theorem further constrains the OPE coefficient in \eqref{OPEgravi} to obey a recursion relation that is solved by
\begin{equation}
    D(\Delta_1,\Delta_2)=-\frac{\kappa}{2}B(\Delta_1-1,\Delta_2-1)\,,
\end{equation}
with $B(x,y)$ the Euler beta function.
The OPE of opposite helicity gravitons can also be derived in a similar way.

\subsection{$w_{1+\infty}$ symmetries}
\label{sec:winfinity}
As \eqref{conf_soft} suggested at the beginning of this section, celestial amplitudes seem to obey a whole tower of Ward identities that are associated with poles of negative values of conformal dimension $\Delta$. Restricting as above to positive-helicity gravitons, a discrete family of conformally soft graviton operators\footnote{Note they have weights $
   (h,\bar h)=\left(\frac{k+2}{2},\frac{k-2}{2}\right)
$.} is defined as\cite{Strominger:2021lvk}
\begin{equation}\label{Hk}
   H^k(z,\bz):=\lim_{\varepsilon \to 0} \varepsilon \mathcal O_{k+\varepsilon,+2} \virg k=2,1,0,-1,\dots
\end{equation}
together with a consistently truncated anti-holomorphic mode expansion
\begin{equation}
   H^k(z,\bz)=\sum_{n=\frac{k-2}{2}}^{\frac{2-k}{2}}\frac{H_n^k(z)}{\bz^{n+\frac{k-2}{2}}}.
\end{equation}
The graviton celestial OPE \eqref{OPEgravi} can be generalized to include anti-holomorphic descendants ($n>0$) \cite{Pate:2019lpp}
\begin{equation}\label{OPEgravi2}
 \badat{2}
        & \mathcal O_{\Delta_1, +2}(z_1,\bz_1) \mathcal O_{\Delta_2, +2}(z_2,\bz_2) \sim -\frac{\kappa}{2} \frac{1}{z_{12}}\sum_{n=0}^\infty B(\Delta_1+n-1,\Delta_2-1)\frac{(\bar z_{12})^{n+1}}{n!}\bar \partial^n\mathcal O_{\Delta_1+\Delta_2, +2}(z_2,\bz_2)\,.
\eadat
\end{equation} 
Using \eqref{OPEgravi2}, one can derive the OPE of the conformally soft gravitons \eqref{Hk} and from there commutators of the modes $H^k_n$ (see \cite{Guevara:2021abz}). This soft current algebra closes but takes a somewhat complicated form. It can be drastically simplified by defining the new soft graviton modes~\cite{Strominger:2021lvk}
\begin{equation}
   w^p_n=\frac{1}{\kappa}(p-n-1)!(p+n-1)!H^{-2p+4}_n\,,
\end{equation}
where $p=1,\frac{3}{2},2,\frac{5}{2},\dots$
The latter form the following algebra
\begin{equation}\label{walg}
  [ w^p_n,w^q_n]=\left[m(q-1)-n(p-1)\right]w^{p+q-2}_{m+n}\,.
\end{equation}
For arbitrary integer $m$, this algebra is known as the `$w_{1+\infty}$ algebra'; see the review \cite{Pope:1991ig}. It includes the $c=0$ Virasoro algebra as the $p=2$ subalgebra. Here $m$  is restricted to the values $1-p\leq m \leq p-1$ (this restriction comes from $\frac{k-2}{2}\leq m \leq \frac{2-k}{2}$), hence the tower of conformally soft currents generated by $w_m^p(z)$ close the wedge subalgebra of $w_{1+\infty}$~\cite{Strominger:2021lvk}. 

The appearance of the $w_{1+\infty}$ algebra is in fact very natural from a twistor theory point of view as it is linked to self-dual gravity via Penrose's non-linear graviton construction \cite{Boyer:1985aj,Park:1989fz,Park:1989vq,Mason,Adamo:2021lrv,Adamo:2021zpw}. 
We thus see that the celestial road has brought us very close to Newman and Penrose's heaven constructions.
The realization of this algebra from an asymptotic gravitational phase space perspective was discussed in \cite{Freidel:2021dfs,Freidel:2021ytz} by means of a truncation at subleading orders in the large $r$ expansion.
While several important questions remain open, these results\footnote{See Refs. \cite{Himwich:2021dau,Jiang:2021csc,Ball:2021tmb,Mago:2021wje,Adamo:2021lrv,Adamo:2021zpw,Ahn:2021erj,Mago:2021wje,Ren:2022sws,Bu:2022iak,Monteiro:2022xwq,Freidel:2022skz,Hu:2022txx,Hu:2023geb,Drozdov:2023qoy,Bittleston:2023bzp} for further works.} raise the hope that $w_{1+\infty}$ symmetries, and possibly their quantized version, provide an organizing principle for a fully-fledged celestial theory dual to quantum gravity.

\section{Outlook and further reading}
\label{chap:conclusion}
In this concluding section, we list some of the topics related to asymptotic symmetries and celestial holography which have not been covered in this review and associated further reading material.
\begin{itemize}[leftmargin=*]
     \item \emph{Lower and higher dimensions\,\,\,} This report focused on the case of spacetimes in $D=4$ dimensions. The status of BMS-type symmetries in higher spacetime dimensions $D>4$ is much less straightforward (and has been subject to some debate). On the one hand, asymptotic symmetries which preserve fall-off conditions where radiation scales as $\mathcal O(r^{\frac{2-D}{2}})$ (and which do not prevent the description of gravitational waves) reduce to Poincar\'e \cite{Hollands:2003ie,Hollands:2016oma}. On the other hand, the existence of soft factorisation theorems in any $D$ called for the existence of Ward identities associated with higher-dimensional generalization of (extended) BMS symmetries. This apparent tension was resolved (in even dimensions) by observing that infinite-dimensional enhancements can occur once the radiative boundary conditions are relaxed to be $\mathcal O(r^{-1})$ for all $D$ \cite{Kapec:2015vwa,Pate:2017fgt,Campoleoni:2020ejn} (see also \cite{Chowdhury:2022gib,Capone:2023roc,Fuentealba:2021yvo,Fuentealba:2022yqt} for related works). On the celestial side, primary wavefunctions and conformally soft operators can be constructed on the $d>2$ celestial sphere \cite{Pasterski:2017kqt,Adamo:2019ipt,Banerjee:2019aoy,Banerjee:2019tam,Pano:2023slc} and the role of the shadow transform is prominent~\cite{Kapec:2017gsg,Kapec:2021eug,Kapec:2022hih,Kapec:2022axw}. The case of $D<4$ is also interesting in its own right and in fact, BMS$_3$ symmetries were studied some time ago~\cite{Barnich:2006av,Barnich:2012aw,Barnich:2012rz,Barnich:2013yka,Bagchi:2015wna,Bagchi:2017cpu}. In $D=2$, top-down celestial holograms were constructed in \cite{Kapec:2022xjw} and in \cite{Kar:2022vqy,Kar:2022sdc,Rosso:2022tsv}, where the role of the holographic system is given by an ensemble of random matrices, namely a celestial matrix model.

     \item \emph{Carrollian holography \,\,\,} A complementary approach to celestial holography has emerged under the name of `Carrollian' holography. Carrollian\footnote{See \cite{Leblond,Duval:2014uoa} for the literary origin of this name.} geometries naturally describe null hypersurfaces, including null infinity (see e.g. \cite{Bekaert:2015xua,Bergshoeff:2017btm,Ciambelli:2019lap,Herfray:2021qmp,Hartong:2015xda,Figueroa-OFarrill:2021sxz,Bergshoeff:2022eog}) but also black hole horizons \cite{Donnay:2016ejv,Donnay:2019jiz,Penna:2018gfx,Adami:2021kvx,Freidel:2022vjq}. It is well-known that the conformal Carroll group is isomorphic to the BMS group \cite{Duval:2014lpa,Duval:2014uva}. Therefore, in the Carrollian approach to flat space holography, the dual theory is proposed to be a conformal Carrollian field theory that lives at $\mathscr I$. Several holographic aspects of such theories have been studied in \cite{Henneaux:1979vn,Bagchi:2009ca,Bagchi:2017cpu,Bagchi:2016geg,Bagchi:2019clu,Bagchi:2019xfx,Gupta:2020dtl,Henneaux:2021yzg,Bekaert:2022oeh,deBoer:2023fnj,Salzer:2023jqv,Nguyen:2023vfz,Campoleoni:2023fug}, including connections with the celestial dictionary \cite{Donnay:2022aba,Donnay:2022wvx,Jorstad:2023ajr} and in the inspiring perspective of a fluid/gravity correspondence~\cite{deBoer:2017ing,Ciambelli:2018wre,Ciambelli:2018xat,Campoleoni:2018ltl,Ciambelli:2020ftk,Petkou:2022bmz,Freidel:2022bai,Campoleoni:2022wmf}. Interestingly, Carroll geometries are not restricted to null infinity, but also naturally emerge at spatial and timelike infinities \cite{Figueroa-OFarrill:2022mcy,Figueroa-OFarrill:2021sxz,Borthwick:2023lye}, and are thus expected to provide a consistent framework in the context of a generic gravitational scattering problem.

    \item \emph{Twistor constructions\,\,\,} As already mentioned, the language of twistor geometry is closely related to the spirit of celestial holography, and the appearance of the $w_{1+\infty}$-algebra is very natural in this context.  It is therefore not surprising that twistor methods have been so fruitful in unveiling certain aspects of celestial amplitudes. 
The frameworks of twistor strings and ambitwistor strings (see \cite{Geyer:2022cey} for a review) adapted to null infinity can be used to derive explicit formulae for the tree-level gravitational $\mathcal S$-matrix. In particular, (conformally) soft limits can be easily studied in this 
    context \cite{Adamo:2014yya, Geyer:2014lca,Adamo:2019ipt,Adamo:2021lrv,Adamo:2021zpw}, and OPE of the CCFT can be directly computed from the worldsheet OPE of vertex operators\cite{Adamo:2021zpw,Bu:2021avc}; see also for\cite{Casali:2020uvr, Monteiro:2022xwq,Brown:2022miw,Mason:2022hly,Bittleston:2023bzp,Bu:2023cef} for further twistor-related constructions. Remarkably, a certain generalization of twisted holography \cite{Costello:2018zrm} has allowed to provide a top-down celestial hologram. The dual pair involves, in the bulk, certain models of self-dual gravitational (or gauge) theories with a $2d$ chiral algebra living on a Riemann sphere \cite{Costello:2022upu,Costello:2022jpg,Costello:2022wso,Costello:2023hmi}.

    \item \emph{Symmetries in the sky \,\,\,} One of the most exciting aspects of asymptotic symmetries of flat spacetimes is that they have associated observable effects! Indeed, a peculiar prediction of GR is that, besides the well-known oscillatory contributions to gravitational waves strain, there exist non-oscillatory contributions, the so-called `gravitational memory' effects \cite{mem1,Christodoulou:1991cr,Blanchet:1992br,gravmem3}. Among those, the displacement memory is in one-to-one with correspondence with the existence of BMS supertranslation symmetries \cite{Ashtekar:2014zsa,Strominger:2014pwa} and has strong chances to be measured in the near future; see \cite{Grant:2022bla,Goncharov:2023woe} for prospects. The enlarged BMS-type of symmetries have predicted novel types of memory effects~\cite{Pasterski:2015tva,Nichols:2017rqr,grantnichols,Compere:2018ylh,Mao:2018xcw,Seraj:2022qyt,Godazgar:2022pbx}, which promises a rich future dialogue with the community of gravitational wave observations.
\end{itemize}

This report has accounted for only a small portion of the considerable research effort that has been undertaken from a wide variety of approaches. The above points are only a few of a much larger list of items. Celestial holography is a rapidly growing research field and many of the fascinating results that have been obtained lie beyond the scope of this report.  We refer the reader to the reviews and lecture notes \cite{Raclariu:2021zjz,Pasterski:2021rjz,McLoughlin:2022ljp,Pasterski:2021raf,Pasterski:2023ikd} for complementary material on celestial holography, including a focus on the amplitude side of the program. \\

While several features of celestial holography can be studied independently of what has been presented here, we have the firm conviction that the `asymptotic symmetry point of view' will never cease to play a prominent role. Since the works of Bondi and collaborators, more than fifty years ago, asymptotic symmetries have repeatedly disconcerted researchers by their subtlety and surprised them by their breadth of applications. By providing a genuine symmetry organizing principle for the sought-for hologram to quantum gravity in flat spacetime, they have paved a way to celestial holography.

\subsection*{Acknowledgments}
I would like to thank Shreyansh Agrawal, Agnese Bissi, Erfan Esmaeili, Adrien Fiorucci, Gaston Giribet, Hern\'an Gonz\'alez,  Carlo Heissenberg, Yannick Herfray, Kevin Nguyen, Sabrina Pasterski, Andrea Puhm, Francisco Rojas, Romain Ruzziconi, Andrew Strominger and Beniamino Valsesia for collaborations on celestial holography. I also thank Shreyansh Agrawal, Yannick Herfray and Beniamino Valsesia for their comments on the manuscript.
L.D. is supported by the European Research Council (ERC) Project 101076737 -- CeleBH. Views and opinions expressed are however those of the author only and do not necessarily reflect those of the European Union or the European Research Council. Neither the European Union nor the granting authority can be held responsible for them.
L.D. is also partially supported by INFN Iniziativa Specifica ST\&FI.
\subsection*{References}
\bibliographystyle{style}
\renewcommand\refname{\vskip -1.3cm}
\bibliography{references}

\end{document}